\documentclass[a4paper]{article}

\usepackage{jheppub} 
                     
\usepackage{graphicx,color}
 \usepackage{bm}
   \usepackage{amsmath}
    \usepackage{amssymb}    
     \usepackage{pifont}
      \usepackage{simplewick}
\usepackage{tikz}
\usetikzlibrary{decorations.pathmorphing}
\usetikzlibrary{decorations.markings}
\usetikzlibrary{calc}
\usepackage[most]{tcolorbox}
\usepackage{rotating}

\usepackage{standalone}
\usepackage{slashed}

\newcommand{\Ds}{\displaystyle}

\newcommand{\nn}{\nonumber}

\newcommand{\Tr}{\mathrm{Tr}}
\newcommand{\sign}{\text{sign}}

\renewcommand{\(}{\left(}
\renewcommand{\)}{\right)}
\renewcommand{\[}{\left[}
\renewcommand{\]}{\right]}

\renewcommand{\vec}[1]{\bm{#1}}
\newcommand{\fnot}[1]{\slashed{#1}}

\title{Transverse momentum dependent operator expansion at next-to-leading power}

\author[a]{Alexey~Vladimirov}
\author[a]{Valentin~Moos}
\author[b]{Ignazio~Scimemi}

\affiliation[a]{Institut f\"ur Theoretische Physik, Universit\"at Regensburg, D-93040 Regensburg, Germany}
\affiliation[b]{Departamento de Física Teórica and IPARCOS, Facultad de Ciencias Físicas, Universidad Complutense Madrid, Plaza Ciencias 1, 28040 Madrid, Spain}

\emailAdd{
alexey.vladimirov@physik.uni-regensburg.de, 
valentin.moos@physik.uni-regensburg.de, 
ignazios@ucm.es}

\abstract{
We develop a method of transverse momentum dependent (TMD) operator expansion that yields the TMD factorization theorem on the operator level. The TMD operators are systematically ordered with respect to TMD-twist, which allows a certain separation of kinematic and genuine power corrections. The process dependence enters via the boundary conditions for the background fields. As a proof of principle, we derive the effective operator for hadronic tensor in TMD factorization up to the next-to-leading power $(\sim q_T/Q)$ at the next-to-leading order for any process with two detected states. 
}

\begin{document} 

\maketitle 

\section{Introduction}
The transverse momentum dependent (TMD) factorization approach has a long history that started from the ``DDT formula'' \cite{Dokshitzer:1978hw}, and Collins-Soper-Sterman resummation \cite{Collins:1981uk,Collins:1984kg}. Nowadays, the TMD factorization approach is a well-developed framework that consistently describes various processes in terms of TMD distribution functions (for a review of the present status and latest development, see \cite{Angeles-Martinez:2015sea,Bacchetta:2016ccz,Scimemi:2019mlf}). Even though the recent achievements, the TMD factorization theorem still lacks the systematicness of collinear factorization theorems. The main reason is that in its heart, the collinear factorization is based on the method of the operator product expansion \cite{Wilson:1972ee}, which allows the first-principles treatment, independently of any hadronic model. The absence of a similar systematic approach for the TMD factorization raises a series of problems, especially when extending the formalism beyond the leading power approximation. In this work, we propose a formal derivation of the TMD operator expansion, overcoming the present {\it status quo}. Structurally, it is similar to the light-cone operator product expansion \cite{Anikin:1978tj,Balitsky:1990ck} for some specific (and broadly known) cross-sections.

The TMD factorization describes inelastic processes with two-detected states (initial or final). The main examples are the Drell-Yan (DY) process, semi-inclusive deep-inelastic scattering (SIDIS), and semi-inclusive annihilation (SIA). The information about the nonperturbative structure is stored in the TMD distribution functions (it could be TMD parton distributions, TMD fragmentation function, TMD jet-function, etc.).  The order parameter of the TMD factorization is $q_T/Q$, where $Q$ is the momentum of the hard probe and $q_T$ is its transverse component. The latest global analyses \cite{Scimemi:2017etj,Scimemi:2019cmh,Bacchetta:2019sam} demonstrate that the  TMD factorization is valid for $q_T/Q\lesssim 0.25$. Beyond that limit, the TMD factorization curve quickly deviates from the experimental measurements. To extend the region of TMD factorization, one needs to incorporate power corrections. 

In the past years, computations of power correction for TMD factorized cross-sections have been made by several groups, see \cite{Balitsky:2017gis,Nefedov:2018vyt,Ebert:2018gsn,Balitsky:2020jzt,Balitsky:2021fer,Inglis-Whalen:2021bea,Hu:2021naj}.  These computations required an update of existing methods. In this work, we go further and present a different approach, which is a natural extension of techniques used for studies of power corrections in the collinear factorization, and which has not been discussed in the framework of TMD factorization yet (as far as we know). Namely, we use the background field method and derive the factorization directly in the space of field functionals (operators). For that reason, we call the technique -- {\it TMD operator expansion}. Starting from the definition of the QCD Lagrangian, we re-derive some known results such as the leading power (LP) TMD factorization, and we proceed extending it to the next-to-leading power (NLP). Since the computation is made at the level of operators, the results can be applied to other cases (partly extending the present discussion), such as processes with jets or factorization with generalized TMD distributions (GTMDs). Many solutions and more minor results derived in this paper are already known or discussed in the literature, but the operational framework presented here is totally new.

There are several sources of power corrections to the factorization theorem. We classify them as follows:
\begin{itemize}
\item power suppressed contributions to the cross-section due to convolution between hadronic and leptonic tensors or phase-space factors. Usually, these corrections can be accounted exactly, and in any case, should not be mixed with QCD corrections. We solely concentrate on the hadronic tensor, and we do not consider these corrections in the present work;
\item corrections in the values of kinematic parameters that flow with the change of the ratio $q_T/Q$. For example, the values of effective momentum fractions in DY process are $x_{1,2}=q^\pm/P_{1,2}^\pm=x^{\text{Bj}}_{1,2}\sqrt{1+q_T^2/Q^2}$. Often, one expands these variables into series, generating a tail of power suppressed contributions. Such expansions are unnecessary and even harmful since the values of kinematic variables are unique for all powers, and their exact accounting supports the frame-independence of the final expression. We make the computation in position space, which allows us to identify values of kinematic parameters unambiguously from the definition of the hadronic tensor;
\item \underline{\textit{kinematic} power corrections}. These are contributions of a higher power with the operator content of lower power terms. The Wandzura-Wilczek relations \cite{Wandzura:1977qf} provide a famous example, which can also be generalized to the TMD case \cite{Bastami:2018xqd}. These corrections inherit the structure of lower power terms. They are essential for the restoration of global properties violated by the factorization theorems, such as electromagnetic (EM) gauge invariance, translation and frame independence, etc. We demonstrate that NLP kinematic corrections restore the EM gauge invariance of hadronic tensor up to $q_T^2/Q^2$, as expected;
\item  \underline{\textit{genuine} power corrections}  incorporate new operators, and hence new TMD distributions. 
\end{itemize}
In our work, we study only \textit{QCD specific} power corrections, i.e., kinematic and genuine. The separation of kinematic and genuine power corrections is an important and non-trivial task \cite{Braun:2012hq}. The most efficient way to solve it consists of ordering the operators with respect to their evolution properties. For the collinear distributions, it implies the separation of operators with respect to the \textit{geometrical twist}. For TMD distributions, we introduce the notion of \underline{{\it TMD-twist}}, such that TMD distributions with different TMD-twist have separate evolution, and thus their matrix elements are independent observables.

The essential feature of the TMD factorization is the appearance of infinite light-like gauge links \cite{Belitsky:2002sm,Boer:2003cm}. They are responsible for all distinctive features of TMD distributions, such as rapidity divergences, nonperturbative evolution, T-odd distributions, and process dependence. The reverse of the medal is that one cannot derive TMD factorization from a local operator expansion. This fact leads to the absence of such an important ordering criterium as the twist of the TMD operator. So far, that has not been a problem, since the leading term of the TMD factorization can be derived with the analysis of Feynman diagrams \cite{Collins:2011zzd}, or using the soft-collinear effective theory (SCET) approach \cite{Echevarria:2011epo}, and it does not need the systematization of TMD operators. However, dealing with power corrections requires the definition of some ordering for the operator basis. In this work, we introduce the twist of the TMD operator (or TMD-twist) given by a pair of numbers, which are geometrical twists of collinear substructures of the TMD operator. This definition of TMD-twist allows for a certain separation of kinematic and genuine contributions of power corrections.

The main aim of this work is the development of a theoretical basis for TMD factorization beyond the leading power. To keep the discussion as general as possible, we perform all computations at the level of operators without specifying a process or referring to cross-sections. The operator representation allows us to keep expressions relatively compact and avoid long algebraic structures that appear at the TMD distributions level. The expressions for  NLP cross-sections will be made in subsequent publication.

The background field method is particularly suitable for the computation of power corrections. For applications of this method to the collinear factorization see e.g. \cite{Balitsky:1987bk,Braun:2011dg,Braun:2012hq,Scimemi:2019gge}. This method computes the operator expansion directly from the function integral, avoiding any matching procedure typical for many approaches. It allows keeping track of the internal structure and the operators' relations at each evaluation step. The computation is naturally performed in position space (although it can easily be turned to momentum space), simplifying the NLP computations. In the case of TMD factorization, the background field Lagrangian \cite{Abbott:1980hw,Abbott:1981ke} must be updated to the case of two independent background fields (we call it a composite background field), which is done in app.~\ref{app:Sint}.

The paper is split into two logical parts. The first part includes sections \ref{sec:gen-gen}, \ref{sec:process}, \ref{sec:general-structure}, and it provides a review of the factorization method in general terms. In particular, sec.~\ref{sec:gen-gen} introduces the main definitions, such as the definition of hadronic tensors, the composite background field, counting rules, and the notion of effective field operators. Sec.~\ref{sec:process} is devoted to the problem of boundary conditions and gauge fixation for background field. We demonstrate that the choice of boundary conditions and an adequate gauge fixing are related to the analytical properties of generating functions, which in turn are tied to the underlying process. Accounting for these properties leads to the famous process dependence at the level of TMD operators. Sec.~\ref{sec:general-structure} is devoted to the general discussion about the structure of power and perturbative expansion. In this section, we introduce the concept of TMD-twist and TMD operator expansion.
The second part consists of sections \ref{sec:tree-gen}, \ref{sec:NLO}, \ref{sec:soft-overlap}, \ref{sec:TMD} and \ref{sec:recombination}, and it is devoted to the computation of TMD factorization at NLP and NLO. For  pedagogical reasons, the computation is presented with many details. The tree order and definition of all operators is given in sec.~\ref{sec:tree-gen}. Sec.~\ref{sec:NLO} presents the NLO computation of the hard coefficient function. The problem of soft overlap and subtraction of overlap region is discussed in sec.~\ref{sec:soft-overlap}. Sec.~\ref{sec:TMD} considers the properties of TMD operators and it is split into subsections \ref{sec:rapdiv}, \ref{sec:UV}, \ref{sec:rap+UV}  that treat respectively  rapidity divergences, ultraviolet (UV) divergences and the renormalization of TMD operators. Finally, in sec.~\ref{sec:recombination}, we demonstrate the cancellation of divergences among elements of TMD factorization, fix the scheme dependence and derive the evolution equations for NLP TMD operators.

The text is supplemented by appendices, which contain additional technical details. Appendix \ref{app:Sint} presents the expression for the QCD Lagrangian in the composite background field. Appendix \ref{app:diag4} demonstrates the technique of computation in the composite background field. App.~\ref{app:evol-momentum} contains the expressions for evolution kernels in momentum space.

One of the difficulties in writing about power corrections comes from the notation. On the one hand, the topic is complicated and requires accurate and exhaustive formulation. On the other hand, all-inclusive writing conceals the general structure of the expression, which is often simple. For that reason, we accept the following convention: we drop the parts of notation that are not important in the present context, such as, color and spinor indices, arguments, etc. However, we keep all essential elements, and a cautious reader should be able to restore all missed components if needed. 

\section{General structure of TMD factorization}
\label{sec:gen-gen}

In this section we provide the notation and the basic definitions that are used in this work.

\subsection*{General setup}
We study the hadronic tensors for Drell-Yan, $h_1+h_2\to \gamma^* +X$, semi-inclusive deep inelastic scattering (SIDIS), $h_1+\gamma^*\to h_2+X$ and semi-inclusive annihilation (SIA), $\gamma^*\to h_1+h_2+X$,
\begin{eqnarray}\label{def:W}
W^{\mu\nu}_{\text{DY}}&=&\int \frac{d^4y}{(2\pi)^4}e^{-i(yq)}\sum_X \langle p_1,p_2|J^{\mu\dagger}(y)|X\rangle\langle X|J^\nu(0)|p_1,p_2\rangle,\\
\label{def:W-SIDIS}
W^{\mu\nu}_{\text{SIDIS}}&=&\int \frac{d^4y}{(2\pi)^4}e^{i(yq)}\sum_X \langle p_1|J^{\mu\dagger}(y)|p_2,X\rangle\langle p_2,X|J^\nu(0)|p_1\rangle,
\\\label{def:W-SIA}
W^{\mu\nu}_{\text{SIA}}&=&\int \frac{d^4y}{(2\pi)^4}e^{i(yq)}\sum_X \langle 0|J^{\mu\dagger}(y)|p_1,p_2,X\rangle\langle p_1,p_2,X|J^\nu(0)|0\rangle.
\end{eqnarray}
where $J_\mu(y)$ is the electro-magnetic (EM) current (for brevity we omit the electric charge)
\begin{eqnarray}\label{def:EM-current}
J^\mu(y)=\bar q \gamma^\mu q(y),
\end{eqnarray}
and $q(y)$ is a quark field. At sufficiently high energies, the current should be replaced by the electro-weak (EW) current. The difference between EM and EW currents is only in the tensor structure and does not impact the factorization procedure, so that we stick to the EM case.

The factorization for all three processes is almost identical, since (in our approach) it is derived at operator level. The main differences appear in the boundary conditions for the fields and the sign of the Fourier exponent (sec.~\ref{sec:process}). For concreteness we center our discussion on the factorization of the DY reaction, commenting necessary modifications for SIDIS and SIA cases.

The kinematics of the process is defined by the photon momentum $q^\mu$, and the hadrons momenta $p_1^\mu$ and $p_2^\mu$. To avoid complications related to the target mass corrections we assume that hadrons are massless,
\begin{eqnarray}
p^2_1=p_2^2=0.
\end{eqnarray}
Hadrons momenta define two light-cone directions, which we traditionally denote as $n^\mu$ and $\bar n^\mu$ with $(n\bar n)=1$,
\begin{eqnarray}\label{def:nnbar}
p_1^\mu=\bar n^\mu p_1^+,\qquad p_2^\mu=n^\mu p_2^-.
\end{eqnarray}
We also introduce the usual notation for components of the light-cone decomposition of a vector,
\begin{eqnarray}
v^\mu=\bar n^\mu v^++ n^\mu v^-+v_T^\mu,
\end{eqnarray}
and $v_T^\mu$ is a component orthogonal to $n^\mu$ and $\bar n^\mu$. The invariant mass of the virtual photon is
\begin{eqnarray}
Q^2=q^2=2q^+q^--\vec q_T^2,
\end{eqnarray}
where $\vec q_T^2=-q_T^\mu q_{T \mu}>0$. In the case, of SIDIS $Q^2=-q^2$. The TMD factorization is derived (as it is demonstrated later) in the limit
\begin{eqnarray}\label{def:factorization-limit}
Q^2\gg \Lambda^2,\qquad Q^2\gg \vec q_T^2=\text{fixed},
\end{eqnarray}
where $\Lambda$ is a typical low-energy QCD scale. It implies that the light-cone components of $q^\mu$ are large $q^+\sim q^-\sim Q$. Additionally, we suppose that $q^+/p_1^+$ and $q^-/p_2^-$ are fixed, which corresponds to a non-small-x regime.

Often the limit in eq.~(\ref{def:factorization-limit}) is quoted as $\vec q_T^2/Q^2\ll 1$, which can lead to some misunderstanding, because $\vec q_T^2/Q^2\ll 1$ can be also interpreted as $\vec q_T^2\to0$ at fixed $Q^2$. In this case, the corrections $\Lambda/Q$ would be present even at $\vec q_T=0$.

The hadronic tensor in eq.~(\ref{def:W}) is symmetric $W^{\mu\nu}=W^{*\nu\mu}$ and transverse to $q^\mu$
\begin{eqnarray}\label{def:qW=0}
q_\mu W^{\mu\nu}=0,
\end{eqnarray}
as a consequence of EM gauge invariance. The transversality relation (\ref{def:qW=0}) is not homogeneous in power counting, because it involves simultaneously large $q^\pm$ and small $q_T^\mu$ components of photon's momentum and  therefore, any truncated power expansion consistent with eq.~(\ref{def:factorization-limit}) unavoidably violates the condition eq.~(\ref{def:qW=0}) up to higher power terms.

\subsection*{Field modes}

In order to apply the background-field method we need to write down the hadronic tensor in eq.~(\ref{def:W}) as a functional integral and to identify the field modes relevant for the task.

The hadronic tensors that we consider have two causally-independent sectors which exchange  real emissions. In this case, the functional integral can be written using Keldysh's method \cite{Keldysh:1964ud}. We introduce two copies of QCD fields, which we address as causal and anti-causal fields (indicated by superscripts $(+)$ and $(-)$ respectively). These fields obey the usual quantization rules with (anti-)time-ordered evolution operator for (anti-)causal fields. The values of fields coincide at the future boundary $\lim_{t\to \infty}q^{(+)}(t,x)=\lim_{t\to \infty} q^{(-)}(t,x)$ and $\lim_{t\to \infty}A_\mu^{(+)}(t,x)=\lim_{t\to \infty} A_\mu^{(-)}(t,x)$. On the perturbative level, it leads to the real-particle propagator connecting $(+)$ and $(-)$ fields, which is equivalent to usual Feynman rules for cut diagrams. More information on this method can be found, f.i. in \cite{Balitsky:1990ck,Belitsky:1997ay,Belitsky:1998tc,Balitsky:2016dgz}. In the present context, the method is necessary because it allows to write a non-time ordered operator as a functional integral. The hadronic tensor reads
\begin{eqnarray}\label{W-as-funInt1}
W^{\mu\nu}_{\text{DY}}&=&\int \frac{d^4y}{(2\pi)^4}e^{-i(yq)}
\int [D\bar q^{(+)}Dq^{(+)}DA^{(+)}]
\int [D\bar q^{(-)}Dq^{(-)}DA^{(-)}]
\\ &&\nn \times
\Psi_{p_1}^{*(-)}\,\Psi_{p_2}^{*(-)}\,
e^{iS_{\text{QCD}}^{(+)}-iS_{\text{QCD}}^{(-)}}\,
J_\mu^{\dagger(-)}(y) J_\nu^{(+)}(0)\,
\Psi_{p_1}^{(+)}\,\Psi_{p_2}^{(+)},
\end{eqnarray}
where $\Psi_{p}$ is the hadron's wave function (formed at the distant past), and $S_{\text{QCD}}$ is the QCD's action. The superscript $(\pm)$ indicates that the element is composed only from causal/anti-causal fields. The functional integration measure incorporates all necessary normalization factors.

In the case of SIDIS or SIA the only change in eq.~(\ref{W-as-funInt1}) is in the hadron wave-functions, that should be replaced by the wave-functions of produced hadrons according to the process.

The next step is to identify the relevant field modes and the ones that are to be integrated, or kept. We suppose that a fast-moving hadron is composed only of fields with momenta along the hadron's one. So, for a hadron with the momentum along $\overline{n}$, the fields which constitute it obey the counting (same for $(+)$ and $(-)$ components)
\begin{eqnarray}\nn
&&\{\partial_+ , \,\partial_- ,\, \partial_T \}\,q_{\bar n}\lesssim Q\{1,\, \lambda^2,\, \lambda\}\, q_{\bar n},
\\\label{collinear-counting}
&&\{\partial_+ , \,\partial_- ,\, \partial_T \}\,A^\mu_{\bar n}\lesssim Q\{1,\, \lambda^2,\, \lambda\}\, A^\mu_{\bar n},
\end{eqnarray}
where $\lambda$ is a generic small scale, $\lambda\sim \Lambda/Q$. Similarly, a hadron with the momentum along $n$ direction is composed out of fields with the counting
\begin{eqnarray}\nn
&&\{\partial_+ , \,\partial_- ,\, \partial_T \}\,q_{n}\lesssim Q\{\lambda^2,\, 1,\, \lambda\}\, q_{n},
\\\label{anti-collinear-counting}
&&\{\partial_+ , \,\partial_- ,\, \partial_T \}\,A^\mu_{n}\lesssim Q\{\lambda^2,\, 1,\, \lambda\}\, A^\mu_{n}.
\end{eqnarray}
In the literature these fields are identified as $n$ and $\bar n$-collinear or collinear and anti-collinear fields (f.i. \cite{Bauer:2002nz,Becher:2010tm}), or target and projectile fields (f.i. \cite{Balitsky:2016dgz}). The background fields, in its own kinematic sector, are ordinary QCD fields and thus satisfy the QCD equation of motion (EOMs). Let us emphasize the ``$\lesssim$''-sign which is used in these formulas. It indicates that the fields incorporate all possible momenta with the corresponding boundary. This is a principal difference of the background approach from  SCET, where the field modes are defined in ``boxes'' of momentum space, alike $\{\partial_+ , \,\partial_- ,\, \partial_T \}\,q_{\bar n}\sim Q\{1,\, \lambda^2,\, \lambda\}\, q_{\bar n}$, i.e. their momentum is localized around a given scale (that labels the fields).
Another difference is that modes scaling as $\{\partial_+ , \,\partial_- ,\, \partial_T \}\,q\sim Q\{\lambda,\, \lambda,\, \lambda\}\, q$ (called "soft" in SCET nomenclature) are not present in our construction.

It is convenient to distinguish ``good'' and ``bad'' spinor components of the quark field
\begin{eqnarray}\label{def:q=xi+eta}
q_{\bar n/n}(x)=\xi_{\bar n/n}(x)+\eta_{\bar n/n}(x).
\end{eqnarray}
They are defined as (same for $(+)$ and $(-)$ components)
\begin{eqnarray}\label{def:xi-eta+}
\xi_{\bar n}(x)&=&\frac{\gamma^-\gamma^+}{2}q_{\bar n}(x),\qquad \eta_{\bar n}(x)=\frac{\gamma^+\gamma^-}{2}q_{\bar n}(x),
\\\label{def:xi-eta-}
\xi_{n}(x)&=&\frac{\gamma^+\gamma^-}{2}q_{n}(x),\qquad \eta_{n}(x)=\frac{\gamma^-\gamma^+}{2}q_{n}(x).
\end{eqnarray}
The (massless) quark EOMs $\fnot Dq(x)=0$, in the terms of these components are
\begin{eqnarray}\label{EOMs-for-nbar}
\gamma^+ D_-[A_{\bar n}]\xi_{\bar n}=-\fnot D_T[A_{\bar n}]\eta_{\bar n},
&\qquad&
\gamma^- D_+[A_{\bar n}]\eta_{\bar n}=-\fnot D_T[A_{\bar n}]\xi_{\bar n},
\\\label{EOMs-for-n}
\gamma^- D_+[A_n]\xi_n=-\fnot D_T [A_n]\eta_n,
&\qquad&
\gamma^+ D_-[A_n]\eta_n=-\fnot D_T[A_n]\xi_n,
\end{eqnarray}
where
\begin{eqnarray}\label{def:cov-derivative}
D_\mu[A]=\partial_\mu-ig A_\mu,
\end{eqnarray}
is the covariant derivative. The EOMs in eq.~(\ref{EOMs-for-nbar}, \ref{EOMs-for-n}) imply that ``good'' and ``bad'' components have different effective counting,
\begin{eqnarray}\label{field-counting-relative}
&&\eta_{\bar n/n}\sim \lambda\,\xi_{\bar n/n}.
\end{eqnarray}
Using tree-level computations (see sec.~\ref{sec:tree-gen}) one finds the power counting for individual components,
\begin{eqnarray}\label{field-counting1}
\xi_{\bar n/n}\sim \lambda,\qquad \eta_{\bar n/n}\sim \lambda^2.
\end{eqnarray}
The counting rules for components of the gluon field also follows from eq.~(\ref{EOMs-for-nbar}, \ref{EOMs-for-n}),
\begin{eqnarray}
\nn
&& A_{\bar n}^+\sim 1,\qquad A_{\bar n}^{\mu_T}\sim \lambda,\qquad A_{\bar n}^-\sim \lambda^2,
\\\label{field-counting2}
&& A_{n}^+\sim \lambda^2,\qquad A_{n}^{\mu_T}\sim \lambda,\qquad A_{n}^-\sim 1.
\end{eqnarray}

\subsection*{Effective operators}

In essence, the background-field method consists in  splitting QCD fields into dynamical and background components with the subsequent (functional) integration of the former. In many aspect, the procedure resembles the Wilsonian renormalization group formulation and it is the simplest way to obtain the operator product expansion (OPE) with any given power counting. It is also a very explicit method to study the power corrections to any OPE (see e.g. power corrections to DIS \cite{Balitsky:1987bk,Balitsky:1990ck}, DVCS \cite{Braun:2012hq}, quasi-distributions \cite{Braun:2021aon}, small-b matching for TMD distributions \cite{Scimemi:2019gge}). 

\begin{figure}[t]
\centering
\includegraphics[width=0.4\textwidth]{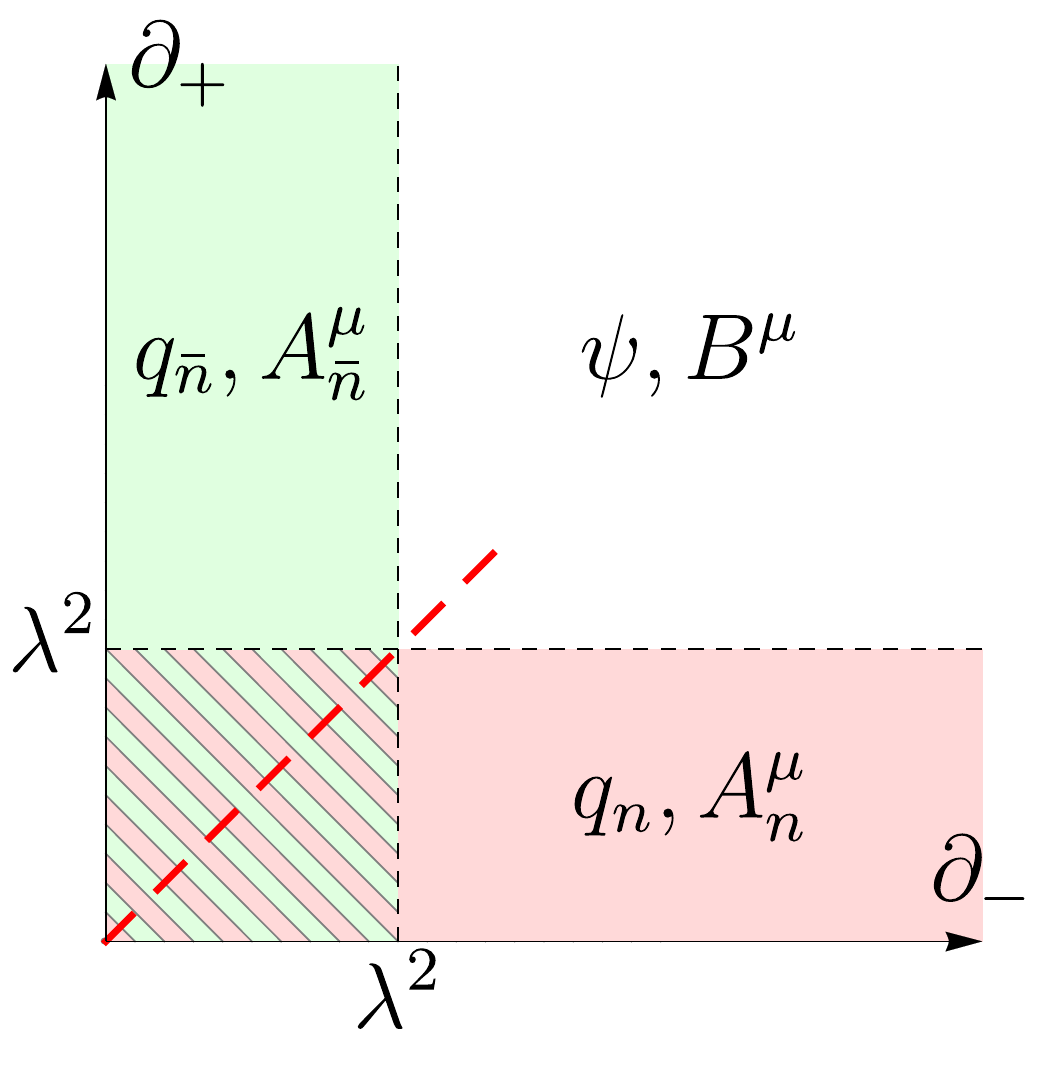}
\caption{Domains of fields in the plane of light-cone momenta components according to eq.~(\ref{collinear-counting}, \ref{anti-collinear-counting}).}
\label{fig:field-regions}
\end{figure}

The distinctive feature of the TMD factorization from cited cases is that the background field has two independent components, collinear and anti-collinear. Thus the causal/anti-causal fields in the functional integral in eq.~(\ref{W-as-funInt1}) split as
\begin{eqnarray}\nn
q^{(\pm)}(x)&=&\psi^{(\pm)}(x)+q^{(\pm)}_{n}(x)+q^{(\pm)}_{\bar n}(x),
\\\label{field-split}
A^{(\pm)}_\mu(x)&=&B^{(\pm)}_\mu(x)+A^{(\pm)}_{n\,\mu}(x)+A^{(\pm)}_{\bar n\,\mu}(x),
\end{eqnarray}
where $\psi$ and $B$ are the dynamical components which cover the remaining part of the Hilbert space, see fig.~\ref{fig:field-regions}. Since the fields represent independent Fourier components \footnote{This statements refers to the fact that the loop-momentum is cut following the counting rules in eq.~(\ref{collinear-counting}, \ref{anti-collinear-counting}). However, practically one uses the dimensional regularization and each field span the whole momentum-space, but each field sector is renormalized at a scale consistent with its counting. Such a mismatch between counting rules and integration regions manifests itself in  ultraviolet power divergences (which are omitted in the dimensional regularization) and renormalon divergences of the perturbative series \cite{Beneke:2000kc}.},
the integration measure can be split as
\begin{eqnarray}\label{functional-measure}
[D\bar q^{(\pm)}Dq^{(\pm)}DA^{(\pm)}]=[D\bar \psi^{(\pm)}D\psi^{(\pm)}DB^{(\pm)}][D\bar q_{\bar n}^{(\pm)}Dq_{\bar n}^{(\pm)}DA_{\bar n}^{(\pm)}][D\bar q_n^{(\pm)}Dq_n^{(\pm)}DA_n^{(\pm)}].
\end{eqnarray}
The separation of the field modes is done respecting  gauge-invariance. The Lagrangian is invariant only under the gauge-transformation of all fields by the same transformation parameter (irrespective of any power counting). Nonetheless, the gauge-transformation for the background field and the dynamical field can be decoupled, if one uses the so-called background gauge \cite{Abbott:1980hw,Abbott:1981ke}. In this case the (covariant-gauge-like) gauge fixation condition for the dynamical field takes the form
\begin{eqnarray}
[\partial_\mu \delta^{AC}+g f^{ABC}(A^{(\pm)B}_{\bar n\,\mu}+A^{(\pm)B}_{n\,\mu})]B^{(\pm)\mu\,C}=D_\mu[A^{(\pm)}_{\bar n}+A^{(\pm)}_{n}]B^{(\pm)\mu}=0,
\end{eqnarray}
where $D_\mu$ is the covariant derivative (\ref{def:cov-derivative}). This gauge fixation modifies vertices of background-to-gluon (ghost) interaction. The advantage of such a choice is that the background fields transform independently from the rest and their gauge can be fixed in any convenient way. This is one of the most profitable features of the background field method, which essentially simplifies the analysis.

Formally, the collinear and anti-collinear background fields transform with the same gauge parameter.  In the region where the momenta of the fields do not overlap (i.e. for $\partial_\mu > \lambda^2 Q$) we treat them as totally independent fields, and thus their gauge transformations are also independent. The problems arise where sectors overlap (see barred part of Fig.~\ref{fig:field-regions}). To avoid this region, we momentarily restrict our discussion to the non-small-x domain with $p_{\text{parton}} \gtrsim \lambda Q$. We will return to the discussion of the mode overlap in sec.~\ref{sec:soft-overlap}.

After implementing these definitions in the functional integral in eq.~(\ref{W-as-funInt1}), the  hadronic tensor reads
\begin{eqnarray}
&&W_{\text{DY(unsub.)}}^{\mu\nu}=\int \frac{d^4y}{(2\pi)^4}e^{-i(yq)}
\\\nn &&
\qquad\times\int [D\bar q_{\bar n}^{(+)}Dq_{\bar n}^{(+)}DA_{\bar n}^{(+)}] [D\bar q_{\bar n}^{(-)}Dq_{\bar n}^{(-)}DA_{\bar n}^{(-)}]e^{i S^{(+)}_{\text{QCD}}[\bar q_{\bar n},q_{\bar n},A_{\bar n}]-i S^{(-)}_{\text{QCD}}[\bar q_{\bar n},q_{\bar n},A_{\bar n}]}
\\\nn &&
\qquad\times\int [D\bar q_{n}^{(+)}Dq_{n}^{(+)}DA_{n}^{(+)}] [D\bar q_{n}^{(-)}Dq_{n}^{(-)}DA_{n}^{(-)}]e^{i S^{(+)}_{\text{QCD}}[\bar q_{n},q_{n},A_{n}]-i S^{(-)}_{\text{QCD}}[\bar q_{n},q_{n},A_{n}]}
\\\nn &&
\qquad\times\int [D\bar \psi^{(+)}D\psi^{(+)}DB^{(+)}] [D\bar \psi^{(-)}D\psi^{(-)}DB^{(-)}]e^{i S^{(+)}_{\text{QCD}}[\bar \psi,\psi,B]-i S^{(-)}_{\text{QCD}}[\bar \psi,\psi,B]}
\\\nn &&
\qquad\times \Psi^{*(-)}_{p_1} \Psi_{p_2}^{*(-)}\,J_\mu^{\dagger(-)}[\bar \psi+\bar q_{\bar n}+\bar q_n,...](y)\,J_\nu^{(+)}[\bar \psi+\bar q_{\bar n}+\bar q_n,...](0)
\Psi^{(+)}_{p_1} \Psi_{p_2}^{(+)} e^{iS^{(+)}_{int}-iS^{(-)}_{int}},
\end{eqnarray}
where in the square brackets we indicate the field content of each term (for brevity we omit superscripts $(\pm)$ on these arguments and indicate similar arguments for currents by dots). The label "unsub." states that this expression has an unsubtracted overlapped region that is discussed in sec.~\ref{sec:soft-overlap}. The gauge fixing terms are included in the $S_{QCD}$ exponents. The cross-modes-interaction term is
\begin{eqnarray}\label{Sint}
S_{int}&=&S_{\text{QCD}}[\bar \psi+\bar q_{\bar n}+\bar q_n,...]
-S_{\text{QCD}}[\bar q_{\bar n},q_{\bar n},A_{\bar n}]
-S_{\text{QCD}}[\bar q_{n},q_{n},A_{n}]-S_{\text{QCD}}[\bar \psi,\psi,B].
\end{eqnarray}
Its explicit form in the background field-gauge is derived in app.~\ref{app:Sint}, eq.~(\ref{app:Sint:main}). The derivation takes into account  the equations of motion (EOMs) for collinear and anti-collinear fields. The cross-modes-interaction term can be split into four terms: $S_{nh}$ ($S_{\bar n h}$) (\ref{app:Sint:S1q}) describing the interaction of (anti-)collinear fields with hard fields; $S_{n\bar n}$ describing the direct interaction of collinear and anti-collinear fields (\ref{app:Sint:S12q}); and  $S_{n\bar n h}$ (\ref{app:Sint:S12q}) describing the interaction of all fields simultaneously. Each action $S_{nh}$ and $S_{\bar nh}$  is equal to the usual QCD action with background field \cite{Abbott:1981ke}, while $S_{n\bar nh}$ and $S_{n\bar n}$ are specific for the composite background case. Let us mention that $S_{n\bar n}=O(\lambda^3)$ due to the power-counting in eq.~(\ref{field-counting1}, \ref{field-counting2}).

Before integrating over the hard modes we specify the content of the hadronic wave functions. The main assumption of the parton model is that the hadron is composed of fields collinear with respect to its momentum, 
\begin{eqnarray}\label{parton-model}
\Psi_{p_1}=\Psi_{p_1}[\bar q_{\bar n},q_{\bar n},A_{\bar n}],\qquad
\Psi_{p_2}=\Psi_{p_2}[\bar q_{n},q_{n},A_{n}].
\end{eqnarray}
Once the wave functions are independent of hard mode, we integrate over it and deduce the expression
\begin{eqnarray}\label{W-1}
&&W_{\text{DY(unsub.)}}^{\mu\nu}=\int \frac{d^4y}{(2\pi)^4}e^{-i(yq)}
\\\nn &&
\quad\times\int [D\bar q_{\bar n}^{(+)}Dq_{\bar n}^{(+)}DA_{\bar n}^{(+)}] [D\bar q_{\bar n}^{(-)}Dq_{\bar n}^{(-)}DA_{\bar n}^{(-)}]e^{i S^{(+)}_{\text{QCD}}[\bar q_{\bar n},q_{\bar n},A_{\bar n}]-i S^{(-)}_{\text{QCD}}[\bar q_{\bar n},q_{\bar n},A_{\bar n}]}
\\\nn &&
\quad\times\int [D\bar q_{n}^{(+)}Dq_{n}^{(+)}DA_{n}^{(+)}] [D\bar q_{n}^{(-)}Dq_{n}^{(-)}DA_{n}^{(-)}]e^{i S^{(+)}_{\text{QCD}}[\bar q_{n},q_{n},A_{n}]-i S^{(-)}_{\text{QCD}}[\bar q_{n},q_{n},A_{n}]}
\\\nn &&
\quad\times \Psi^{*(-)}_{p_1}[\bar q_{\bar n},q_{\bar n},A_{\bar n}] \Psi_{p_2}^{*(-)}[\bar q_{n},q_{n},A_{n}] \,
\mathcal{J}^{\mu\nu}_{eff}[\bar q_{\bar n},\bar q_{n},...](y) \,\Psi^{(+)}_{p_1}[\bar q_{\bar n},q_{\bar n},A_{\bar n}] \Psi_{p_2}^{(+)}[\bar q_{n},q_{n},A_{n}],
\end{eqnarray}
where $\mathcal{J}^{\mu\nu}_{eff}[\bar q_{\bar n},\bar q_{n},\bar q_{s},...]$ depends on all background (causal and anti-causal, collinear and anti-collinear) modes and is defined as
\begin{eqnarray}\label{Jefftot}
&&\mathcal{J}^{\mu\nu}_{eff}[\bar q_{\bar n},\bar q_{n},...](y)=
\int [D\bar \psi^{(+)}D\psi^{(+)}DB^{(+)}] [D\bar \psi^{(-)}D\psi^{(-)}DB^{(-)}]
\\ \nn  && 
\qquad\times
J_\mu^{\dagger(-)}[\bar \psi+\bar q_{\bar n}+\bar q_n,...](y)J_\nu^{(+)}[\bar \psi+\bar q_{\bar n}+\bar q_n,...](0)
e^{i S^{(+)}_{\text{QCD}}[\bar \psi,\psi,B]-i S^{(-)}_{\text{QCD}}[\bar \psi,\psi,B]}
 e^{iS^{(+)}_{int}-iS^{(-)}_{int}}.
\end{eqnarray}
The effective operator satisfies
\begin{eqnarray}\label{symmetries-of-J}
&&\mathcal{J}^{\mu\nu}(y)=\mathcal{J}^{\dagger\nu\mu}(-y),
\\
&&\frac{\partial}{\partial y^\mu}\mathcal{J}^{\mu\nu}(y)=0,
\end{eqnarray}
which are consequences of symmetry and transversality of the hadronic tensor.

In the end of the section let us sketch the further steps of the factorization procedure, which are discussed in detail in sec.~\ref{sec:general-structure}. The effective operator is an infinite series of individually gauge-invariant terms. This sum can be ordered with respect to $\lambda$,
\begin{eqnarray}\label{Jeff}
\mathcal{J}^{\mu\nu}_{eff}[\bar q_{\bar n},\bar q_{n},...](y)=\sum_{N=0}^\infty \sum_k \mathcal{J}^{\mu\nu}_{N,k}[\bar q_{\bar n},\bar q_{n},...](y),
\end{eqnarray}
where $N$ represents power order ($\mathcal{J}_{N,k}^{\mu\nu}\sim \lambda^{N+4}$), and $k$ enumerates the operators with the same power counting. Generally, each $\mathcal{J}_{N,k}^{\mu\nu}$ is a convolution of background fields, and it can be written in the schematic form
\begin{eqnarray}\label{JN=COO}
\mathcal{J}_{N,(a,b)}^{\mu\nu}[\bar q_{\bar n},\bar q_{n},...](y)=C_{N,(a,b)}^{\mu\nu}(y)\otimes \mathcal{O}_{a}[\bar q_{\bar n},q_{\bar n},A_{\bar n}]\otimes \mathcal{O}_{b}[\bar q_n,q_n,A_n],
\end{eqnarray}
where $C_{N,(a,b)}^{\mu\nu}$ is a coefficient function, $\mathcal{O}_{a}$ are some operators (the possible multi-index and multi-position structure is encoded in the single label $a$) that also depend on $y$, and $\otimes$ represents a convolution in variables and indices.

Let us note, that ordering of operators $\mathcal{J}_{N,k}^{\mu\nu}$ by power counting implies the expansion of components along ``slow''-direction, f.i. $q_{\bar n}(x+n x_0)= q_{\bar n}(x)+O(\lambda^2)$ (for non-extreme $x_0$). The resulting derivatives contribute to the operators at higher powers. Expansions that involve $y^\mu$ need special care. The counting rules for the components of $y^\mu$ are $(yq)\sim 1$, that gives
\begin{eqnarray}\label{y-counting}
\{y^+,y^-,y_T\}\sim Q^{-1}\{1,1,\lambda^{-1}\}.
\end{eqnarray}
Therefore, the transverse derivatives which involve $y$ have the compensating factor $y_T\sim\lambda^{-1}$. The combination $y_T^\mu\partial_\mu\sim 1$  is not suppressed, in contrast to other transverse derivatives. Due to it, all dynamics in the transverse plane in the final expression is tied to $y$. 

Changing the (functional) integration and summation order in eq.~(\ref{W-1}) we get
\begin{eqnarray}\label{Wunsub}
W_{DY\text{(unsub.)}}^{\mu\nu}&=&
\sum_{N=0}^\infty\sum_{(a,b)}
\int \frac{d^4y}{(2\pi)^4}e^{-i(yq)} C_{N,(a,b)}^{\mu\nu}(y)\otimes \Phi^{\text{unsub}}_{a}(y,p_1) \otimes \Phi^{\text{unsub}}_{b}(y,p_2),
\end{eqnarray}
where
\begin{eqnarray}
\Phi^{\text{unsub}}_{a}(y,p)&=&\langle p|\mathcal{O}_{a}|p\rangle=
\int [D\bar q^{(+)}Dq^{(+)}DA^{(+)}] [D\bar q^{(-)}Dq^{(-)}DA^{(-)}]
\\\nn && \qquad\qquad\qquad\times e^{i S^{(+)}_{\text{QCD}}[\bar q,q,A]-i S^{(-)}_{\text{QCD}}[\bar q,q,A]}
\Psi^{*(-)}_{p}[\bar q,q,A]\mathcal{O}_{a}[\bar q,q,A]\Psi^{(+)}_{p}[\bar q,q,A],
\end{eqnarray}
and the fields in definition of $\Phi$'s are just collinear or anti-collinear. In these expressions the functions $\Phi^{\text{unsub}}$ (that are unsubtracted TMD distributions) are nonperturbative in the sense that they contain unknown information on the hadronic structure and low-energy QCD interactions. 

The sketched derivation remains the same for the cases of SIDIS and SIA. The only difference among effective operators in different processes is due to different boundary conditions prescribed to collinear fields for initial and final state hadrons (see the next section). In the rest of the paper, the effective operator is the same for all cases.

\section{Process dependence and gauge fixation}
\label{sec:process}

The effective operator is gauge invariant term-by-term in the series in eq.~(\ref{Jeff}), as a consequence of  the gauge invariant definition in eq.~(\ref{Jefftot}). Fixing the gauges for collinear and anti-collinear fields in a convenient way, we can  simplify  calculations in the intermediate steps restoring  gauge invariant expressions at the end of the computations. 
We use the light-cone gauges for background fields,
\begin{eqnarray}\label{light-cone-gauge}
n^\mu A_{\bar n,\mu}^{(\pm)}(z)=0,\qquad \bar n^\mu A_{n,\mu}^{(\pm)}(z)=0.
\end{eqnarray}
This choice removes $ {\cal O}(1)$ gluon components in eq.~(\ref{field-counting2}). Therefore, the power counting for operators increases with the number of fields in the operator, which essentially simplifies the computation. In any other gauge, there would be an infinite set of operators of the same order in power counting but with different numbers of $A_{\bar n}^+$ or $A_n^-$, which eventually sums into Wilson lines but makes the computation cumbersome. 

The light-cone gauge conditions in eq.~(\ref{light-cone-gauge}) alone do not remove all gauge freedom, leaving possible $z^\mp$-independent transformations (for $A^\pm=0$ gauge). The detailed discussion of this can be found in ref.~\cite{Belitsky:2002sm}. To fix the redundant gauge freedom, one imposes specific boundary conditions on the components of the gluon field. The final expression is independent of the choice of boundary conditions.  However, an inappropriate choice can lead to unnecessary complications during the computation, and to avoid them we  specify convenient boundary conditions for each process.

To deduce the proper boundary conditions, we study a generic expression (for regular graphs) contributing to an effective operator,
\begin{eqnarray}\label{I1}
I=\int_{-\infty}^\infty dz^+ dz^- \frac{f_{\bar n}(z^-)f_n(z^+)}{[-2z^+z^-+i0]^\alpha},
\end{eqnarray}
where $f_{\bar n(n)}$ is a combination of (anti-)collinear fields and their derivatives that are (eventually) integrated along the light-cone positions with some weights. The only important property is that $f_{\bar n(n)}$ depends only on $z^{-(+)}$. The dependence of $f_{\bar n(n)}$ on $z^{+(-)}$ would violate the counting (there is also dependence on $y_T$, but it is external and does not modify analytical properties). The power $\alpha$ is non-integer in the dimensional regularization.  Examples of such expressions can be found in eq.~(\ref{diag1}, \ref{diag2},  \ref{diag3}, \ref{NLO:3point-diagrams}). The $+i0$ prescription is the appropriate one for the causal-sector of interactions. For the anti-causal sector it is replaced by $-i0$, but simultaneously all analytical properties are reverted, leaving the final result unchanged.

\begin{figure}[t]
\centering
\includegraphics[width=0.45\textwidth]{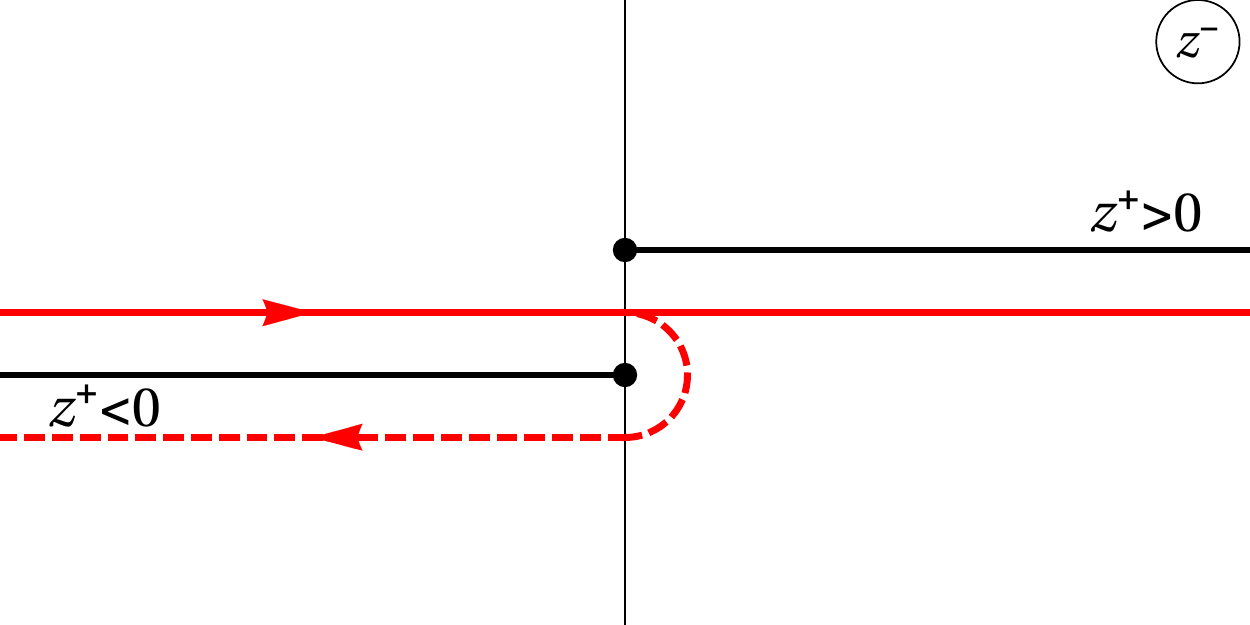}
\caption{Analytical structure of the integral (\ref{I1}) in the complex plane of $z^-$ for the case of DY reaction. The solid red line is the original integration contour. The dashed red line is the modified integration contour.}
\label{fig:DY_contour}
\end{figure}

The denominator of eq.~(\ref{I1}) possesses a branch cut (for definiteness we place the branch cut of $x^\alpha$ along the line $(-\infty,0)$) along the line
\begin{eqnarray}
z^\mp = \left\{
\begin{array}{cc}
(+i0, +\infty+i0)& \text{for }z^\pm>0, \\
(-\infty-i0, -i0)& \text{for }z^\pm<0 .
\end{array}\right.
\end{eqnarray}
This is illustrated in fig.~\ref{fig:DY_contour}. The field combinations $f$ in eq.~(\ref{I1}) give raise to parton distributions and thus we must assume some analytical properties for them to guarantee the existence of Fourier integrals. For the case of incoming partons (i.e. PDFs) $f$ is analytical in the lower half-plane, whereas for the case of outgoing partons (i.e. FFs) $f$ is analytical in the upper half plane. Therefore, depending on the process we deal with different combination of analytical properties, summarized in the following table
\begin{equation}\label{anal-prop}
\text{
\begin{tabular}{c|c|c|c|c}
& for DY & for SIDIS & for SIA & \\\cline{2-4}
$f_{\bar n}(z^-)$ is analytical in
& lower & lower & upper & half-plane.\\ 
$f_{n}(z^+)$ is analytical in
& lower & upper & upper & half-plane.
\end{tabular}
}    
\end{equation}
Given these properties we can deform the integration contour placing it on the sides of the branch cut as it is shown in fig.~\ref{fig:DY_contour}. Finally, the integral splits into several contributions.  The contribution at infinity is troublesome, since it produces  ill-defined operators with gluons concentrated at infinity. A properly selected boundary condition nullifies it.

We illustrate this discussion for the case of DY reaction. In this case, we can close the integration in the lower half-plane, and deform the contour as  shown in fig.~\ref{fig:DY_contour}. We get
\begin{eqnarray}
I=\int_{-\infty}^0 dz^+ \frac{f_{n}(z^+)}{(-2z^+)^\alpha}\(I_0+I_1+I_2+I_\infty\),
\end{eqnarray}
where $I_k$ are integrals of $f_{\bar n}(z^-)/(z^-)^{\alpha}$ along elements of the contour. $I_1$ and $I_2$ are integrals along sides for the branch cut, $I_0$ is a semi-circle at $z^-=-i0$ and $I_\infty$ is a semi-circle at $z^-=-\infty-i0$. The dimensional regularization regularizes a possible ultraviolet pole at $z\to0$ (demanding $\alpha<1$), and thus $I_0$ vanishes. However, it also implies that $I_\infty$ contributes to the integral, unless the fields $f$ vanishes at $z^-\to-\infty$. We can use the freedom to fix the redundant gauge condition such that $I_\infty$ vanishes\footnote{If we use the improper boundary condition the contribution $I_\infty$ remains. It is singular and independent on $z^-$, and to be cancelled by the interactions with the transverse gauge links located at the light-cone infinity \cite{Belitsky:2002sm,Idilbi:2010im}. The interaction with transverse links  automatically vanishes with the proper boundary condition.}. The same analysis can be done for $z^+$ variable. So, we conclude that for the DY reaction the convenient set of boundary conditions is such that fields $f_{\bar n(n)}$ vanish at $z^{-(+)}\to-\infty$. 

Repeating the same analysis for the cases of SIDIS and SIA, we arrive to the following set of appropriate boundary conditions
\begin{eqnarray}\nn
\text{for DY:}&\qquad& \lim_{z^-\to -\infty}A_{\bar n}^\mu(z)=0,\qquad \lim_{z^+\to -\infty}A_{n}^\mu(z)=0,
\\\label{gauge-fix-boundary}
\text{for SIDIS:}&\qquad& \lim_{z^-\to +\infty}A_{\bar n}^\mu(z)=0,\qquad \lim_{z^+\to -\infty}A_{n}^\mu(z)=0,
\\\nn
\text{for SIA:}&\qquad& \lim_{z^-\to +\infty}A_{\bar n}^\mu(z)=0,\qquad \lim_{z^+\to +\infty}A_{n}^\mu(z)=0.
\end{eqnarray}
This set guarantees the absence of an $I_\infty$ contribution in the loop integrals. For future convenience, we introduce the variables $L$ and $\bar L$ which can take the values $\pm \infty$, and summarize boundary conditions as
\begin{eqnarray}
\lim_{z^-\to L}A_{\bar n}^\mu(z)=0,\qquad \lim_{z^+\to \bar L}A_{n}^\mu(z)=0,
\end{eqnarray}
with
\begin{eqnarray}
(L, \bar L) = \left\{
\begin{array}{cc}
(-\infty,-\infty),& \text{for DY}, \\
(+\infty,-\infty),& \text{for SIDIS}, \\
(+\infty,+\infty),& \text{for SIA}.
\end{array}\right.
\end{eqnarray}
The value of $(L,\bar L)$ is the only difference between processes at the operator level.

Notice  that for DY and SIA cases the integrals in eq.~(\ref{I1}) can be written in the factorized form
\begin{eqnarray}\label{I1-for-DY}
I_{\text{DY}}&=&\frac{-i\pi 2^{1-\alpha}}{\Gamma(\alpha)\Gamma(1-\alpha)}\int_{-\infty}^0 dz^+\frac{f_n(z^+)}{(-z^+)^\alpha}\int_{-\infty}^0 dz^-\frac{f_{\bar n}(z^-)}{(-z^-)^\alpha},
\\\label{I1-for-SIA}
I_{\text{SIA}}&=&\frac{-i\pi 2^{1-\alpha}}{\Gamma(\alpha)\Gamma(1-\alpha)}\int_{0}^\infty dz^+\frac{f_n(z^+)}{(z^+)^\alpha}\int_0^{\infty} dz^-\frac{f_{\bar n}(z^-)}{(z^-)^\alpha},
\end{eqnarray}
whereas for SIDIS case both integral representations $I_{\text{DY}}$ and $I_{\text{SIA}}$ are valid.

Once the boundary conditions are specified, the gauge-fixing condition in eq.~(\ref{light-cone-gauge}) can be inverted. It gives
\begin{eqnarray}\label{def:A->F}
A_{\bar n}^\mu(z)=-g\int_{L}^0 d\sigma F^{\mu +}_{\bar n}(z+n \sigma),
\qquad
A_{n}^\mu(z)=-g\int_{\bar L}^0 d\sigma F^{\mu -}_{n}(z+\bar n \sigma),
\end{eqnarray}
where $F^{\mu\nu}$ is the gluon field-strength tensor.

After the computation is complete, and the result is written in  terms of $F_{\mu\nu}$'s, one can restore the explicit form of the gauge-invariant operator by multiplying fields with gauge links. The rules are
\begin{eqnarray}\nn
q_{\bar n}(z)&\to& [L n+\infty_T,L n+z] [L n+z,z]q_{\bar n}(z),
\\\label{def:LC->regular}
\bar q_{\bar n}(z)&\to& \bar q_{\bar n}(z)[z,L n+z][L n+z,L n+\infty_T],
\\\nn 
F_{\bar n}^{\mu\nu}(z)&\to& [L n+\infty_T,L n+z] [L n+z,z]F^{\mu\nu}_{\bar n}(z)[z,L n+z][L n+z, L n+\infty_T],
\end{eqnarray}
where $\infty_T$ in an infinitely distant point in the transverse plane, and $[a,b]$ is a straight Wilson line
\begin{eqnarray}
[a,b]=P\exp\Big(-ig\int_a^b dz_\mu A^\mu_{\bar n}(z)\Big).
\end{eqnarray}
For the anti-collinear fields the expressions are analogous.

Finally,  we mention that in SCET one often uses the operator $\mathcal{A}_\perp^\mu$ that  (for a collinear field) is defined as  \cite{Bauer:2001ct,Bauer:2002nz,Beneke:2017ztn}
\begin{eqnarray}\label{def:A->D}
\mathcal{A}_\perp^\mu(z)&=&i\,[L n+z,z]\overrightarrow{D}_\mu [z,L n +z]
-i\,[L n+z,z]\overleftarrow{D}_\mu [z,L n +z]
.
\end{eqnarray}
In the explicit form this operator reads
\begin{eqnarray}
\mathcal{A}_\perp^\mu(z)=-g \int_{-\infty}^0 d\sigma [L n+z,z+\sigma n]F^{\mu+}_{\bar n}(z+\sigma n)  [z+\sigma n,L n +z]).
\end{eqnarray}
Comparing with eq.~(\ref{def:A->F}, \ref{def:LC->regular}) we find that $\mathcal{A}_\perp^\mu=A^\mu$ in  light-cone gauge, which gives a map between our expressions and the ones written in the terms of $\mathcal{A}_\perp^\mu$.

\section{General structure of the TMD operator expansion}
\label{sec:general-structure}

The computation of the effective operator in the composite background follows the common pattern of computation in a single background field, see, e.g., refs.~\cite{Braun:2012hq,Braun:2021aon,Scimemi:2019gge}. The logical steps following in the computation are:
\begin{enumerate}
\item Expansion of eq.~(\ref{Jefftot}) into monomials of collinear and anti-collinear fields.
\item Multipole expansion of collinear and anti-collinear fields according to eq.~(\ref{collinear-counting}, \ref{anti-collinear-counting}, \ref{y-counting}).
\item Rewriting of fields in terms of ``good'' and ``bad'' components.
\item Evaluation of necessary (loop-)integrals.
\item Reduction of operators to a given basis, using algebra and EOMs.
\item Renormalization/Recombination of divergences.
\item Fiertz transformation into TMD operators.
\end{enumerate}
During the evaluation, one should keep in mind that in the very end, the operators are inserted into matrix elements, and some of them vanish (e.g., due to non-zero fermion number or due to non-singlet color representation). Such operators can be eliminated without full consideration. At each step, we find a series of terms with increasing power, and the ordering of power counting is preserved. Thus only terms with the desired power counting are finally kept.

\subsection*{General structure of operators}

The starting point for the effective operator expansion is the product of two EM currents separated by a distance $y$. The fact that $y_T\sim (\lambda Q)^{-1}$ spoils the usual intuition about the computation of the effective operator because it is not allowed to expand over $y_T$, since $(y_T^\mu\partial_\mu)\sim 1$, and the effective operator splits into two parts separated by $y_T$. These parts are independent in the sense that they have separate anomalous dimensions and can be separately expanded in powers (this, however, does not mean that the coefficient function is a product of coefficient functions).

\begin{figure}[t]
\centering
\includegraphics[width=0.55\textwidth]{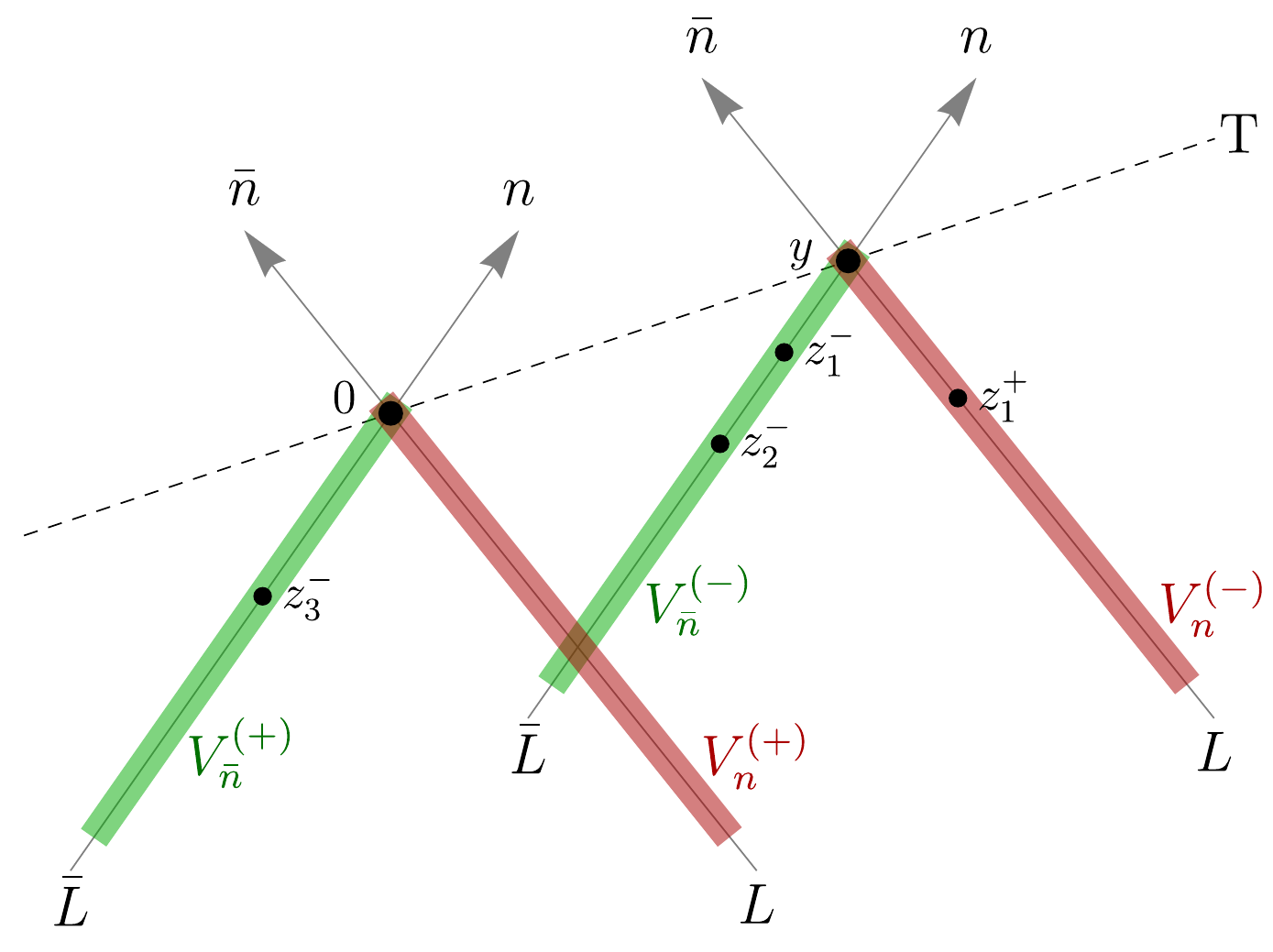}
\caption{Spatial configuration of the operators contributing to an effective operator (\ref{a-term}) in the case of DY kinematics. The dots represent insertion of collinear (green) and anti-collinear (red) fields. The thick colored lines are light-rays along which the operators are localized.}
\label{fig:geometry}
\end{figure}

After step 2, a contribution to the effective operator has a general form
\begin{eqnarray}\label{a-term}
C^{\mu\nu}(\{z^\pm\},y)
\otimes \Big[
V^{(-)}_{\bar n}(\{z_1^-\},y_T)
\cdot V^{(-)}_{n}(\{z_1^+\},y_T)
\cdot  V^{(+)}_{\bar n}(\{z_2^-\},0_T)
\cdot V^{(+)}_{n}(\{z_2^+\},0_T)\Big],
\end{eqnarray}
where $\{z_i^\pm\}$ indicate a set of coordinates, and $\{z^\pm\}=\{z_1^\pm\}\cup \{z_2^\pm\}$. In eq.~(\ref{a-term}) $V^{(+)}_{\bar n}(\{z^-\},y_T)$ is a light-cone operator composed from causal collinear fields, that are positioned at the light-ray with transverse coordinate $y_T$ with coordinates $\{z^-\}$, and similar for other $V$'s. $C^{\mu\nu}$ is the differential operator that possibly contains a (loop-)integral to be evaluated at step 4. The symbol $\otimes$ indicates the integral convolution in positions $z^\pm$ and contraction in Lorentz and color indices between coefficient function and operators. The illustration for spatial configuration of fields in eq.~(\ref{a-term}) is given in fig.\ref{fig:geometry}.

Each operator $V_{\bar n}(\{z^-\},r_T)$  in eq.~(\ref{a-term}) consists of some number of fields located on the light-ray pointing from $r_T$ to $L\bar n$ (and analogously for $V_{n}(\{z^+\},r_T)$). The positions $\{z^-\}$ are distributed along this light-ray. Each field within $V_n$ is accompanied by a semi-infinite Wilson line in eq.~(\ref{def:LC->regular}). These Wilson lines recombine with each other and partially cancel. However, in general, a Wilson line does not vanish beyond the position $\max\{z^-\}$ and continues till $L\bar n$. In this sense, \textit{the operator $V$ is semi-compact}. It also means that the operator $V$ is not entirely gauge-invariant because, under the gauge-transformation, it receives a gauge-rotation factor at $(L\bar n+\infty_T)$. These properties are in contrast to usual DIS-like OPE, where resulting operators are compact (i.e., localized in the finite volume) and entirely gauge-invariant. On the other hand, the semi-compact operator basis is the only difference between DIS-like and TMD operator expansions. Therefore, we can apply the powerful machinery of OPE, correcting it only for semi-compact operators.

Step 3 decomposes operators with respect to power counting of the components of the fields in eq.~(\ref{field-counting1}, \ref{field-counting2}).  The resulting operators have the same form as in eq.~(\ref{a-term}), with the difference that operators $V$ have a  ``pure'' power counting and can not be decomposed further. 

\subsection*{Twist-decomposition}

The aim of step 5 is to decompose each operator $V$ with respect to a convenient basis. The most convenient and physically motivated decomposition is the decomposition with respect to \textit{geometrical twist}, which is the ``dimension-minus-spin'' of the operator. One has
\begin{eqnarray}\label{twist-dec}
V_{\bar n}(\{z^-\},r_T)&=&\sum_N c_N\otimes U_{\bar n,N}(\{z^-\},r_T),
\end{eqnarray}
where $U$ is an operator with a definite twist (for simplicity, we set it equal $N$, although there can be several terms with the same twist in the decomposition), $c_N$ is an integral-differential operator, and $\otimes$ is the integral and matrix convolution. An operator $U_N$ with a definite geometrical twist belongs to an irreducible Lorentz group and thus does not mix with operators of other geometrical twists. This important property is preserved by perturbation theory, and thus operators with definite geometrical twists have an independent evolution, and their matrix elements are separate physical observables.

The twist-decomposition in eq.~(\ref{twist-dec}) is an algebraic procedure (consisting of symmetrization and anti-symmetrization of indices) originally defined for local operators. It can be also generalized for non-local operators by means of generating functionals of local operators (see e.g.\cite{Geyer:1999uq}), or by implication of differential operators (see e.g.~\cite{Balitsky:1987bk}), or by considering the conformal transformation properties (see e.g.~\cite{Braun:2003rp}).  All these methods cannot be applied directly to semi-compact operators $V$ and should be revised. In particular, the method of reconstruction of non-local form of operators via generating functions has been applied to semi-compact operators in ref.~\cite{Moos:2020wvd}. The main idea used in ref.~\cite{Moos:2020wvd} is to set the parameter $L$ finite. It allows to write down a formal local expansion for semi-compact operators, make the twist-decomposition and resum the result into a generating functional. Lastly, the limit $L\to\infty$ is taken. This method correctly reproduces known lower-power properties of semi-compact operators, and it can be applied to operators of any power. In ref.~\cite{Moos:2020wvd}, the twist-decomposition for several cases of semi-compact operators has been made, including P-odd operators, which are specific for the semi-compact case.

\textit{The lowest twist for semi-compact operators is twist-1}. The twist-1 operator is just a single ``good'' component of quark $\xi$ or gluon field $F_{\mu+}$ (with $\mu$ being transverse index) with attached semi-infinite Wilson line. Although such numbering looks unusual, it also follows from the formal counting of geometrical twist by adding up conformal spins \cite{Braun:2009vc}.

\subsection*{Recombination of divergences}

At this stage a contribution to the effective operator has the form
\begin{eqnarray}\label{a-term1}
\widetilde{C}_{NM,\bar N\bar M}^{\mu\nu}(\{z^\pm\},y^\pm)
\otimes \Big[
U^{(-)}_{\bar n, N}(\{z_1^-\},y_T)
\cdot U^{(-)}_{n, \bar N}(\{z_1^+\},y_T)
\cdot U^{(+)}_{\bar n, M}(\{z_2^-\},0_T)
\cdot U^{(+)}_{n, \bar M}(\{z_2^+\},0_T)\Big],
\end{eqnarray}
where $\widetilde{C}$ is a combination of $C$ from eq.~(\ref{a-term}) and $c$'s from eq.~(\ref{twist-dec}). The coefficient function $\widetilde{C}$ has divergences in $\epsilon$ (in the dimensional regularization) that are IR. These divergences match the UV divergences of operators $U$. Let $Z_N$ be the renormalization factor for the operator $U_N$, $U_N=Z_N(\mu)\otimes U_{N}$, where the convolution is in the position of operators, and $\mu$ is the renormalization scale. Then inserting unit factors $1=Z_N^{-1}\otimes Z_N$ into eq.~(\ref{a-term1}) we obtain an expression of the form
\begin{eqnarray}\label{a-term2}
C_{NM,\bar N\bar M}^{\mu\nu}(\{z^\pm\},y^\pm;\mu)\!
\otimes\! \Big[
U^{(-)}_{\bar n, N}(\{z_1^-\},y_T)
\!\cdot\! U^{(-)}_{n, \bar N}(\{z_1^+\},y_T)
\!\cdot\! U^{(+)}_{\bar n, M}(\{z_2^-\},0_T)
\!\cdot\! U^{(+)}_{n, \bar M}(\{z_2^+\},0_T)\!\Big],
\end{eqnarray}
where operators $U$ are renormalized at the scale $\mu$, and 
\begin{eqnarray}\label{C-finite}
C_{NM,\bar N\bar M}^{\mu\nu}(\{z^\pm\},y^\pm;\mu)=
\widetilde{C}_{NM,\bar N\bar M}^{\mu\nu}(\{z^\pm\},y^\pm;\mu)\otimes [
Z_N^{-1}(\mu)\cdot 
Z_{\bar N}^{-1}(\mu)\cdot
Z_M^{-1}(\mu)\cdot
Z_{\bar M}^{-1}(\mu)],
\end{eqnarray}
is finite. To demonstrate that eq.~(\ref{C-finite}) is finite is a non-trivial task and the object of the factorization theorem.

Let us note that the rapidity divergences do not appear in the effective operator and have no traces in the coefficient function. The origin and the factorization of rapidity divergences are discussed in the sec.~\ref{sec:soft-overlap}.

\subsection*{TMD operators}

The Fierz transformation at step 7 recouples the color indices and groups operators $U$ into color-neutral TMD operators
\begin{eqnarray}\label{UU->O}
U_{\bar n,N}^{A(-)}(\{z_1^-\},y_T)
U_{\bar n,M}^{B(+)}(\{z_2^-\},0_T)=\frac{\delta^{AB}}{\text{dim}(R_{AB})}\mathcal{O}_{\bar n,NM}(\{z^-\},y_T),
\end{eqnarray}
where we explicitly indicate the color indices $A$ and $B$ (that belong to representation $R_{AB}$ of $SU(N_c)$), and $\text{dim}(R_{AB})$ is the dimension of their representation. After this operation a contribution to the effective operator has a general form
\begin{eqnarray}\label{a-term3}
\widetilde{C}^{\mu\nu}_{NM,\bar N\bar M}(\{z^\pm\},y^\pm,\mu)\otimes \Big[
\mathcal{O}_{\bar n,NM}(\{z^-\},y_T)
\cdot \mathcal{O}_{n,\bar N\bar M}(\{z^+\},y_T)\Big]
\end{eqnarray}
where $\widetilde{C}$ contains also derivatives that act on $\mathcal{O}$'s. This is the final form, see also eq.~(\ref{JN=COO}), of the effective operator.

\subsection*{TMD-twist}

Each term of the effective operators in eq.~(\ref{a-term2}) is labeled by four numbers $(NM,\bar N\bar M)$, which indicate the geometrical twists of its internal components. Therefore, the terms with different labels $(NM,\bar N\bar M)$ do not mix, and their matrix elements are unique combinations of independent nonperturbative functions (TMD distributions). Each TMD operator $\mathcal{O}$ (and consequently each TMD distribution) is labeled by a pair of numbers $(NM)$. This pair labels the \textit{TMD-twist} of the operator. 

For convenience, we define the TMD-twist of the operator $\mathcal{O}_{NM}$ equal to (N+M) (``N-plus-M''). Such notation matches the usual jargon. In particular, the leading power TMD distributions that are often referred to as twist-2 TMD distributions (without specification of the meaning of twist for TMD operator) are twist-(1+1) TMD distributions within our formalism. The sub-leading power TMD distributions are referred to as twist-3 and have TMD-twist-(1+2) or TMD-twist-(2+1). In principle, the operators with twist-(1+2) or twist-(2+1) are different and define two separate TMD distributions with different evolution equations, although the C-conjugation relates them. The real profit from this notation comes from the higher powers. So, the twist-4 TMD distributions (in usual terminology) can be twist-(3+1), twist-(2+2) and twist-(1+3). The properties of twist-(2+2) TMD distributions are very different from twist-(3+1) distributions, and they do not relate to each other by any means.

In the limit of small transverse separation $y_T\to 0$, TMD operators turn to collinear operators. Consistently, this limit is computed by the light-cone OPE, see e.g. \cite{Scimemi:2019gge,Moos:2020wvd}. The leading term is equal to the product of $U$'s at $y_T=0$,
\begin{equation}
\mathcal{O}_{\bar n,NM}(\{z^-\},y_T)=U_{\bar n,N}^{(-)}(\{z_1^-\},0)U_{\bar n,M}^{(+)}(\{z_2^-\},0)+\mathcal{O}(\alpha_s)+\mathcal{O}(y_T).
\end{equation}
The smallest possible geometrical twist for the operator on r.h.s. is $(N+M)$, since it is a product of spin $N$ and $M$ tensors. Therefore, at high-$q_T$, where $\sim y_T$ contributions are small, and TMD factorization turns into the resummation approach, TMD distributions of twist-(N+M) match collinear distributions of twist $(N+M)$ or higher. In this way, computing contributions of only TMD-twist-(1+1) operators (at all powers of TMD operator expansion), one should be able to reconstruct all leading twist terms of collinear factorization, including the fixed order computations, such as in ref.~\cite{Ellis:1981hk}.

\section{Effective operator at NLP/LO}
\label{sec:tree-gen}

Using the scheme depicted in the previous section, we compute the effective operator to  leading  and next-to-leading power (LP and NLP, respectively) and up NLO in  perturbation theory. In this section, we derive the tree order of NLP effective operator and introduce necessary definitions. Although this result is (partially) known, see e.g.\cite{Boer:2003cm,Bacchetta:2006tn,Hu:2021naj}, our derivation is novel in many aspects because it is made at the operator level and with no explicit reference to a specific process. For this reason, we give a detailed explanation for each step of the computation. The NLO computation is given in the next section.

\subsection*{Tree order for LP}

We start with the decomposition of the EM current
\begin{eqnarray}\label{tree:EM1}
J^\mu[\bar \psi+\bar q_{\bar n}+\bar q_n,...]
&=& \bar q_{\bar n}\gamma^\mu q_{n}
+
\bar q_{n}\gamma^\mu q_{\bar n}
\\\nn && +
\bar \psi\gamma^\mu \psi
+
\bar q_{\bar n}\gamma^\mu \psi
+
\bar q_n\gamma^\mu \psi
+
\bar \psi\gamma^\mu q_{\bar n}
+
\bar \psi\gamma^\mu q_n
\\\nn &&
+
\bar q_{\bar n}\gamma^\mu q_{\bar n}
+
\bar q_{n}\gamma^\mu q_{n}
\end{eqnarray}
Here,  the first line provides the leading tree-order contribution.
The second line contains fields $\psi$ that are to be contracted with $\psi$ from $S_{QCD/int}$. These terms contribute to NLP and NLO, and they are considered in the next section. The terms in the third line have two fields from the same collinear sector, and thus they produce disconnected contributions to matrix elements, unless extra fields are taken from $S_{int}$. The first possible non-vanishing contributions of the terms in the third line happen only at N$^4$LP. 

To get the LP term, we consider the first line of eq.~(\ref{tree:EM1}), and perform the multipole expansion. We get
\begin{eqnarray}\label{tree:EM2}
&&\bar q_{\bar n}\gamma^\mu q_{n}(y)
+
\bar q_{n}\gamma^\mu q_{\bar n}(y)
\\\nn &&\qquad\qquad=
\bar q_{\bar n}(y^-n+y_T)\gamma^\mu q_{n}(y^+\bar n+y_T)
+
\bar q_{n}(y^+\bar n+y_T)\gamma^\mu q_{\bar n}(y^-n+y_T)+\mathcal{O}(\lambda^4).
\end{eqnarray}
The terms in  $\mathcal{O}(\lambda^4)$ contains derivatives, such as $y^+\bar q_{\bar n}\overleftarrow{\partial_-}\gamma^\mu q_{n}$. Decomposing the field $q$ into ``good'' and ``bad'' components as in eq.~(\ref{def:q=xi+eta}) we obtain
\begin{eqnarray}\label{tree:EM3}
\bar q_{\bar n}\gamma^\mu q_{n}
+
\bar q_{n}\gamma^\mu q_{\bar n}
&=&
\bar \xi_{\bar n}\gamma^\mu_T \xi_{n}
+
\bar \xi_{n}\gamma^\mu_T \xi_{\bar n}
\\\nn &&+
\bar n^\mu \bar \xi_{\bar n}\gamma^+ \eta_{n}
+
n^\mu \bar \xi_{n}\gamma^- \eta_{\bar n}
+
n^\mu \bar \eta_{\bar n}\gamma^- \xi_{n}
+
\bar n^\mu \bar \eta_{n}\gamma^+ \xi_{\bar n}
\\\nn &&+
\bar \eta_{\bar n}\gamma^\mu_T \eta_{n}
+
\bar \eta_{n}\gamma^\mu_T \eta_{\bar n},
\end{eqnarray}
where we omit the arguments understanding implicitly that each collinear field depends on $(y^-n+y_T)$ and each anti-collinear field depends on $(y^+ \bar n+y_T)$. The first, second, and third lines  in eq.~(\ref{tree:EM3}) are $\mathcal{O}(\lambda^2)$, $\mathcal{O}(\lambda^3)$, and $\mathcal{O}(\lambda^4)$, respectively. Eq.~(\ref{tree:EM3}) has the form of eq.~(\ref{a-term}), with operators having pure counting, in the sense that no further expansion is needed. This term represents the LP term of EM current. The second and the third lines of eq.~(\ref{tree:EM3}) have indefinite twist. They are discussed in the following subsection.
Thus, the LP term is a composition of twist-1 and twist-1 operators
\begin{eqnarray}\label{tree:JLP}
J^\mu_{\text{LP}}(y)=
\bar \xi_{\bar n}(y^-n+y_T)\gamma^\mu_T \xi_{n}(y^+\bar n+y_T)
+
\bar \xi_{n}(y^+\bar n+y_T)\gamma^\mu_T \xi_{\bar n}(y^-n+y_T).
\end{eqnarray}
Let us note that the LP current  in eq.~(\ref{tree:JLP}) violates the EM charge conservation. Indeed,
\begin{eqnarray}\label{EM-conservation:LP}
\partial_\mu J^\mu_{\text{LP}}=\bar \xi_{\bar n}(\overleftarrow{\fnot \partial_T}+\overrightarrow{\fnot \partial_T})\xi_{n}
+
\bar \xi_{n}(\overleftarrow{\fnot \partial_T}+\overrightarrow{\fnot \partial_T}) \xi_{\bar n}\neq 0.
\end{eqnarray}
However, the expression in r.h.s. of eq.~(\ref{EM-conservation:LP}) is $\mathcal{O}(\lambda^3)$, and thus formally the charge is conserved at LP.

Combining together the LP currents, eq.~(\ref{tree:JLP}), we get the LP expression for the effective operator
\begin{eqnarray}
\mathcal{J}_{\text{LP}}^{\mu\nu}(y)
&=&
[\bar \xi^{(-)}_{\bar n}(y^-n+y_T)\gamma^\mu_T \xi^{(-)}_{n}(y^+\bar n+y_T)
+
\bar \xi^{(-)}_{n}(y^+\bar n+y_T)\gamma^\mu_T \xi^{(-)}_{\bar n}(y^-n+y_T)]
\\\nn && \times [\bar \xi^{(+)}_{\bar n}(0)\gamma^\nu_T \xi^{(+)}_{n}(0)
+
\bar \xi^{(+)}_{n}(0)\gamma^\nu_T \xi^{(+)}_{\bar n}(0)].
\end{eqnarray}
The quark fields $\xi$ are operators of twist-1, and thus $\mathcal{J}_{\text{LP}}^{\mu\nu}(y)$ is of $(1+1)\times(1+1)$-twist in our nomenclature. Combining the fields into TMD operators, eq.~(\ref{UU->O}), we get 
\begin{eqnarray}\label{tree:LP}
\mathcal{J}_{\text{LP}}^{\mu\nu}(y)
&=&
\frac{\gamma^\mu_{T,ij}\gamma^\nu_{T,kl}}{N_c}\Big(
\mathcal{O}_{11,\bar n}^{li}\,
\overline{\mathcal{O}}_{11,n}^{jk}
+
\overline{\mathcal{O}}_{11,\bar n}^{jk}\,
\mathcal{O}_{11,n}^{li}
\Big)
\end{eqnarray}
where $(i,j,k,l)$ are spinor indices, and the TMD twist-(1+1) operators have the argument $(\{y^-,0\},y_T)$. They are defined as
\begin{eqnarray}\label{def:O11-position}
\mathcal{O}_{11,\bar n}^{ji}(\{y^-,0\},y_T)
&=&\bar \xi^{(-)}_{\bar n,i}(y^-n+y_T)\, \xi^{(+)}_{\bar n,j}(0),
\\\label{def:O11bar-position}
\overline{\mathcal{O}}_{11,\bar n}^{ji}(\{y^-,0\},y_T)
&=&\xi^{(-)}_{\bar n,j}(y^-n+y_T)\,\bar \xi^{(+)}_{\bar n,i}(0).
\end{eqnarray}
The operator $\mathcal{O}_{11,n}$ and $\overline{\mathcal{O}}_{11,n}$ are obtained by replacing $n\leftarrow \bar n$ everywhere, including the subscripts of the fields. This expression is well known and the basis for the LP TMD factorization, see e.g.\cite{Boer:2003cm,Bacchetta:2006tn,Echevarria:2012js}. The matrix elements of operators $\mathcal{O}_{11,\bar n}$ and $\overline{\mathcal{O}}_{11,\bar n}$ give rise to the quark and anti-quark TMD distributions, respectively.

\subsection*{Tree order for NLP}

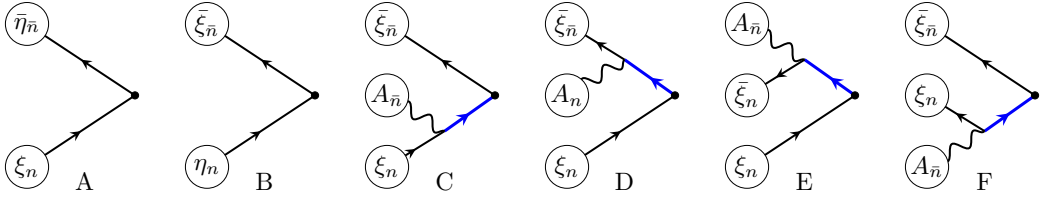
\begin{figure}
\begin{center}
\begin{tikzpicture}

\coordinate (A) at (2, 1.5);

\draw[thick,postaction={decorate, decoration={
    markings,
    mark=at position 0.5 with {\arrow{stealth}}}}] ($(A)$)--($(A)+(-1.5,1.) $);
\draw[thick,postaction={decorate, decoration={
    markings,
    mark=at position 0.5 with {\arrow{stealth}}}}] ($(A)+(-1.5,-1.)$) -- ($(A)$);

\draw[fill=black] ($(A)$) circle (0.05cm);    
\draw[fill=white] ($(A)+(-1.5,-1.)$) circle (0.3cm);
\draw[fill=white] ($(A)+(-1.5,1.)$) circle (0.3cm);
\node[] at ($(A)+(-1.5,1.)$) {$\Ds\bar \eta_{\bar n}$};
\node[] at ($(A)+(-1.5,-1.)$) {$\Ds \xi_{n}$};
\node[] at ($(A)+(-0.7,-1.2)$) {A};

\coordinate (A) at (4.5, 1.5);

\draw[thick,postaction={decorate, decoration={
    markings,
    mark=at position 0.5 with {\arrow{stealth}}}}] ($(A)$)--($(A)+(-1.5,1.) $);
\draw[thick,postaction={decorate, decoration={
    markings,
    mark=at position 0.5 with {\arrow{stealth}}}}] ($(A)+(-1.5,-1.)$) -- ($(A)$);

\draw[fill=black] ($(A)$) circle (0.05cm);    
\draw[fill=white] ($(A)+(-1.5,-1.)$) circle (0.3cm);
\draw[fill=white] ($(A)+(-1.5,1.)$) circle (0.3cm);
\node[] at ($(A)+(-1.5,1.)$) {$\Ds\bar \xi_{\bar n}$};
\node[] at ($(A)+(-1.5,-1.)$) {$\Ds \eta_{n}$};
\node[] at ($(A)+(-0.7,-1.2)$) {B};

\coordinate (A) at (7, 1.5);

\draw[thick, postaction={decorate, decoration={
    markings,
    mark=at position 0.5 with {\arrow{stealth}}}}] ($(A)$)--($(A)+(-1.5,1.)$);
\draw[very thick, blue!99!black, postaction={decorate, decoration={
    markings,
    mark=at position 0.5 with {\arrow{stealth}}}}] ($(A)+(-0.7,-0.5)$)--($(A) $);
\draw[thick,postaction={decorate, decoration={
    markings,
    mark=at position 0.5 with {\arrow{stealth}}}}] ($(A)+(-1.5,-1.)$) -- ($(A)+(-0.7,-0.5)$);
\draw[thick,style={decorate, decoration=snake}] ($(A)+(-0.7,-0.5)$)--($(A)+(-1.5,0)$);

\draw[fill=black] ($(A)$) circle (0.05cm);    
\draw[fill=white] ($(A)+(-1.5,-1.)$) circle (0.3cm);
\draw[fill=white] ($(A)+(-1.5,+1)$) circle (0.3cm);
\draw[fill=white] ($(A)+(-1.5,0)$) circle (0.3cm);

\node[] at ($(A)+(-1.5,1.)$) {$\Ds\bar \xi_{\bar n}$};
\node[] at ($(A)+(-1.5,-1.)$) {$\Ds \xi_{n}$};
\node[] at ($(A)+(-1.5,0.)$) {$\Ds A_{\bar n}$};
\node[] at ($(A)+(-0.7,-1.2)$) {C};

\coordinate (A) at (9.5, 1.5);

\draw[very thick, blue!99!black, postaction={decorate, decoration={
    markings,
    mark=at position 0.5 with {\arrow{stealth}}}}] ($(A)$)--($(A)+(-0.7,0.5)$);
\draw[thick,postaction={decorate, decoration={
    markings,
    mark=at position 0.5 with {\arrow{stealth}}}}] ($(A)+(-0.7,0.5)$)--($(A)+(-1.5,1.) $);
\draw[thick,postaction={decorate, decoration={
    markings,
    mark=at position 0.5 with {\arrow{stealth}}}}] ($(A)+(-1.5,-1.)$) -- ($(A)$);
\draw[thick,style={decorate, decoration=snake}] ($(A)+(-0.7,0.5)$)--($(A)+(-1.5,0)$);

\draw[fill=black] ($(A)$) circle (0.05cm);    
\draw[fill=white] ($(A)+(-1.5,-1.)$) circle (0.3cm);
\draw[fill=white] ($(A)+(-1.5,+1)$) circle (0.3cm);
\draw[fill=white] ($(A)+(-1.5,0)$) circle (0.3cm);

\node[] at ($(A)+(-1.5,1.)$) {$\Ds\bar \xi_{\bar n}$};
\node[] at ($(A)+(-1.5,-1.)$) {$\Ds \xi_{n}$};
\node[] at ($(A)+(-1.5,0.)$) {$\Ds A_{n}$};
\node[] at ($(A)+(-0.7,-1.2)$) {D};

\coordinate (A) at (12, 1.5);

\draw[very thick, blue!99!black, postaction={decorate, decoration={
    markings,
    mark=at position 0.5 with {\arrow{stealth}}}}] ($(A)$)--($(A)+(-0.7,0.5)$);
\draw[thick,style={decorate, decoration=snake}] ($(A)+(-0.7,0.5)$)--($(A)+(-1.5,1.) $);
\draw[thick,postaction={decorate, decoration={
    markings,
    mark=at position 0.5 with {\arrow{stealth}}}}] ($(A)+(-1.5,-1.)$) -- ($(A)$);
\draw[thick,postaction={decorate, decoration={
    markings,
    mark=at position 0.5 with {\arrow{stealth}}}}] ($(A)+(-0.7,0.5)$)--($(A)+(-1.5,0)$);

\draw[fill=black] ($(A)$) circle (0.05cm);    
\draw[fill=white] ($(A)+(-1.5,-1.)$) circle (0.3cm);
\draw[fill=white] ($(A)+(-1.5,+1)$) circle (0.3cm);
\draw[fill=white] ($(A)+(-1.5,0)$) circle (0.3cm);

\node[] at ($(A)+(-1.5,1.)$) {$\Ds A_{\bar n}$};
\node[] at ($(A)+(-1.5,-1.)$) {$\Ds \xi_{n}$};
\node[] at ($(A)+(-1.5,0.)$) {$\Ds\bar \xi_{n}$};
\node[] at ($(A)+(-0.7,-1.2)$) {E};

\coordinate (A) at (14.5, 1.5);

\draw[thick, postaction={decorate, decoration={
    markings,
    mark=at position 0.5 with {\arrow{stealth}}}}] ($(A)$)--($(A)+(-1.5,1.)$);
\draw[very thick, blue!99!black, postaction={decorate, decoration={
    markings,
    mark=at position 0.5 with {\arrow{stealth}}}}] ($(A)+(-0.7,-0.5)$)--($(A) $);
\draw[thick,style={decorate, decoration=snake}] ($(A)+(-1.5,-1.)$) -- ($(A)+(-0.7,-0.5)$);
\draw[thick,postaction={decorate, decoration={
    markings,
    mark=at position 0.5 with {\arrow{stealth}}}}] ($(A)+(-0.7,-0.5)$)--($(A)+(-1.5,0)$);

\draw[fill=black] ($(A)$) circle (0.05cm);    
\draw[fill=white] ($(A)+(-1.5,-1.)$) circle (0.3cm);
\draw[fill=white] ($(A)+(-1.5,+1)$) circle (0.3cm);
\draw[fill=white] ($(A)+(-1.5,0)$) circle (0.3cm);

\node[] at ($(A)+(-1.5,1.)$) {$\Ds\bar \xi_{\bar n}$};
\node[] at ($(A)+(-1.5,-1.)$) {$\Ds A_{\bar n}$};
\node[] at ($(A)+(-1.5,0.)$) {$\Ds \xi_{n}$};
\node[] at ($(A)+(-0.7,-1.2)$) {F};

\end{tikzpicture}
\caption{The diagrams contributing to the tree order of vector current at NLP. The diagrams with $n\leftrightarrow \bar n$ should be added. The blobs indicate the type of background field. The blue lines are the dynamical fields.}
\label{fig:tree-NLP}    
\end{center}
\end{figure}

The NLP part of the effective operator can appear only via the combination of a LP current,  eq.~(\ref{tree:JLP}), with NLP part of another EM current. The NLP contribution of order $\sim 
\lambda^3$ to $J^\mu$ can be composed in two ways. The first one is combining a ``good'' and a ``bad'' component of the quark fields from the second line of eq.~(\ref{tree:EM3}). The second possibility is to have three ``good'' components of fields, e.g. 
$\bar \xi A_T \xi$. To get such a term, one needs to pull down an interaction term from $e^{iS_{int}}$ and couple it to the first line of eq.~(\ref{tree:JLP}). The diagrams representing the NLP contributions are shown in fig.\ref{fig:tree-NLP}. The diagrams A and B correspond to the second line of (eq.~\ref{tree:EM3}). The remaining diagrams represent the interaction contribution. 

The diagrams C, D, E and F are specific for the computation in the composite background field, and would be absent for the case of ordinary background field. The reason is that an ordinary background field does not couple to the dynamical fields via 1PI vertex. In other words, there are no vertices with a single dynamical field and background field(s). Such diagrams (in the sum) compose EOM for the background field, and therefore, vanish (e.g. a diagram $C$ with $A_{\bar n}$ replaced by $A_{n}$ is not present). This fact is already taken into account in the construction of the effective background action \cite{Abbott:1980hw}, and thus the corresponding vertices are not present in $S_{int}$. In the case of the composite background, a vertex with a single dynamical field and \textit{different} background fields is not forbidden, and actually it is provided  by $S_{\bar n nh}$ in the effective action, eq.~(\ref{app:Sint:S12q}). These vertices do not contribute to the EOM\footnote{The subtraction of EOMs from the effective action is straightforward for the dynamical-to-background part of the interaction. However, this procedure is ambiguous for the contact part of the effective action $S_{12}$, eq.~(\ref{app:Sint:S12}), where the EOM terms can be subtracted in different proportions. The contact terms that represent background-to-background interaction are specific for the composite background case, and have effective counting of $\mathcal{O}(\lambda^3)$. The ambiguity in the definition of these terms disappears once matrix elements are taken.}. As a result the diagrammatic expansion is not (explicitly) symmetric with respect to $n^\mu \leftrightarrow \bar n^\mu$. This symmetry is restored once the operators are rewritten in a unique basis. As an example, we present in detail the evaluation of diagrams A and C, which form a symmetric pair.

Diagram C reads (we set the global position of the current to $0$ for brevity)
\begin{eqnarray}
\text{diag}_C&=&g\frac{\Gamma(2-\epsilon)}{2 \pi^{d/2}}\int d^dz
\,
\bar \xi_{\bar n}(0)\frac{\gamma^\mu \fnot z \gamma^\nu}{[-z^2+i0]^{2-\epsilon}}
A^\nu_{\bar n}(z)\xi_{n}(z),
\end{eqnarray}
where we used the propagators in  dimensional regularization, eq.~(\ref{app:prop-q}) with $d=4-2\epsilon$. The multipole expansion sets $A_{\bar n}^\nu(z)\xi_n(z)\to A_n^\nu(z^- n)\xi_n(z^+ \bar n)$, and the integral over $(d-2)$ transverse components can be evaluated, eq.~(\ref{app:transverse-integral}). The result is
\begin{eqnarray}\label{tree:NLP:C2}
\text{diag}_C&=&g \bar n^\mu \int \frac{dz^+dz^-}{\pi} \frac{z^+}{[-2z^+z^-+i0]}\bar \xi_{\bar n}(0)\fnot A_{\bar n,T}(z^- n)\xi_n(z^+ \bar n),
\end{eqnarray}
where we take into account that only the transverse components of the gluon field contribute to LP term. Note that in this expression one cannot cancel $z^+$ between the denominator and the numerator, because it would spoil the analytical properties of the integrand. To evaluate eq.~(\ref{tree:NLP:C2}), we recall the analytical properties of $A_{\bar n}$ summarized in eq.~(\ref{anal-prop}), then we close the contour of the $z^-$-integration in the lower (DY and SIDIS cases) or upper (SIA case) half-plane, shrink it to the pole at $z^-=i0/(2z^+)$, and evaluate the residue. We obtain the expression
\begin{eqnarray}\label{tree:NLP:C3}
\text{diag}_C&=&ig \bar n^\mu \int_{\bar L}^0 dz^+ \bar \xi_{\bar n}(0)\fnot A_{\bar n,T}(0)\xi_n(z^+ \bar n).
\end{eqnarray}
In this expression the anti-collinear fields $\bar \xi_{\bar n}\fnot A_{\bar n,T}$ form an operator of twist-2.

The diagram A is given by a simple expression
\begin{eqnarray}\label{tree:NLP:A1}
\text{diag}_A=n^\mu \bar \eta_{\bar n}(0)\gamma^-\xi_n(0).
\end{eqnarray}
Here, the field $\bar \eta_{\bar n}$ has indefinite geometrical twist, which can be checked e.g. by conformal transformation \cite{Braun:2011dg}. Practically, it implies that the field $\eta$ mixes with a quark-gluon pair during the evolution. To rewrite $\eta$ in the terms of definite twist operators, we apply EOMs, eq.~(\ref{EOMs-for-nbar}). In the the present case, we need the EOM for $\bar \eta_{\bar n}$,  eq.~(\ref{EOMs-for-nbar}). In the light-cone gauge it can be rewritten 
\begin{eqnarray}\label{EOM-integral}
\bar \eta_{\bar n}(0)\gamma^-=-\int_{L}^0 dz^- \Big( 
\partial_\mu \bar \xi_{\bar n}(z^- n )\gamma^\mu_T
+ig \bar \xi_{\bar n}(z^- n)\fnot A_{\bar n,T}(z^- n)
\Big).
\end{eqnarray}
Here, the first term on r.h.s. is the total derivative of twist-1 operator, and the second term is the twist-2 operator. The twist-counting can be confirmed by expanding the operators in a series of local operators (setting $L$ finite), and computing the twist of each term in the series. Inserting EOM into eq.~(\ref{tree:NLP:A1}) we get
\begin{eqnarray}
\text{diag}_A=
-n^\mu  \int_{L}^0 dz^- \bar \xi_{\bar n}(z^- n)\overleftarrow{\fnot \partial_T}\xi_n(0)
-ig n^\mu \int_L^0 dz^- \bar \xi_{\bar n}(z^- n)\fnot A_{\bar n,T}(z^- n)\xi_n(0).
\end{eqnarray}
The second term is complementary to diagram C, eq.~(\ref{tree:NLP:C3}). Together they from a transverse expression. To make it explicit we rewrite diagram A in the form of eq.~(\ref{tree:NLP:C2}) and sum the diagrams together. We obtain
\begin{eqnarray}
\text{diag}_{A+C}&=&-n^\mu  \int_{L}^0 dz^- \bar \xi_{\bar n}(z^- n)\overleftarrow{\fnot \partial_T}\xi_n(0)
\\\nn &&
+g \int \frac{dz^+dz^-}{\pi} \frac{\bar n^\mu z^+-n^\mu z^-}{[-2z^+z^-+i0]}
\bar \xi_{\bar n}(z^- n)\fnot A_{\bar n,T}(z^-n)\xi_n(z^+ \bar n).
\end{eqnarray}

Evaluating the rest of the diagrams in the same manner we obtain the expression for the EM current at NLP. We split the result into the following parts
\begin{eqnarray}\label{tree:JNLP}
J^{\mu}_{\text{NLP}}(0)=J^{\mu}_{1'1}(0)+J^{\mu}_{11'}(0)
+J^{\mu}_{21}(0)+J^{\mu}_{12}(0)
+J^{\mu}_{21;\mathbf{8}}(0)+J^{\mu}_{12;\mathbf{8}}(0).
\end{eqnarray}
The first term contains the derivatives of twist-1 operators
\begin{eqnarray}\label{NLP:tree:1'1}
J^{\mu}_{1'1}(0)&=&
-n^\mu  \int_{L}^0 dz^- \Big[
\bar \xi_{\bar n}(z^- n)\overleftarrow{\fnot\partial_T}\xi_n(0)
+
\bar \xi_{n}(0)\overrightarrow{\fnot\partial_T}\xi_{\bar n}(z^-n)\Big].
\end{eqnarray}
The terms $J_{21}^\mu$ and $J_{21,\mathbf{8}}^\mu$ contain operators of twist-2 and twist-1 
\begin{eqnarray}\label{NLP:tree:21}
J^\mu_{21}(0)&=&
ig \bar n^\mu\int_{\bar L}^0 dz^+\Big[
\bar \xi_{\bar n}(0)\fnot A_{\bar n,T}(0)\xi_n(z^+ \bar n)
-
\bar \xi_{n}(z^+\bar n)\fnot A_{\bar n,T}(0)\xi_{\bar n}(0)\Big]
\\\nn &&-
ig n^\mu\int_{L}^0 dz^-\Big[
\bar \xi_{\bar n}(z^- n)\fnot A_{\bar n,T}(z^- n)\xi_n(0)
-
\bar \xi_{n}(0)\fnot A_{\bar n,T}(z^- n)\xi_{\bar n}(z^- n)\Big],
\\
J^\mu_{21,\mathbf{8}}(0)&=&
\frac{ig}{2} \int_{L}^0 dz^-\Big[
\bar \xi_{\bar n}(z^-n)\gamma^+\gamma^\nu_T\gamma^\mu_T t^A\xi_{\bar n}(0)
-
\bar \xi_{\bar n}(0)\gamma^+\gamma^\mu_T\gamma^\nu_T t^A\xi_{\bar n}(z^-n)\Big]
A_{n,\nu}^A(0),
\end{eqnarray}
where $A$ is the color index in the adjoint representation and $t^A$ is the generator of $SU(N_c)$. The expression for $J_{11'}^\mu$, $J_{12}^\mu$ and $J_{12,\mathbf{8}}^\mu$ are obtained from $J_{1'1}^\mu$, $J_{21}^\mu$ and $J_{21,\mathbf{8}}^\mu$ by exchanging $n\leftrightarrow\bar n$ (also in the subscripts of fields).
The contributions to $J_{21,\mathbf{8}}^\mu$ and $J_{12,\mathbf{8}}^\mu$ arise from the diagrams of E and F (plus $n \leftrightarrow \bar n$ diagrams). The collinear and anti-collinear parts of $J_{21,\mathbf{8}}^\mu$ are in the adjoint representation of the color group. These parts of the EM current do not contribute at NLP, because to form a color neutral TMD operator, eq.~(\ref{UU->O}), an another operator in the adjoint representation is required. The first non-zero contribution from $J_{21,\mathbf{8}}^\mu$ and $J_{12,\mathbf{8}}^\mu$ into effective operator takes place at N$^2$LP.

Let us define the inverse derivative operator, as
\begin{eqnarray}\label{def:inverse-derivative}
\frac{1}{\partial_+}f(x)=\int_{L}^0 dz^- f(x+z^-n),
\qquad
\frac{1}{\partial_-}f(x)=\int_{\bar L}^0 dz^+ f(x+z^+\bar n).
\end{eqnarray}
With this operator the expressions in eq.~(\ref{NLP:tree:1'1}, \ref{NLP:tree:21}) have a simpler form
\begin{eqnarray}\label{NLP:tree:1'1+}
J^{\mu}_{1'1}&=&
-n^\mu  \bar \xi_{\bar n}\frac{\overleftarrow{\fnot\partial_T}}{\overleftarrow{\partial_+}}\xi_n
-n^\mu
\bar \xi_{n}\frac{\overrightarrow{\fnot\partial_T}}{\overrightarrow{\partial_+}}\xi_{\bar n},
\\\label{NLP:tree:21+}
J^\mu_{21}&=&
ig 
\bar \xi_{\bar n}\fnot A_{\bar n,T}\(
\frac{\bar n^\mu}{\overrightarrow{\partial_-}}-\frac{n^\mu}{\overleftarrow{\partial_+}}\)\xi_n
-
ig \bar \xi_{n}\(\frac{\bar n^\mu}{\overleftarrow{\partial_-}}-\frac{n^\mu}{\overrightarrow{\partial_+}}\)\fnot A_{\bar n,T}\xi_{\bar n},
\end{eqnarray}
where the positions of fields on r.h.s. and l.h.s. are formally the same. Note that the inverse derivatives in the last line acts on the quark and gluon fields together. 

The operators $J_{1'1}^\mu$ and $J_{11'}^\mu$ combine with the LP current eq.~(\ref{tree:JLP}) and form
\begin{eqnarray}\label{def:J11}
J_{11}^\mu &=& J_{\text{LP}}+J_{1'1}^\mu+J_{11'}^\mu
\\\nn &=& 
\bar \xi_{\bar n}\gamma_T^\mu \xi_n
+\bar \xi_{n}\gamma_T^\mu \xi_{\bar n}
-n^\mu  \bar \xi_{\bar n}\frac{\overleftarrow{\fnot\partial_T}}{\overleftarrow{\partial_+}}\xi_n
-n^\mu \bar \xi_{n}\frac{\overrightarrow{\fnot\partial_T}}{\overrightarrow{\partial_+}}\xi_{\bar n}
-\bar n^\mu  \bar \xi_{n}\frac{\overleftarrow{\fnot\partial_T}}{\overleftarrow{\partial_-}}\xi_{\bar n}
-\bar n^\mu \bar \xi_{\bar n}\frac{\overrightarrow{\fnot\partial_T}}{\overrightarrow{\partial_-}}\xi_{n}.
\end{eqnarray}
The terms $J_{1'1}^\mu$ and $J_{11'}^\mu$ restore the electric charge-conservation at $\sim\lambda^3$ and $\sim \lambda^4$ orders. Indeed,
\begin{eqnarray}\label{current-conserve-NLP}
\partial_\mu J_{11}^\mu&=&-\bar \xi_{\bar n}\(\frac{\overleftarrow{\fnot \partial_T}}{\overleftarrow{\partial_+}}\overrightarrow{\partial_+}
+
\overleftarrow{\partial_-}\frac{\overrightarrow{\fnot \partial_T}}{\overrightarrow{\partial_-}}\)\xi_{n}
-\bar \xi_{n}\(\frac{\overleftarrow{\fnot \partial_T}}{\overleftarrow{\partial_-}}\overrightarrow{\partial_-}
+
\overleftarrow{\partial_+}\frac{\overrightarrow{\fnot \partial_T}}{\overrightarrow{\partial_+}}\)\xi_{\bar n}=\mathcal{O}(\lambda^5).
\end{eqnarray}
Similarly, for $J_{21}^\mu$ terms
\begin{eqnarray}
\partial_\mu J^\mu_{21}&=&
ig 
\bar \xi_{\bar n}\fnot A_{\bar n,T}\(
\frac{\overleftarrow{\partial_-}}{\overrightarrow{\partial_-}}-\frac{\overrightarrow{\partial_+}}{\overleftarrow{\partial_+}}\)\xi_n
-
ig \bar \xi_{n}\(\frac{\overrightarrow{\partial_-}}{\overleftarrow{\partial_-}}-\frac{\overleftarrow{\partial_+}}{\overrightarrow{\partial_+}}\)\fnot A_{\bar n,T}\xi_{\bar n}=\mathcal{O}(\lambda^5).
\end{eqnarray}
Clearly, the operators with a particular twist combinations form series where each next term restores the charge conservation to a higher power. These series are known as series of \textit{kinematic power corrections}. So, the operators $J^\mu_{1'1}$ and $J^\mu_{11'}$ are the kinematic power corrections to the LP current. Due to the that fact the kinematic power corrections restore the global properties of current (and hadronic tensor), all operators in the series must have the same coefficient function. We demonstrate it explicitly at NLO in the next section.

The effective operator at NLP is obtained by composing LP and NLP terms of EM currents. We obtain 
\begin{eqnarray}\label{tree:NLP}
\mathcal{J}_{\text{NLP}}^{\mu\nu}&=&
-\frac{n^\mu \gamma^\rho_{T,ij}\gamma^\nu_{T,kl}+n^\nu \gamma^\mu_{T,ij}\gamma^\rho_{T,kl}}{N_c}
\Big(
\frac{\partial_{\rho}}{\partial_{+}}\mathcal{O}_{11,\bar n}^{li}
\overline{\mathcal{O}}_{11,n}^{jk}
+
\frac{\partial_{\rho}}{\partial_{+}}\overline{\mathcal{O}}_{11,\bar n}^{jk}
\mathcal{O}_{11,n}^{li}
\Big)
\\\nn &&
-\frac{\bar n^\mu \gamma^\rho_{T,ij}\gamma^\nu_{T,kl}+\bar n^\nu \gamma^\mu_{T,ij}\gamma^\rho_{T,kl}}{N_c}
\Big(
\mathcal{O}_{11,\bar n}^{li}
\frac{\partial_{\rho}}{\partial_{-}}\overline{\mathcal{O}}_{11,n}^{jk}
+
\overline{\mathcal{O}}_{11,\bar n}^{jk}
\frac{\partial_{\rho}}{\partial_{-}}\mathcal{O}_{11,n}^{li}
\Big)
\\\nn &&
+ig \frac{ \delta_{ij}\gamma^\nu_{T,kl}}{N_c}\Bigg\{
\mathbb{O}_{21,\bar n}^{li}\(\frac{\bar n^\mu}{\overrightarrow{\partial_-}}-\frac{n^\mu}{\overleftarrow{\partial_+}}\)\overline{\mathcal{O}}_{11,n}^{jk}
-
\overline{\mathbb{O}}_{21,\bar n}^{jk}\(\frac{\bar n^\mu}{\overrightarrow{\partial_-}}-\frac{n^\mu}{\overleftarrow{\partial_+}}\)\mathcal{O}_{11,n}^{li}
\\\nn && \qquad\qquad \qquad+
\mathcal{O}_{11,\bar n}^{li}\(\frac{\bar n^\mu}{\overrightarrow{\partial_-}}-\frac{n^\mu}{\overleftarrow{\partial_+}}\)\overline{\mathbb{O}}_{21,n}^{jk}
-
\overline{\mathcal{O}}_{11,\bar n}^{jk}\(\frac{\bar n^\mu}{\overrightarrow{\partial_-}}-\frac{n^\mu}{\overleftarrow{\partial_+}}\)\mathbb{O}_{21,n}^{li}
\Bigg\}
\\\nn &&
+ig \frac{\gamma^\mu_{T,ij}\delta_{kl}}{N_c}\Bigg\{
\mathbb{O}_{12,\bar n}^{li}
\(\frac{\bar n^\nu}{\overrightarrow{\partial_-}}-\frac{n^\nu}{\overleftarrow{\partial_+}}\)
\overline{\mathcal{O}}_{11,n}^{jk}
-
\overline{\mathbb{O}}_{12,\bar n}^{jk}\(\frac{\bar n^\nu}{\overrightarrow{\partial_-}}-\frac{n^\nu}{\overleftarrow{\partial_+}}\)\mathcal{O}_{11,n}^{li}
\\\nn && \qquad\qquad \qquad+
\mathcal{O}_{11,\bar n}^{li}\(\frac{\bar n^\nu}{\overrightarrow{\partial_-}}-\frac{n^\nu}{\overleftarrow{\partial_+}}\)\overline{\mathbb{O}}_{12,n}^{jk}
-
\overline{\mathcal{O}}_{11,\bar n}^{jk}\(\frac{\bar n^\nu}{\overrightarrow{\partial_-}}-\frac{n^\nu}{\overleftarrow{\partial_+}}\)\mathbb{O}_{12,n}^{li}
\Bigg\},
\end{eqnarray}
where $\delta_{ij}$ is the Kronecker delta for spinor indices. All derivatives are with respect to the coordinate $y$, and act only on the subsequent operator. To derive this expression we have used the identity
\begin{eqnarray}
\bar \xi_{\bar n,j}(y^-n+y_T)\frac{\overleftarrow{\partial_{\rho}}}{\overleftarrow{\partial_+}}\xi_{\bar n,i}(0)
=
\bar \xi_{\bar n,j}(y^-n+y_T)\frac{\overrightarrow{\partial_{\rho}}}{\overrightarrow{\partial_+}}\xi_{\bar n,i}(0)=\frac{\partial_{\rho}}{\partial_+}O_{11,\bar n}^{ij}(\{y^-,0\},y_T),
\end{eqnarray}
and similar for anti-collinear fields. To derive this identity we assume that the total derivatives of TMD operators can be eliminated, since they do not contribute to the forward matrix elements.  Here, TMD operators of twist-(1+1) have the arguments $(\{y^-,0\},y_T)$. All TMD operators of twist-(1+2) and (2+1) have argument $(\{y^-,y^-,0\},y_T)$ and $(\{y^-,0,0\},y_T)$ respectively.  The TMD operators of twist-(1+2) and twist-(2+1) are
\begin{align}\label{def:O21bb-position}
\mathbb{O}_{21,\bar n}^{ji}(\{y^-,y^-,0\},y_T)&= 
[\bar \xi_{\bar n}^{(-)}\fnot A^{(-)}_{\bar n,T}(y^-n+y_T)]_i\xi_{\bar n,j}^{(+)}(0),
\\\label{def:O12bb-position}
\mathbb{O}_{12,\bar n}^{ji}(\{y^-,0\phantom{^-},0\},y_T)&= 
\bar \xi_{\bar n,i}^{(-)}(y^-n+y_T)[\fnot A^{(+)}_{\bar n,T}\xi_{\bar n}^{(+)}(0)]_{j},
\\\label{def:O21bbbar-position}
\overline{\mathbb{O}}_{21,\bar n}^{ji}(\{y^-,y^-,0\},y_T)&= 
[\fnot A^{(-)}_{\bar n,T} \xi^{(-)}_{\bar n}(y^-n+y_T)]_j \bar \xi^{(+)}_{\bar n,i}(0),
\\\label{def:O12bbbar-position}
\overline{\mathbb{O}}_{12,\bar n}^{ji}(\{y^-,0\phantom{^-},0\},y_T)&= 
\xi^{(-)}_{\bar n,j}(y^-n+y_T)[\bar \xi_{\bar n}^{(+)}\fnot A^{(+)}_{\bar n,T}(0) ]_{i}.
\end{align}

\subsection*{Gauge invariant expressions for TMD operators}

The definitions in eq.~(\ref{def:O21bb-position}-\ref{def:O12bbbar-position}) are given in  light-cone gauge. They can be written in  gauge invariant form using the relations in eq.~(\ref{def:A->F}, \ref{def:A->D}). There are two usual ways to write the twist-(2+1) operators, the one typical for the SCET literature e.g. \cite{Beneke:2017ztn,Inglis-Whalen:2021bea}, and the one typical for direct QCD computations e.g. \cite{Boer:2003cm,Hu:2021naj}. Both have advantages and disadvantages. In particular, the SCET-like notation is convenient for the computation of hard coefficient function, whereas the traditional-like is convenient for computation of the evolution properties. Nicely, they are related by a simple transformation. In the present work we use both kinds of notations, designating them by different fonts.

The elementary building blocks are the semi-compact operators of twist-1 and twist-2. They are
\begin{eqnarray}\label{def:U1}
U_{1,\bar n}(z,b)&=&[Ln+b,zn+b]\xi_{\bar n}(zn+b),
\\\label{def:U2}
U^\mu_{2,\bar n}(\{z_1,z_2\},b)&=&g[Ln+b,z_1n+b]F_{\bar n}^{\mu+}[z_1n+b,z_2n+b]\xi_{\bar n}(z_2n+b),
\end{eqnarray}
where index $\mu$ is transverse and we omit the transverse links. Both operators have an open spinor and color indices. The operator $U^\mu_{2,\bar n}$ has open spinor and vector indices, and is more general than the operator appearing at NLP. The latter reads
\begin{eqnarray}\label{def:U2-i}
U_{2,\bar n}(\{z_1,z_2\},b)=\gamma_{T\mu} U^\mu_{2,\bar n}(\{z_1,z_2\},b),
\end{eqnarray}
where both sides are spinors. The semi-compact operator built from anti-collinear fields are obtained from $U_{\bar n}$ with replacement $n\leftrightarrow \bar n$. In addition, we define the operators
\begin{eqnarray}\label{def:U1-C}
\overline{U}_{1,\bar n}(z,b)&=&\bar \xi_{\bar n}(zn+b)[zn+b,Ln+b],
\\\label{def:U2-C}
\overline{U}^\mu_{2,\bar n}(\{z_1,z_2\},b)&=&g \,\bar \xi_{\bar n}(z_1n+b)[z_1n+b,z_2n+b]F_{\bar n}^{\mu+}[z_2n+b,Ln+b].
\end{eqnarray}
Analogously, the reduced twist-2 operator $\overline{U}_{2}$ is
\begin{eqnarray}\label{def:U2bar-i}
\overline{U}_{2,\bar n}(\{z_1,z_2\},b)= \overline{U}^\mu_{2,\bar n}(\{z_1,z_2\},b)\gamma_{T\mu}.
\end{eqnarray}

The TMD operators are products of semi-compact operators. So, the twist-(1+1) operators in eq.~(\ref{def:O11-position},\ref{def:O11bar-position}) are simply
\begin{eqnarray}\label{O=UU:11}
\mathcal{O}_{11,\bar n}(\{z_1,z_2\},b)&=&\overline{U}^{(-)}_{1,\bar n}(z_1,b)U^{(+)}_{1,\bar n}(z_2,0),
\\
\overline{\mathcal{O}}_{11,\bar n}(\{z_1,z_2\},b)&=&U^{(-)}_{1,\bar n}(z_1,b)\overline{U}^{(+)}_{1,\bar n}(z_2,0).
\end{eqnarray}
The twist-(2+1) and twist-(1+2) operators are
\begin{eqnarray}\label{def:O21-position}
\mathcal{O}_{21,\bar n}(\{z_1,z_2,z_3\},b)&=&\overline{U}^{(-)}_{2,\bar n}(\{z_1,z_2\},b)U^{(+)}_{1,\bar n}(z_3,0),
\\\label{def:O12-position}
\mathcal{O}_{12,\bar n}(\{z_1,z_2,z_3\},b)&=&\overline{U}^{(-)}_{1,\bar n}(z_1,b)U^{(+)}_{1,\bar n}(\{z_2,z_3\},0),
\\\label{def:O21bar-position}
\overline{\mathcal{O}}_{21,\bar n}(\{z_1,z_2,z_3\},b)&=&U^{(-)}_{2,\bar n}(\{z_2,z_1\},b)\overline{U}^{(+)}_{1,\bar n}(z_3,0),
\\\label{def:O12bar-position}
\overline{\mathcal{O}}_{12,\bar n}(\{z_1,z_2,z_3\},b)&=&U^{(-)}_{1,\bar n}(z_1,b)\overline{U}^{(+)}_{2,\bar n}(\{z_3,z_2\},0).
\end{eqnarray}
The superscripts $(\pm)$ indicate that semi-compact operators are made out of corresponding causal or anti-causal fields. Notice the enumeration of positions in operators $\overline{\mathcal{O}}_{21,\bar n}$ and $\overline{\mathcal{O}}_{12,\bar n}$. It is adjusted such that the gluon field has uniformly coordinate $z_2$. All these operators are matrices in spinor space, and singlets in  color space. The operators $\overline{\mathcal{O}}$ are related to $\mathcal{O}$ by the charge-conjugation. The matrix elements of these pairs define quark and anti-quark TMD distributions, such as in ref.~\cite{Bacchetta:2006tn}.

The definitions of eq.~(\ref{def:U1} - \ref{def:U2-C}) are given in the traditional QCD basis. In the SCET literature instead, one would use
\begin{eqnarray}
\mathbb{U}^\mu_{2,\bar n}(\{z_1,z_2\},b)&=&-i[Ln+b,z_1n+b]\overleftarrow{D}^\mu_{\bar n}[z_1n+b,z_2n+b]\xi_{\bar n}(z_2n+b),
\\
\overline{\mathbb{U}}^\mu_{2,\bar n}(\{z_1,z_2\},b)&=&
i \,\bar \xi_{\bar n}(z_1n+b)[z_2n+b,z_2n+b]\overrightarrow{D}^\mu_{\bar n}[z_2n+b,Ln+b],
\end{eqnarray}
where $D$ is the covariant derivative.  The operators $\mathbb{O}$ in eq.~(\ref{def:O21bb-position}-\ref{def:O12bbbar-position}) are defined with replacement of $U\to\mathbb{U}$, e.g.
\begin{eqnarray}
\mathbb{O}_{21,\bar n}(\{z_1,z_2,z_3\},b)&=&\overline{\mathbb{U}}^{(-)}_{2,\bar n}(\{z_1,z_2\},b)U^{(+)}_{1,\bar n}(z_3,0).
\end{eqnarray}
The relations between the operators $\mathbb{O}$ and $\mathcal{O}$ is
\begin{eqnarray}\label{O->Obb}
\mathcal{O}_{NM,\bar n}(\{z_1,z_2,z_3\},b)=-\frac{\partial}{\partial z_2}\mathbb{O}_{NM,\bar n}(\{z_1,z_2,z_3\},b),
\end{eqnarray}
where $N$ and $M$ are 1 or 2. The relation (\ref{O->Obb}) can be inverted
\begin{eqnarray}
\mathbb{O}_{NM,\bar n}(\{z_1,z_2,z_3\},b)=-\int_{L}^{z_2}\mathcal{O}_{NM,\bar n}(\{z_1,\sigma,z_3\},b).
\end{eqnarray}
Similar expressions hold for $\overline{\mathbb{O}}$-type operators.

\subsection*{TMD operators in momentum space}

Ordinary, the TMD distributions are defined in the the mixed momentum-coordinate representation. Namely, they are Fourier-transformed just in light-cone coordinates. Such a transformation provides a similarity with the parton densities (which are defined in  terms of fractions of momentum) and it preserves a simple structure of the TMD evolution (which is diagonal in  transverse-position space). We also follow this practice and define
\begin{eqnarray}\label{def:U1-momentum}
U_{1,\bar n}(z,b)&=&p_+ \int dx e^{ixzp_+}U_{1,\bar n}(x,b),
\\\label{def:U2-momentum}
U_{2,\bar n}(\{z_1,z_2\},b)&=&p_+^2 \int dx_1dx_2 e^{i(x_1z_1+x_2z_2)p_+}U_{2,\bar n}(x_{1,2},b),
\end{eqnarray}
where $x_{1,2}$ is a shorthand notation for $(x_1,x_2)$, which can be interpreted as the fraction parton's momentum, and $p_+$ is the hadron's momentum. The $\overline{U}_{1,\bar n}(x,b)$ and $\overline{U}_{2,\bar n}(x_{1,2},b)$ have similar definitions.

The TMD distributions are defined by  forward matrix elements and they are insensitive to the global positioning of the operator,
\begin{eqnarray}\label{translation-inv}
\langle p|\mathcal{O}(z)|p\rangle = \langle p|\mathcal{O}(z+a)|p\rangle.
\end{eqnarray}
Taking into account the translation invariance, we define
\begin{eqnarray}\label{def:O11}
\mathcal{O}_{11,\bar n}(\{z,0\},b)&=&p_+ \int dx \, e^{izxp_+}\mathcal{O}_{11,\bar n}(x,b),
\\\label{def:O11bar}
\overline{\mathcal{O}}_{11,\bar n}(\{z,0\},b)&=&p_+ \int dx \, e^{izxp_+}\overline{\mathcal{O}}_{11,\bar n}(x,b),
\\\label{def:OMN}
\mathcal{O}_{MN,\bar n}(\{z_1,z_2,z_3\},b)&=&p_+^2 \int [dx] \, e^{i(z_1x_1+z_2x_2+z_3x_3) p_+}\mathcal{O}_{MN,\bar n}(x_{1,2,3},b),
\\\label{def:OMNbar}
\overline{\mathcal{O}}_{MN,\bar n}(\{z_1,z_2,z_3\},b)&=&p_+^2 \int [dx] \, e^{i(z_1x_1+z_2x_2+z_3x_3) p_+}\overline{\mathcal{O}}_{MN,\bar n}(x_{1,2,3},b),
\end{eqnarray}
where $M+N=3$, and
\begin{eqnarray}\label{eq:im}
\int [dx]=\int dx_1dx_2dx_3\,\delta(x_1+x_2+x_3).
\end{eqnarray}
At this point, we cannot not specify the domain of integration over $x$. However, once the matrix elements are taken, the values of $x$ are restricted $x\in[-1,1]$ for parton distributions, and $x\in(-\infty,-1]\cup[1,\infty)$ for fragmentation functions. The integral measure in eq.~(\ref{eq:im}) takes into account the translation invariance of the TMD operator defined in eq.~(\ref{translation-inv}) \footnote{
The twist-(1+1) operator can be defined in translation invariant way as well,
$$\mathcal{O}_{11,\bar n}(\{z_1,z_2\},b)=p_+ \int dx_1 dx_2 \,\delta(x_1+x_2)
\,e^{i(z_1x_1+z_2x_2)p_+}\mathcal{O}_{11,\bar n}(x_{1,2},b).$$}.
In principle, our computation (since it is done in  position space) does not imply any simplification that comes from the translation invariance. Nonetheless, we use it to make notation somewhat lighter.

The relation between the operators $\mathbb{O}$ and $\mathcal{O}$ is simplified in  momentum space. According to eq.~(\ref{O->Obb}), we have
\begin{eqnarray}
\mathbb{O}_{MN,\bar n}(\{z_1,z_2,z_3\},b)&=&p_+^2 \int [dx] \, e^{i(z_1x_1+z_2x_2+z_3x_3) p_+}\frac{i}{x_2p_+}\mathcal{O}_{MN,\bar n}(x_{1,2,3},b).
\end{eqnarray}

\subsection*{EM current in momentum space}

Using the definitions in eq.~(\ref{def:U1-momentum}, \ref{def:U2-momentum}) we present the EM current eq.~(\ref{def:J11}, \ref{tree:JNLP}) in the form
\begin{eqnarray}\label{tree:EM-current-momentum}
J^\mu(y)&=&p_1^+p_2^-\int d x d\tilde x \,e^{ixp_1^+y^-+i\tilde xp_2^-y^+} J^\mu_{11}(x,\tilde x,y_T)
\\\nn &&
+(p_1^+)^2p_2^-\int d x_1 dx_2 d\tilde x \,e^{i(x_1+x_2)p_1^+y^-+i\tilde xp^-y^+} J^\mu_{21}(x_{1,2},\tilde x,y_T)
\\\nn &&
+p_1^+(p_2^-)^2\int d x d\tilde x_1 d\tilde x_2 \,e^{ixp^+_1y^-+i(\tilde x_1+\tilde x_2)p^-_2y^+}J^\mu_{12}(x,\tilde x_{1,2},y_T)+...~,
\end{eqnarray}
where all partons momenta are incoming. Performing the transformation for eq.(\ref{tree:JLP}, \ref{tree:JNLP}) we get
\begin{eqnarray}\label{def:J11-momentum}
J^\mu_{11}(x,\tilde x)&=&
\overline{U}_{1,\bar n}(x)\gamma^\mu_T U_{1,n}(\tilde x)
+\overline{U}_{1,n}(\tilde x)\gamma^\mu_T U_{1,\bar n}(x)
\\\nn &&
+\frac{in^\mu}{x p_1^+}\Big(
\overline{U}_{1,\bar n}(x)\overleftarrow{\fnot \partial_T}U_{1,n}(\tilde x)
+\overline{U}_{1,n}(\tilde x)\fnot \partial_TU_{1,\bar n}(x)\Big)
\\\nn &&
+\frac{i\bar n^\mu}{\tilde xp_2^-}\Big(
\overline{U}_{1,n}(\tilde x)\overleftarrow{\fnot \partial_T}U_{1,\bar n}(x)
+\overline{U}_{1,\bar n}(x)\fnot \partial_TU_{1,n}(\tilde x)\Big),
\\\label{def:J21-momentum}
J^\mu_{21}(x_{1,2},\tilde x)&=&
\frac{i}{x_2p_1^+}\(\frac{\bar n^\mu}{\tilde xp_2^-}-\frac{n^\mu}{(x_1+x_2)p_1^+}\)
\(\overline{U}_{2,\bar n}(x_{1,2})U_{1,n}(\tilde x)
-
\overline{U}_{1,n}(\tilde x)U_{2,\bar n}(x_{2,1})
\),
\\\label{def:J12-momentum}
J^\mu_{12}(x,\tilde x_{1,2})&=&
\frac{i}{\tilde x_2p_2^-}\(\frac{n^\mu}{x p_1^+}-\frac{\bar n^\mu}{(\tilde x_1+\tilde x_2)p_2^-}\)
\(\overline{U}_{2,n}(\tilde x_{1,2})U_{1,\bar n}(x)
-
\overline{U}_{1,\bar n}(x)U_{2,n}(\tilde x_{2,1})
\),
\end{eqnarray}
where $x_{2,1}=(x_2,x_1)$ and the repeating argument $y_T$ is suppressed for  brevity. Note that the arguments $x_1$ and $x_2$ are adjusted such that $x_1$ is the momentum fraction of quark or anti-quark, and $x_2 $ is the momentum fraction of the gluon.

\subsection*{Effective operator in momentum space}

Similar expressions can be written for the effective currents eq.~(\ref{tree:LP}, \ref{tree:NLP}). In this case, it is convenient to take into account the Fourier integral that is present in the hadronic tensor, eq.~(\ref{def:W}). Defining
\begin{eqnarray}\label{tree:def:J-momentum}
\mathcal{J}^{\mu\nu}_{\text{eff}}(q)=\int \frac{d^4 y}{(2\pi)^4}e^{-i(qy)}\mathcal{J}^{\mu\nu}_{\text{eff}}(y),
\end{eqnarray}
we derive
\begin{eqnarray}\label{tree:J-momentum-deltas}
&&\mathcal{J}^{\mu\nu}_{\text{eff}}(q)=\int \frac{d^2b}{(2\pi)^2}
e^{-i(qb)}\Bigg\{
\int dx d\tilde x
\delta\(x-\frac{q^+}{p_1^+}\)
\delta\(\tilde x-\frac{q^-}{p_2^-}\)
\mathcal{J}^{\mu\nu}_{1111}(x,\tilde x,b)
\\\nn &&\qquad
+\int [dx]d\tilde x
\delta\(\tilde x-\frac{q^-}{p_2^-}\)
\(
\delta\(x_1-\frac{q_1^+}{p_1^+}\)
\mathcal{J}^{\mu\nu}_{1211}(x,\tilde x,b)
+
\delta\(x_3+\frac{q_1^+}{p_1^+}\)
\mathcal{J}^{\mu\nu}_{2111}(x,\tilde x,b)\)
\\\nn &&\qquad
+\int dx [d\tilde x]
\delta\(x-\frac{q^+}{p_1^+}\)
\(
\delta\(\tilde x_1-\frac{q^-}{p_2^-}\)
\mathcal{J}^{\mu\nu}_{1112}(x,\tilde x,b)
+
\delta\(\tilde x_3+\frac{q^-}{p_2^-}\)
\mathcal{J}^{\mu\nu}_{1121}(x,\tilde x,b)
\)
\\\nn &&\qquad
+...
\Bigg\},
\end{eqnarray}
where
\begin{eqnarray} \label{def:J1111}
&&\mathcal{J}^{\mu\nu}_{1111}(x,\tilde x,b)=
\frac{\gamma^\mu_{T,ij}\gamma^\nu_{T,kl}}{N_c}\Big(
\mathcal{O}_{11,\bar n}^{li}(x,b)
\overline{\mathcal{O}}_{11,n}^{jk}(\tilde x,b)
+
\overline{\mathcal{O}}_{11,\bar n}^{jk}(x,b)
\mathcal{O}_{11,n}^{li}(\tilde x,b)
\Big)
\\\nn &&\quad
+i\frac{n^\mu \gamma^\rho_{T,ij}\gamma^\nu_{T,kl}+n^\nu \gamma^\mu_{T,ij}\gamma^\rho_{T,kl}}{q^+\,N_c}
\Big(
\partial_{\rho}\mathcal{O}_{11,\bar n}^{li}(x,b)
\overline{\mathcal{O}}_{11,n}^{jk}(\tilde x,b)
+
\partial_{\rho}\overline{\mathcal{O}}_{11,\bar n}^{jk}(x,b)
\mathcal{O}_{11,n}^{li}(\tilde x,b)
\Big)
\\\nn &&\quad
+i\frac{\bar n^\mu \gamma^\rho_{T,ij}\gamma^\nu_{T,kl}+\bar n^\nu \gamma^\mu_{T,ij}\gamma^\rho_{T,kl}}{q^-\,N_c}
\Big(
\mathcal{O}_{11,\bar n}^{li}(x,b)
\partial_{\rho}\overline{\mathcal{O}}_{11,n}^{jk}(\tilde x,b)
+
\overline{\mathcal{O}}_{11,\bar n}^{jk}(x,b)
\partial_{\rho}\mathcal{O}_{11,n}^{li}(\tilde x,b)
\Big),
\end{eqnarray}
\begin{eqnarray}
&&\mathcal{J}^{\mu\nu}_{1211}(x,\tilde x,b)=
\\\nn &&\qquad
\frac{ig}{x_2}\(\frac{\bar n^\nu}{q^-}-\frac{n^\nu}{q^+}\) \frac{\gamma^\mu_{T,ij}\delta_{kl}}{N_c}\Big(
\mathcal{O}_{12,\bar n}^{li}(x,b)\overline{\mathcal{O}}_{11,n}^{jk}(\tilde x,b)
-
\overline{\mathcal{O}}_{12,\bar n}^{jk}(x,b)\mathcal{O}_{11,n}^{li}(\tilde x,b)
\Big),
\\&&\label{def:2111}
\mathcal{J}^{\mu\nu}_{2111}(x,\tilde x,b)=
\\\nn &&\qquad
\frac{ig}{x_2}\(\frac{\bar n^\mu}{q^-}-\frac{n^\mu}{q^+}\)\frac{\delta_{ij}\gamma^\nu_{T,kl}}{N_c}\Big(
\mathcal{O}_{21,\bar n}^{li}(x,b)\overline{\mathcal{O}}_{11,n}^{jk}(\tilde x,b)
-
\overline{\mathcal{O}}_{21,\bar n}^{jk}(x,b)\mathcal{O}_{11,n}^{li}(\tilde x,b)
\Big),
\\
&&\mathcal{J}^{\mu\nu}_{1112}(x,\tilde x,b)=
\\\nn &&\qquad
\frac{ig}{\tilde x_2}\(\frac{\bar n^\nu}{q^-}-\frac{n^\nu}{q^+}\) \frac{\gamma^\mu_{T,ij}\delta_{kl}}{N_c}\Big(
\mathcal{O}_{11,\bar n}^{li}(x,b)\overline{\mathcal{O}}_{12,n}^{jk}(\tilde x,b)
-
\overline{\mathcal{O}}_{11,\bar n}^{jk}(x,b)\mathcal{O}_{12,n}^{li}(\tilde x,b)
\Big),
\\ &&\label{def:J1121}
\mathcal{J}^{\mu\nu}_{1121}(x,\tilde x,b)=
\\\nn &&\qquad
\frac{ig}{\tilde x_2}\(\frac{\bar n^\mu}{q^-}-\frac{n^\mu}{q^+}\)\frac{\delta_{ij}\gamma^\nu_{T,kl}}{N_c}\Big(
\mathcal{O}_{11,\bar n}^{li}(x,b)\overline{\mathcal{O}}_{21,n}^{jk}(\tilde x,b)
-
\overline{\mathcal{O}}_{11,\bar n}^{jk}(x,b)\mathcal{O}_{21,n}^{li}(\tilde x,b)
\Big).
\end{eqnarray}
In the twist-(1+2) and twist-(2+1) operators the argument is $x=(x_1, x_2,x_3)$. Note that the operators $\mathcal{J}_{1211}^{\mu\nu}$ and $\mathcal{J}_{2111}^{\mu\nu}$ are $\mathcal{J}_{1112}^{\mu\nu}$ and $\mathcal{J}_{1121}^{\mu\nu}$ with $n\leftrightarrow\bar n$. 

In the case of twist-(1+2) TMD operators, the values of $x_{1,2,3}$ are not sign definite. The momentum conservation delta-function fixes the value (and the sign) of only one of them. The other two variables are integrated and can be positive or negative.

To derive these expressions we took into account that the global position of the currents is irrelevant for DY, SIDIS and SIA processes. Also note, that the definition of the Fourier transform in eq.~(\ref{tree:def:J-momentum}) is taken similar to the DY case in eq.~(\ref{def:W}). According to eq.~(\ref{def:W-SIDIS}, \ref{def:W-SIA}) the sign of $q$ is the opposite for SIDIS and SIA.

The expressions for the currents (\ref{def:J1111} - \ref{def:J1121}) are simple to generalize to the case of electro-weak interaction. In this case, one needs to replace $\gamma^\mu_T\to \gamma_T^\mu(g_V\pm g_A\gamma^5)$, and $\delta\to g_V\pm g_A\gamma^5$ (the sign of $g_A$ depends on the term). The general structure and other factors is the same.

\section{NLO perturbative correction to NLP operator}
\label{sec:NLO}

The computation of perturbative corrections to effective operators follows the same path as  in sec.~\ref{sec:general-structure}. Here we find another difference between TMD factorization and collinear factorization. Namely, the loop integrals can produce extra suppressing factors. That happens due to the unhomogeneous counting rule for the vector $y$ in eq.~(\ref{y-counting}). In particular, the diagrams with the interaction of causal and anti-causal fields are suppressed by a factor $\lambda^2$. So, the twist-(1+1)$\times$(1+1) contribution of the exchange configuration is N$^2$LP (see ref.~\cite{Ebert:2018gsn,Inglis-Whalen:2021bea} for a discussion about these contributions). Therefore, at NLP we can consider only the interactions within a single causal sector, i.e. for each EM current independently.

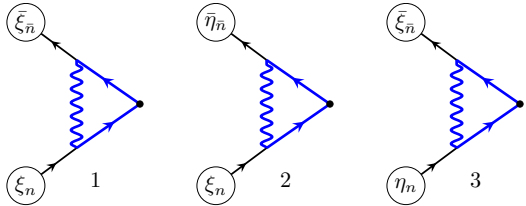
\begin{figure}[t]
\begin{center}
\begin{tikzpicture}

\coordinate (A) at (2, 1.5);

\draw[very thick, blue,postaction={decorate, decoration={
    markings,
    mark=at position 0.6 with {\arrow{stealth}}}}] ($(A)$)--($(A)+(-1.,0.7) $);
\draw[thick,postaction={decorate, decoration={
    markings,
    mark=at position 0.5 with {\arrow{stealth}}}}] ($(A)+(-1.,0.7) $)--($(A)+(-1.8,1.3)$);
\draw[very thick, blue,postaction={decorate, decoration={
    markings,
    mark=at position 0.6 with {\arrow{stealth}}}}] ($(A)+(-1.,-0.7)$) -- ($(A)$);
\draw[thick,postaction={decorate, decoration={
    markings,
    mark=at position 0.6 with {\arrow{stealth}}}}] ($(A)+(-1.8,-1.3)$)--($(A)+(-1.,-0.7) $);
\draw[very thick, blue,style={decorate, decoration={snake,segment length=6.2}}] ($(A)+(-1.,-0.7)$)--($(A)+(-1.,0.7)$);

\draw[fill=black] ($(A)$) circle (0.05cm);    
\draw[fill=white] ($(A)+(-1.8,-1.3)$) circle (0.3cm);
\draw[fill=white] ($(A)+(-1.8,1.3)$) circle (0.3cm);
\node[] at ($(A)+(-1.8,1.3)$) {$\Ds\bar \xi_{\bar n}$};
\node[] at ($(A)+(-1.8,-1.3)$) {$\Ds \xi_{n}$};
\node[] at ($(A)+(-0.7,-1.2)$) {1};

\coordinate (A) at (5, 1.5);

\draw[very thick, blue,postaction={decorate, decoration={
    markings,
    mark=at position 0.6 with {\arrow{stealth}}}}] ($(A)$)--($(A)+(-1.,0.7) $);
\draw[thick,postaction={decorate, decoration={
    markings,
    mark=at position 0.5 with {\arrow{stealth}}}}] ($(A)+(-1.,0.7) $)--($(A)+(-1.8,1.3)$);
\draw[very thick, blue,postaction={decorate, decoration={
    markings,
    mark=at position 0.6 with {\arrow{stealth}}}}] ($(A)+(-1.,-0.7)$) -- ($(A)$);
\draw[thick,postaction={decorate, decoration={
    markings,
    mark=at position 0.6 with {\arrow{stealth}}}}] ($(A)+(-1.8,-1.3)$)--($(A)+(-1.,-0.7) $);
\draw[very thick, blue,style={decorate, decoration={snake,segment length=6.2}}] ($(A)+(-1.,-0.7)$)--($(A)+(-1.,0.7)$);

\draw[fill=black] ($(A)$) circle (0.05cm);    
\draw[fill=white] ($(A)+(-1.8,-1.3)$) circle (0.3cm);
\draw[fill=white] ($(A)+(-1.8,1.3)$) circle (0.3cm);
\node[] at ($(A)+(-1.8,1.3)$) {$\Ds\bar \eta_{\bar n}$};
\node[] at ($(A)+(-1.8,-1.3)$) {$\Ds \xi_{n}$};
\node[] at ($(A)+(-0.7,-1.2)$) {2};

\coordinate (A) at (8, 1.5);

\draw[very thick, blue,postaction={decorate, decoration={
    markings,
    mark=at position 0.6 with {\arrow{stealth}}}}] ($(A)$)--($(A)+(-1.,0.7) $);
\draw[thick,postaction={decorate, decoration={
    markings,
    mark=at position 0.5 with {\arrow{stealth}}}}] ($(A)+(-1.,0.7) $)--($(A)+(-1.8,1.3)$);
\draw[very thick, blue,postaction={decorate, decoration={
    markings,
    mark=at position 0.6 with {\arrow{stealth}}}}] ($(A)+(-1.,-0.7)$) -- ($(A)$);
\draw[thick,postaction={decorate, decoration={
    markings,
    mark=at position 0.6 with {\arrow{stealth}}}}] ($(A)+(-1.8,-1.3)$)--($(A)+(-1.,-0.7) $);
\draw[very thick, blue,style={decorate, decoration={snake,segment length=6.2}}] ($(A)+(-1.,-0.7)$)--($(A)+(-1.,0.7)$);

\draw[fill=black] ($(A)$) circle (0.05cm);    
\draw[fill=white] ($(A)+(-1.8,-1.3)$) circle (0.3cm);
\draw[fill=white] ($(A)+(-1.8,1.3)$) circle (0.3cm);
\node[] at ($(A)+(-1.8,1.3)$) {$\Ds\bar \xi_{\bar n}$};
\node[] at ($(A)+(-1.8,-1.3)$) {$\Ds \eta_{n}$};
\node[] at ($(A)+(-0.7,-1.2)$) {3};

\end{tikzpicture}
\caption{\label{fig:loop-2point}   The two-points diagrams contributing to the NLO of the effective operator. The diagram 1 contributes to LP, and the diagrams 2 \& 3 to NLP. The diagrams with $n\leftrightarrow \bar n$ should be added. The blobs indicate the type of background field. The blue lines are the dynamical fields.} 
\end{center}
\end{figure}

At NLO and  NLP we have only two-point and three-point relevant diagrams, that are shown in fig.~\ref{fig:loop-2point}  and fig.~\ref{fig:loop-3point}, respectively.  The two-point diagrams couple operators of twist-1 and operators of twist-1 or 2, while  
the three point diagrams couple only operators of twist-1 and twist-2. 

The computation of the diagrams with the background field is slightly different from the ordinary computation of amplitudes because a part of the loop-variables are also arguments of the background fields. Therefore, the integration over these arguments cannot be computed and we obtain the integral convolution in eq.~(\ref{a-term}). To compute the integrals over the rest of the loop variables we used a technique, which has been developed in ref.~\cite{Scimemi:2019gge}. This technique is based on a series of shifts for integration variables and leads to the usual Lorentz-invariant loop-integrals. The technique works equally well for any type of power-suppressed diagrams. For pedagogical purposes in appendix~\ref{app:diag4} we present a detailed computation of the three-point diagram 5 in fig.~\ref{fig:loop-3point}. 

Alternatively, one can present the background field as a Fourier image and perform the computation in  momentum space, using standard methods. However, the loop-integral in momentum space becomes complicated once a large number of background fields participate. For example, three-point diagrams in  momentum space can contain polylogarithms already at one-loop level, whereas one-loop expressions in  position space are polynomial for any number of external fields. 

Another complication of a momentum space computation is the necessity to specify the direction of the parton's momentum, which is crucial for the computation of Feynman variable integrals. This is not a problem for two-point diagrams, where one has a unique choice (however, different for DY, SIDIS and SIA cases), but for the three-point diagrams one has already three possible combinations of momenta directions. This is because the individual momenta of the quark-gluon pair can have different sign, and the momentum conservation fixes only the sign of their sum, see second and third lines of eq.~(\ref{tree:J-momentum-deltas}).

The computation presented here  has been done in position space. It is the first computation of the Sudakov form factor at NLP and the first one made in position space. As a cross-check, we have performed also the NLO computation in momentum (for DY kinematics), and checked that the results coincide with each other. Also we have checked that the LP coefficient coincides with the known one, and NLP coincides with the recent computation\footnote{We thank M.Beneke for sharing the results of their computation with us.} in ref.~\cite{Strohm:2021, Beneke:2021}.

\subsection*{Two point diagrams}

The two-point diagrams, shown in fig.~\ref{fig:loop-2point}, contribute to LP and NLP effective operator. Note, that in fig.~\ref{fig:loop-2point} we have already split the quark field into $\xi$ and $\eta$ components for convenience. The diagram 1 is the only diagram that has a LP contribution.  Explicitly, the diagram 1 reads
\begin{eqnarray}\label{loop:1}
\text{diag}_1=
g^2C_F\frac{\Gamma^2(2-\epsilon)\Gamma(1-\epsilon)}{16\pi^{3d/2}}\int d^dx d^dz 
\frac{
\bar \xi_{\bar n}(x) \gamma^\nu \fnot x \gamma^\mu \fnot z \gamma_\nu \xi_n(z)}{
[-x^2+i0]^{2-\epsilon}
[-z^2+i0]^{2-\epsilon}
[-(x-z)^2+i0]^{1-\epsilon}},
\end{eqnarray}
where $C_F=(N_c^2-1)/2N_c$ is the eigenvalue of the quadratic Casimir operator for the fundamental representation of $SU(N_c)$. 

The background fields in eq.~(\ref{loop:1}) are expanded along the light-cone, as
\begin{eqnarray}\label{loop:2}
\bar \xi_{\bar n}(x)=\bar \xi_{\bar n}(n x^-)+x_T^\mu \partial_\mu\bar \xi_{\bar n}(n x^-)
+\frac{x_T^\mu x_T^\nu}{2} \partial_\mu\partial_\nu\bar \xi_{\bar n}(n x^-)
+x^+ \partial_-\bar \xi_{\bar n}(n x^-)
+...~,
\end{eqnarray}
and similar for other fields. The expansion eq.~(\ref{loop:2}) can be safely done under the sign of the loop integration. Each next term increases the counting of the operator. At NLP only the operators with one transverse derivative contributes. 

Substituting the decomposition (\ref{loop:2}) into eq.~(\ref{loop:1}) we obtain a $2d$-dimensional integral. Two variables $x^-$ and $z^+$ are arguments of background fields and thus the integral with respect to them cannot be computed. The integration over the rest $(2d-2)$ loop-variables is done by the method explained in appendix \ref{app:diag4}. We get
\begin{eqnarray}\label{diag1}
\text{diag}_1&=&2ia_sC_F\frac{\Gamma(-\epsilon)\Gamma(1-\epsilon)\Gamma(2-\epsilon)}{\Gamma(3-2\epsilon)}
\int \frac{dz^+ dz^-}{4^\epsilon \pi}\frac{1}{[-2z^+z^-+i0]^{1-\epsilon}}\Bigg\{
\\\nn &&\qquad
(2-\epsilon+2\epsilon^2)\bar \xi_{\bar n}(z^- n)\gamma_T^\mu \xi_{n}(z^+\bar n)
\\\nn &&\qquad
+\bar \xi_{\bar n}(z^- n)\overleftarrow{\fnot \partial_T} \xi_{n}(z^+\bar n)
\[z^-n^\mu \frac{2+\epsilon^2}{\epsilon}-z^+\bar n^\mu(1-\epsilon)\]
\\\nn && \qquad
+\bar \xi_{\bar n}(z^- n)\overrightarrow{\fnot \partial_T} \xi_{n}(z^+\bar n)
\[z^+\bar n^\mu \frac{2+\epsilon^2}{\epsilon}-z^- n^\mu(1-\epsilon)\]+...\Bigg\},
\end{eqnarray}
where 
$$a_s=\frac{g^2}{(4\pi)^{d/2}},$$
and dots denote the higher-power terms. Similarly we compute the diagrams 2 and 3,
\begin{eqnarray}\label{diag2}
\text{diag}_2&=&2ia_sC_F\frac{\Gamma(-\epsilon)\Gamma(1-\epsilon)\Gamma(2-\epsilon)}{\Gamma(3-2\epsilon)}
\int \frac{dz^+ dz^-}{4^\epsilon \pi}\frac{1}{[-2z^+z^-+i0]^{1-\epsilon}}\Big\{
\\\nn &&\qquad 
\bar \eta_{\bar n}(z^- n)\gamma^- \xi_{n}(z^+\bar n)\[-\epsilon(1-\epsilon)n^\mu-(1-\epsilon)^2 \bar n^\mu \frac{z^+}{z^-}\]+...\Big\},
\\\label{diag3}
\text{diag}_3&=&2ia_sC_F\frac{\Gamma(-\epsilon)\Gamma(1-\epsilon)\Gamma(2-\epsilon)}{\Gamma(3-2\epsilon)}
\int \frac{dz^+ dz^-}{4^\epsilon \pi}\frac{1}{[-2z^+z^-+i0]^{1-\epsilon}}\Big\{
\\\nn &&\qquad 
\bar \xi_{\bar n}(z^- n)\gamma^+ \eta_{n}(z^+\bar n)\[-\epsilon(1-\epsilon)\bar n^\mu-(1-\epsilon)^2  n^\mu \frac{z^-}{z^+}\]+...\Big\}.
\end{eqnarray}
Eq.~(\ref{diag1}, \ref{diag2}, \ref{diag3}) are structured as in  eq.~(\ref{I1}), and thus they can be rewritten in  factorized form as in eq.~(\ref{I1-for-DY}) or eq.~(\ref{I1-for-SIA}). For example, the LP part of the diagram 1 in the case of DY process reads
\begin{eqnarray}\label{diag1-fac}
\text{diag}^{\text{LP;DY}}_1&=&\frac{2a_sC_F}{2^\epsilon}(2-\epsilon+2\epsilon^2)\frac{\Gamma(-\epsilon)\Gamma(2-\epsilon)}{\Gamma(\epsilon)\Gamma(3-2\epsilon)}
\int_{-\infty}^0 dz^-\frac{\bar \xi_{\bar n}(z^- n)}{(-z^-)^{1-\epsilon}} \gamma_T^\mu \int_{-\infty}^0 dz^+ \frac{\xi_{n}(z^+\bar n)}{(-z^+)^{1-\epsilon}}.
\end{eqnarray}
The prefactor of eq.~(\ref{diag1-fac}) is finite for $\epsilon\to0$. However, both integrals are UV divergent at $z^\pm\to0$, and regularized by $\epsilon>0$. As it is shown below, these poles exactly reproduce the Sudakov double pole. At $z^\pm\to\infty$ the loop integrals are regularized by the natural field decay. Other diagrams also can be written in this process-dependent form. However, it is more convenient to keep expression as in eq.~(\ref{I1}), which makes many properties transparent.

The diagrams 2 and 3 are to be rewritten using EOMs eq.~(\ref{EOMs-for-nbar}, \ref{EOMs-for-n}). To do so, one should rewrite EOMs in the integral form eq.~(\ref{EOM-integral}), substitute it into the diagrams written as in eq.~(\ref{diag1-fac}), exchange the order of integration, and integrate the remaining integrals. This computation can be simplified using the inverse derivative operators eq.~(\ref{def:inverse-derivative}), and the relations
\begin{eqnarray}\label{relations2}
\int_{-\infty}^{\infty} dz^+dz^- \frac{f_{n}(z^+)\partial_+^{-1}f_{\bar n}(z^-)}{[-2z^+z^-+i0]^{1-\epsilon}}&=&
\int_{-\infty}^{\infty} dz^+dz^- \frac{-z^-}{\epsilon}\frac{f_{n}(z^+)f_{\bar n}(z^-)}{[-2z^+z^-+i0]^{1-\epsilon}},
\\\nn 
\int_{-\infty}^{\infty} dz^+dz^- \frac{z^+}{z^-}\frac{f_{n}(z^+)\partial_+^{-1}f_{\bar n}(z^-)}{[-2z^+z^-+i0]^{1-\epsilon}}&=&
\int_{-\infty}^{\infty} dz^+dz^- \frac{z^+}{1-\epsilon}\frac{f_{n}(z^+)f_{\bar n}(z^-)}{[-2z^+z^-+i0]^{1-\epsilon}}.
\end{eqnarray}
As a result, the sum of the two-point diagrams is
\begin{eqnarray}\label{diag123}
&&\text{diag}_{1+2+3}=2ia_sC_F\frac{\Gamma(-\epsilon)\Gamma(1-\epsilon)\Gamma(2-\epsilon)}{\Gamma(3-2\epsilon)}
\int \frac{dz^+ dz^-}{4^\epsilon \pi}\frac{1}{[-2z^+z^-+i0]^{1-\epsilon}}\Big\{
\\\nn &&\qquad
(2-\epsilon+2\epsilon^2)\Big[
\bar \xi_{\bar n}(z^-n)\gamma_T^\mu\xi_n(z^+\bar n)
-n^\mu \bar \xi_{\bar n}(z^-n)\frac{\overleftarrow{\fnot \partial_T}}{\overleftarrow{\partial_+}}\xi_n(z^+\bar n)
-\bar n^\mu \bar \xi_{\bar n}(z^-n)\frac{\overrightarrow{\fnot \partial_T}}{\overrightarrow{\partial_-}}\xi_n(z^+\bar n)\Big]
\\\nn &&\qquad
-ig(1-\epsilon)\(z^-n^\mu-z^+\bar n^\mu\)\bar \xi_{\bar n}(z^- n)\fnot A_{\bar n,T}(z^-n)\xi_{n}(z^+\bar n)
\\\nn &&\qquad
+ig(1-\epsilon)\(z^+\bar n^\mu-z^- n^\mu\)\bar \xi_{\bar n}(z^- n)\fnot A_{n,T}(z^+\bar n)\xi_{n}(z^+\bar n)+...\Big\}.
\end{eqnarray}
The expression for mirror diagrams is equal to eq.~(\ref{diag123}) with $n\leftrightarrow\bar n$. Note, that the last two lines can be rewritten with inverse derivatives, reproducing the operator $J_{21}^\mu$ eq.~(\ref{NLP:tree:21+}).

The operator in the second line of eq.~(\ref{diag123}) is the $J_{11}^\mu$ current of eq.~(\ref{def:J11}). As expected, the coefficient functions for all terms of $J_{11}^\mu$ are the same, such that the current conservation eq.~(\ref{current-conserve-NLP}) is preserved. Since the contribution to NLP part is the result of the combination of  several diagrams, it gives a strong check of our computation.

\subsection*{Three-point diagrams}

\begin{figure}[t]
\begin{center}
\begin{tikzpicture}

\coordinate (A) at (2, 1.5);

\draw[very thick, blue,postaction={decorate, decoration={
    markings,
    mark=at position 0.5 with {\arrow{stealth}}}}] ($(A)$)--($(A)+(-1.2,0.85) $);
\draw[thick,postaction={decorate, decoration={
    markings,
    mark=at position 0.5 with {\arrow{stealth}}}}] ($(A)+(-1.2,0.85) $)--($(A)+(-1.8,1.3)$);
\draw[very thick, blue,postaction={decorate, decoration={
    markings,
    mark=at position 0.5 with {\arrow{stealth}}}}] ($(A)+(-1.2,-0.85)$) -- ($(A)$);
\draw[thick,postaction={decorate, decoration={
    markings,
    mark=at position 0.5 with {\arrow{stealth}}}}] ($(A)+(-1.8,-1.3)$)--($(A)+(-1.2,-0.85) $);
\draw[very thick, blue,style={decorate, decoration={snake,segment length=6.2}}] ($(A)+(-1.2,-0.85)$)--($(A)+(-1.2,0.85)$);

\draw[thick,style={decorate, decoration=snake}] ($(A)+(-0.7,1.3)$)--($(A)+(-0.7,0.5)$);

\draw[fill=black] ($(A)$) circle (0.05cm);    
\draw[fill=white] ($(A)+(-1.8,-1.3)$) circle (0.3cm);
\draw[fill=white] ($(A)+(-1.8,1.3)$) circle (0.3cm);
\draw[fill=white] ($(A)+(-0.7,1.3)$) circle (0.3cm);
\node[] at ($(A)+(-1.8,1.3)$) {$\Ds\bar \xi_{\bar n}$};
\node[] at ($(A)+(-1.8,-1.3)$) {$\Ds \xi_{n}$};
\node[] at ($(A)+(-0.7,1.3)$) {$\Ds A_{\bar n}$};
\node[] at ($(A)+(-0.7,-1.2)$) {4};

\coordinate (A) at (5, 1.5);

\draw[very thick, blue,postaction={decorate, decoration={
    markings,
    mark=at position 0.5 with {\arrow{stealth}}}}] ($(A)$)--($(A)+(-1.2,0.85) $);
\draw[thick,postaction={decorate, decoration={
    markings,
    mark=at position 0.5 with {\arrow{stealth}}}}] ($(A)+(-1.2,0.85) $)--($(A)+(-1.8,1.3)$);
\draw[very thick, blue,postaction={decorate, decoration={
    markings,
    mark=at position 0.6 with {\arrow{stealth}}}}] ($(A)+(-1.2,-0.85)$) -- ($(A)$);
\draw[thick,postaction={decorate, decoration={
    markings,
    mark=at position 0.5 with {\arrow{stealth}}}}] ($(A)+(-1.8,-1.3)$)--($(A)+(-1.2,-0.85) $);
\draw[very thick, blue,style={decorate, decoration={snake,segment length=6.2}}] ($(A)+(-1.2,-0.85)$)--($(A)+(-1.2,0.85)$);

\draw[thick,style={decorate, decoration=snake}] ($(A)+(-0.7,-1.3)$)--($(A)+(-0.7,-0.5)$);

\draw[fill=black] ($(A)$) circle (0.05cm);    
\draw[fill=white] ($(A)+(-1.8,-1.3)$) circle (0.3cm);
\draw[fill=white] ($(A)+(-1.8,1.3)$) circle (0.3cm);
\draw[fill=white] ($(A)+(-0.7,-1.3)$) circle (0.3cm);
\node[] at ($(A)+(-1.8,1.3)$) {$\Ds\bar \xi_{\bar n}$};
\node[] at ($(A)+(-1.8,-1.3)$) {$\Ds \xi_{n}$};
\node[] at ($(A)+(-0.7,-1.3)$) {$\Ds A_{\bar n}$};
\node[] at ($(A)+(-0.2,-1.2)$) {5};

\coordinate (A) at (8, 1.5);

\draw[very thick, blue,postaction={decorate, decoration={
    markings,
    mark=at position 0.5 with {\arrow{stealth}}}}] ($(A)$)--($(A)+(-1.2,0.85) $);
\draw[thick,postaction={decorate, decoration={
    markings,
    mark=at position 0.5 with {\arrow{stealth}}}}] ($(A)+(-1.2,0.85) $)--($(A)+(-1.8,1.3)$);
\draw[very thick, blue,postaction={decorate, decoration={
    markings,
    mark=at position 0.5 with {\arrow{stealth}}}}] ($(A)+(-1.2,-0.85)$) -- ($(A)$);
\draw[thick,postaction={decorate, decoration={
    markings,
    mark=at position 0.5 with {\arrow{stealth}}}}] ($(A)+(-1.8,-1.3)$)--($(A)+(-1.2,-0.85) $);
\draw[very thick, blue,style={decorate, decoration={snake,segment length=6.2}}] ($(A)+(-1.2,-0.85)$)--($(A)+(-1.2,0.85)$);

\draw[thick,style={decorate, decoration=snake}] ($(A)+(-1.15,0)$)--($(A)+(-2,0)$);
\draw[fill=blue] ($(A)+(-1.15,0)$) circle (0.05cm);  

\draw[fill=black] ($(A)$) circle (0.05cm);    
\draw[fill=white] ($(A)+(-1.8,-1.3)$) circle (0.3cm);
\draw[fill=white] ($(A)+(-1.8,1.3)$) circle (0.3cm);
\draw[fill=white] ($(A)+(-2,0)$) circle (0.3cm);
\node[] at ($(A)+(-1.8,1.3)$) {$\Ds\bar \xi_{\bar n}$};
\node[] at ($(A)+(-1.8,-1.3)$) {$\Ds \xi_{n}$};
\node[] at ($(A)+(-2,0)$) {$\Ds A_{\bar n}$};
\node[] at ($(A)+(-0.7,-1.2)$) {6};

\end{tikzpicture}
\\\vspace{0.5cm}
\begin{tikzpicture}

\coordinate (A) at (2, 1.5);

\draw[very thick, blue,postaction={decorate, decoration={
    markings,
    mark=at position 0.6 with {\arrow{stealth}}}}] ($(A)$)--($(A)+(-0.7,0.5) $);
\draw[thick,postaction={decorate, decoration={
    markings,
    mark=at position 0.5 with {\arrow{stealth}}}}] ($(A)+(-0.7,0.5) $)--($(A)+(-1.8,1.3)$);
\draw[very thick, blue,postaction={decorate, decoration={
    markings,
    mark=at position 0.6 with {\arrow{stealth}}}}] ($(A)+(-0.7,-0.5)$) -- ($(A)$);
\draw[very thick, blue,postaction={decorate, decoration={
    markings,
    mark=at position 0.5 with {\arrow{stealth}}}}] ($(A)+(-1.2,-0.85)$) -- ($(A)+(-0.7,-0.5)$);
\draw[thick,postaction={decorate, decoration={
    markings,
    mark=at position 0.7 with {\arrow{stealth}}}}] ($(A)+(-1.8,-1.3)$)--($(A)+(-1.2,-0.85) $);
\draw[very thick, blue,style={decorate, decoration={snake,segment length=6.2}}] ($(A)+(-0.7,0.5)$)--($(A)+(-0.7,-0.5)$);

\draw[thick,style={decorate, decoration=snake}] ($(A)+(-1.2,-0.85)$)--($(A)+(-2.0,-0.2)$);

\draw[fill=black] ($(A)$) circle (0.05cm);    
\draw[fill=white] ($(A)+(-1.8,-1.3)$) circle (0.3cm);
\draw[fill=white] ($(A)+(-1.8,1.3)$) circle (0.3cm);
\draw[fill=white] ($(A)+(-2.0,-0.2)$) circle (0.3cm);
\node[] at ($(A)+(-1.8,1.3)$) {$\Ds\bar \xi_{\bar n}$};
\node[] at ($(A)+(-1.8,-1.3)$) {$\Ds \xi_{n}$};
\node[] at ($(A)+(-2.0,-0.2)$) {$\Ds A_{\bar n}$};
\node[] at ($(A)+(-0.7,-1.2)$) {7};

\coordinate (A) at (5, 1.5);

\draw[thick,postaction={decorate, decoration={
    markings,
    mark=at position 0.5 with {\arrow{stealth}}}}] ($(A)$)--($(A)+(-1.8,1.3) $);

\draw[very thick, blue,postaction={decorate, decoration={
    markings,
    mark=at position 0.5 with {\arrow{stealth}}}}] ($(A)+(-0.4,-0.3)$) -- ($(A)$);

\draw[very thick, blue,style={decorate, decoration={snake,segment length=6.2}}] ($(A)+(-0.4,-0.3)$)--($(A)+(-1.2,-0.85)$);

\draw[very thick, blue,postaction={decorate, decoration={
    markings,
    mark=at position 0.6 with {\arrow{stealth}}}}] ($(A)+(-1.2,-0.85)$) -- ($(A)+(-1.1,-0.)$);
\draw[very thick, blue,postaction={decorate, decoration={
    markings,
    mark=at position 0.6 with {\arrow{stealth}}}}] ($(A)+(-1.1,-0.)$) -- ($(A)+(-0.4,-0.3)$);
    
\draw[thick,postaction={decorate, decoration={
    markings,
    mark=at position 0.7 with {\arrow{stealth}}}}] ($(A)+(-1.8,-1.3)$)--($(A)+(-1.2,-0.85) $);

\draw[thick,style={decorate, decoration=snake}] ($(A)+(-1.1,-0.)$)--($(A)+(-2.0,0.3)$);

\draw[fill=black] ($(A)$) circle (0.05cm);    
\draw[fill=white] ($(A)+(-1.8,-1.3)$) circle (0.3cm);
\draw[fill=white] ($(A)+(-1.8,1.3)$) circle (0.3cm);
\draw[fill=white] ($(A)+(-2.0,0.3)$) circle (0.3cm);
\node[] at ($(A)+(-1.8,1.3)$) {$\Ds\bar \xi_{\bar n}$};
\node[] at ($(A)+(-1.8,-1.3)$) {$\Ds \xi_{n}$};
\node[] at ($(A)+(-2.0,0.3)$) {$\Ds A_{\bar n}$};
\node[] at ($(A)+(-0.7,-1.2)$) {8};

\coordinate (A) at (8, 1.5);

\draw[thick,postaction={decorate, decoration={
    markings,
    mark=at position 0.5 with {\arrow{stealth}}}}] ($(A)$)--($(A)+(-1.8,1.3) $);

\draw[very thick, blue,postaction={decorate, decoration={
    markings,
    mark=at position 0.5 with {\arrow{stealth}}}}] ($(A)+(-0.4,-0.3)$) -- ($(A)$);

\draw[very thick, blue,postaction={decorate, decoration={
    markings,
    mark=at position 0.6 with {\arrow{stealth}}}}] ($(A)+(-1.2,-0.85)$)--($(A)+(-0.4,-0.3)$);

\draw[very thick, blue,style={decorate, decoration={snake,segment length=6.2}}] ($(A)+(-1.2,-0.85)$) -- ($(A)+(-1.1,-0.)$);
\draw[very thick, blue,style={decorate, decoration={snake,segment length=6.2}}] ($(A)+(-0.4,-0.3)$) -- ($(A)+(-1.1,-0.)$);
    
\draw[thick,postaction={decorate, decoration={
    markings,
    mark=at position 0.7 with {\arrow{stealth}}}}] ($(A)+(-1.8,-1.3)$)--($(A)+(-1.2,-0.85) $);

\draw[thick,style={decorate, decoration=snake}] ($(A)+(-1.1,-0.)$)--($(A)+(-2.0,0.3)$);
\draw[fill=blue] ($(A)+(-1.1,-0.)$) circle (0.05cm);  

\draw[fill=black] ($(A)$) circle (0.05cm);    
\draw[fill=white] ($(A)+(-1.8,-1.3)$) circle (0.3cm);
\draw[fill=white] ($(A)+(-1.8,1.3)$) circle (0.3cm);
\draw[fill=white] ($(A)+(-2.0,0.3)$) circle (0.3cm);
\node[] at ($(A)+(-1.8,1.3)$) {$\Ds\bar \xi_{\bar n}$};
\node[] at ($(A)+(-1.8,-1.3)$) {$\Ds \xi_{n}$};
\node[] at ($(A)+(-2.0,0.3)$) {$\Ds A_{\bar n}$};
\node[] at ($(A)+(-0.7,-1.2)$) {9};

\coordinate (A) at (11, 1.5);

\draw[thick,postaction={decorate, decoration={
    markings,
    mark=at position 0.5 with {\arrow{stealth}}}}] ($(A) $)--($(A)+(-1.8,1.3)$);
\draw[very thick, blue,postaction={decorate, decoration={
    markings,
    mark=at position 0.5 with {\arrow{stealth}}}}] ($(A)+(-1.2,-0.85) $) -- ($(A)$);
\draw[thick,postaction={decorate, decoration={
    markings,
    mark=at position 0.7 with {\arrow{stealth}}}}] ($(A)+(-1.8,-1.3)$)--($(A)+(-1.2,-0.85) $);

\draw[very thick, blue,style={decorate, decoration={snake,segment length=6.2}}] ($(A)+(-1,-0.7)$) arc (-120 : 10 : 0.5 );

\draw[thick,style={decorate, decoration=snake}] ($(A)+(-1.2,-0.85)$)--($(A)+(-2.0,-0.2)$);

\draw[fill=black] ($(A)$) circle (0.05cm);    
\draw[fill=white] ($(A)+(-1.8,-1.3)$) circle (0.3cm);
\draw[fill=white] ($(A)+(-1.8,1.3)$) circle (0.3cm);
\draw[fill=white] ($(A)+(-2.0,-0.2)$) circle (0.3cm);
\node[] at ($(A)+(-1.8,1.3)$) {$\Ds\bar \xi_{\bar n}$};
\node[] at ($(A)+(-1.8,-1.3)$) {$\Ds \xi_{n}$};
\node[] at ($(A)+(-2.0,-0.2)$) {$\Ds A_{\bar n}$};
\node[] at ($(A)+(-0.7,-1.2)$) {10};

\end{tikzpicture}
\caption{\label{fig:loop-3point} The three-point diagrams contributing to the NLO of the effective operator. The diagrams with $A_{\bar n}\to A_{n}$ and then with $n\leftrightarrow \bar n$ should be added. The blobs indicate the type of background field. The blue lines are the dynamical fields.}
\end{center}
\end{figure}
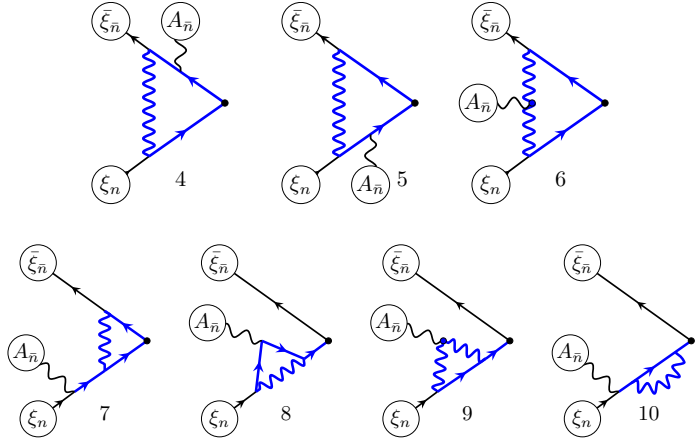

The three-point diagrams are shown in fig.~\ref{fig:loop-3point}. The diagrams 7-10 are specific for the composite background field and would be absent in the usual background field computation. Note, that the three-gluon vertex that appears in diagrams 6 and 9 is not equal to three-gluon vertex in QCD but has a modification that comes from the background-gauge gauge-fixing condition. 

The computation of these diagrams is straightforward and described in details in appendix~\ref{app:diag4}. Here we present the final expression for the sum of three-point diagrams. For convenience we add the $\bar \xi A \xi$-part of the two-point  diagrams, such that the result is the full expression for the coefficient function of $J_{21}^\mu$ operator. It reads
\begin{eqnarray}\label{NLO:3point-diagrams}
&&\text{diag}_{4}+...+\text{diag}_{10}+\text{diag}_2^{\bar \xi A\xi-\text{part}}
=
ga_s \frac{\Gamma(-\epsilon)\Gamma(1-\epsilon)\Gamma(2-\epsilon)}{\Gamma(2-2\epsilon)}
\\\nn &&\qquad
\int \frac{dz^+ dz^-}{4^\epsilon \pi}\frac{1}{[-2z^+z^-+i0]^{1-\epsilon}}\int_0^1 ds\Big\{
(z^+\bar n^\mu -z^- n^\mu)C_F\frac{2-\epsilon}{\epsilon}\,\mathcal{K}(1,1)
\\\nn &&\qquad
-\(
C_F\frac{\epsilon(1+\epsilon)}{(1-\epsilon)^2}+C_A\frac{1-\epsilon-\epsilon^2}{(1-\epsilon)^2}\)
\Big[
(\epsilon z^-n^\mu +(1-\epsilon)z^+\bar n^\mu)\,\mathcal{K}(s,1)
-z^+\bar n^\mu\, \mathcal{K}(0,1)\Big]
\\\nn &&\qquad
+\(C_F-\frac{C_A}{2}\)\frac{2(1-\epsilon-\epsilon^2)}{\epsilon(1-\epsilon)^2}\Big[
(\epsilon z^-n^\mu+(1-\epsilon)z^+\bar n^\mu)\mathcal{K}(1,s)
-z^+\bar n^\mu\,\mathcal{K}(1,0)\Big]
\Big\},
\end{eqnarray}
where $C_A=N_c$ and
\begin{eqnarray}
\mathcal{K}(s,t)=\bar \xi_{\bar n}(s z^- n)\fnot A_{\bar n,T}(tz^- n)\xi_{n}(z^+ \bar n).
\end{eqnarray}
The terms in eq.~(\ref{NLO:3point-diagrams}) are grouped such that each line forms a transverse combination.

\subsection*{NLO expressions in  momentum space}

The passage to momentum space is straightforward. The expression for EM current in eq.~(\ref{tree:EM-current-momentum}) takes the form
\begin{eqnarray}\label{NLO:EM-current-momentum}
\widetilde{J}^\mu(y)&=&p_1^+p_2^-\int d x d\tilde x \,e^{ixp_1^+y^-+i\tilde xp_2^-y^+} \widetilde{C}_{1}J^\mu_{11}(x,\tilde x,y_T)
\\\nn &&
+(p_1^+)^2p_2^-\int d x_1 dx_2 d\tilde x \,e^{i(x_1+x_2)p_1^+y^-+i\tilde xp_2^-y^+} \widetilde{C}_{2}(x_{1,2}) J^\mu_{21}(x_{1,2},\tilde x,y_T)
\\\nn &&
+p_1^+(p_2^-)^2\int d x d\tilde x_1 d\tilde x_2 \,e^{i xp^+_1y^-+i(\tilde x_1+\tilde x_2)p^-_2y^+}\widetilde{C}_{2}(\tilde x_{1,2}) J^\mu_{12}(x,\tilde x_{2,1},y_T)+...~,
\end{eqnarray}
where $C$'s are the coefficient functions. Their expression up to NLO are
\begin{eqnarray}\label{NLO:C1}
\widetilde{C}_{1}&=&1+2a_sC_F\frac{2-\epsilon+2\epsilon^2}{[-2q^+q^--i0]^\epsilon}\frac{\Gamma(\epsilon)\Gamma(-\epsilon)\Gamma(2-\epsilon)}{\Gamma(3-2\epsilon)},
\\\label{NLO:C2}
\widetilde{C}_{2}(x_{1,2})&=&
1+2ga_s \frac{\Gamma(\epsilon)\Gamma(-\epsilon)\Gamma(1-\epsilon)}{\Gamma(3-2\epsilon)}\frac{1}{[-2q^+q^--i0]^\epsilon}
\\\nn && \times\Big\{ C_F(1-\epsilon)^2(2-\epsilon)
-2\(C_F-\frac{C_A}{2}\)(1-\epsilon-\epsilon^2)\frac{x_1+x_2}{x_2}\[1-\(\frac{x_1+x_2}{x_1}\)^\epsilon\]
\\\nn && +\[C_F \epsilon^2(1+\epsilon)+C_A \epsilon (1-\epsilon-\epsilon^2)\]\frac{x_1+x_2}{x_1}
\[1-\(\frac{x_1+x_2}{x_2}\)^\epsilon\]\Big\},
\end{eqnarray}
In these expressions the momenta $q^\pm$ are set in accordance to eq.~(\ref{tree:J-momentum-deltas}). I.e. for the $J_{11}^\mu$ current $q^+q^-=x_1x_2p_1^+p_2^-$, for $J_{21}^\mu$ current $q^+q^-=(x_1+x_2)x_3p_1^+p_2^-$ and for $J_{12}^\mu$ current $q^+q^-=x_1(x_2+x_3)p_1^+p_2^-$. 

Eqs.~(\ref{NLO:C1}, \ref{NLO:C2}) are the bare form of the coefficient functions, that contains the IR poles. These poles are removed by the operation in eq.~(\ref{C-finite}). In the present case, this operation can be made for EM currents individually before recombining them into the effective operator. The renormalization constants $Z_1$ and $Z_2$ are derived in sec.~\ref{sec:TMD}, and they exactly remove the pole part of $\tilde C$'s (see eqns.~(\ref{pole-cancel:C1}, \ref{pole-cancel:C2}) and discussion there). So, we obtain in the $\overline{\text{MS}}$-scheme\footnote{\label{def:MS} $\overline{\text{MS}}$-scheme is defined with an extra factor $\mu^{2\epsilon}e^{\epsilon\gamma_E}$ for each $a_s$. Here, $\gamma_E$ is the Euler–Mascheroni constant.}
\begin{eqnarray}
C_{1}&=&1+a_sC_F\(-\mathbf{L}_Q^2+3\mathbf{L}_Q-8+\frac{\pi^2}{6}\)+O(a_s),
\\
C_{2}(x_{1,2})&=&1+a_s\Big[
C_F\(-\mathbf{L}_Q^2+\mathbf{L}_Q-3+\frac{\pi^2}{6}\)
+C_A\frac{x_1+x_2}{x_1}\ln\(\frac{x_1+x_2}{x_2}\)
\\\nn &&
+\(C_F-\frac{C_A}{2}\)\frac{x_1+x_2}{x_2}\ln\(\frac{x_1+x_2}{x_1}\)\(
2\mathbf{L}_Q-\ln\(\frac{x_1+x_2}{x_1}\)-4\)\Big]+O(a_s^2),
\end{eqnarray}
where
\begin{eqnarray}
\mathbf{L}_Q=\ln\(\frac{-2q^+q^--i0}{\mu^2}\).
\end{eqnarray}
Here and in Eqs.~(\ref{NLO:C1}, \ref{NLO:C2}), the notation is adopted such that $x_2$ is the momentum fraction of the gluon field. The expression for $C_{1}$ coincides with earlier computations, see e.g. ref.~\cite{Mueller:1989hs,Manohar:2003vb,Echevarria:2011epo,Collins:2011zzd,Becher:2010tm}. Nowadays, the coefficient $C_{1}$ is known up N$^3$LO order \cite{Gehrmann:2010ue}. The coefficient functions for the anti-causal sector are obtained from the causal ones with  complex conjugation.

The expression for $\mathcal{J}^{\mu\nu}$ in eq.~(\ref{tree:J-momentum-deltas}) became decorated by the coefficient functions,
\begin{eqnarray}\label{NLO:J-momentum-deltas}
&&\mathcal{J}^{\mu\nu}_{\text{eff}}(q)=\int \frac{d^2b}{(2\pi)^2}
e^{-i(qb)}\Bigg\{
\int dx d\tilde x
\delta\(x-\frac{q^+}{p_1^+}\)
\delta\(\tilde x-\frac{q^-}{p_2^-}\) |C_{1}|^2
\mathcal{J}^{\mu\nu}_{1111}(x,\tilde x,b)
\\\nn &&\qquad
+\int [dx]d\tilde x
\delta\(\tilde x-\frac{q^-}{p_2^-}\)
\\\nn && \qquad\qquad\times
\(\delta\(x_1-\frac{q_1^+}{p_1^+}\)
C_1^*C_2(x_{3,2})\mathcal{J}^{\mu\nu}_{1211}(x,\tilde x,b)
+
\delta\(x_3+\frac{q_1^+}{p_1^+}\)
C_2^*(x_{1,2})C_1
\mathcal{J}^{\mu\nu}_{2111}(x,\tilde x,b)\)
\\\nn &&\qquad
+\int dx [d\tilde x]
\delta\(x-\frac{q^+}{p_1^+}\)
\\\nn && \qquad\qquad\times
\(
C_1^* C_2(\tilde x_{3,2})
\delta\(\tilde x_1-\frac{q^-}{p_2^-}\)
\mathcal{J}^{\mu\nu}_{1112}(x,\tilde x,b)
+
C_2^*(\tilde x_{1,2}) C_1
\delta\(\tilde x_3+\frac{q^-}{p_2^-}\)
\mathcal{J}^{\mu\nu}_{1121}(x,\tilde x,b)
\)
\\\nn &&\qquad
+...
\Bigg\}
\end{eqnarray}
where asterisks denote the complex conjugation, which, in fact, applies only to $\mathbf{L}_Q$.

\section{Mode overlap and the soft factor}
\label{sec:soft-overlap}

So far, we have assumed that the collinear and anti-collinear fields are entirely independent. It allows us to impose individual gauge-fixing conditions and separate fields into independent gauge-invariant TMD operators. However, it should be kept in mind that there is a part of functional integration phase space where collinear and anti-collinear fields are a single background field. In fig.~\ref{fig:field-regions}, this region is covered by diagonal shading. This is the so-called soft region (or glauber region in the SCET nomenclature). Fields in this region (marked by $s$) satisfy the counting
\begin{eqnarray}\label{s-counting-q}
&&\{\partial_+, \partial_-,\partial_T\}q_{s}\lesssim Q \{\lambda^2,\lambda^2,\lambda\}q_{s},
\\\label{s-counting-A}
&&\{\partial_+, \partial_-,\partial_T\}A_{s}^\mu\lesssim Q \{\lambda^2,\lambda^2,\lambda\}A_{s}^\mu.
\end{eqnarray}
The soft region is double-counted in the functional integral with the measure of eq.~(\ref{functional-measure}). 

Let us stress that the double-counting of the soft region does not effect the TMD factorization procedure described in previous sections. Instead, each TMD operator has an uncompensated rapidity divergence. In fact, the rapidity divergences should cancel between collinear and anti-collinear TMD operators, but they cannot due to the double counting of the soft region. There are several solutions of this problem. Let us list some of them:
\begin{itemize}
\item The definition of collinear and anti-collinear fields can be modified in the soft region, such that there is no overlap. For example, by introducing a cut in rapidity for each field as it is shown by the red-dashed line in fig.\ref{fig:field-regions}. Then each TMD operator depend on the cut parameter, such that this dependence is compensated in the product. See e.g. discussion in ref.~\cite{Echevarria:2012js}. The same idea is used in the rapidity factorization approach \cite{Balitsky:2017gis,Balitsky:2017flc}.
\item The product nature of the functional integral allows to remove double-counting simply dividing by the (functional) integral over soft modes. It is possible if the hadron states do not contain soft fields, which is valid in a non-small-x regime. The resulting factor is known as the soft factor \cite{Collins:2011zzd,Echevarria:2011epo} or zero-bin subtraction \cite{Manohar:2006nz}. It is the most popular procedure nowadays. However, there is no general approach to determine the soft factor operators at higher powers. Most plausible, such a simple multiplicative structure does not hold for higher power operators.
\item One can ignore problems of overlapping modes entirely, and reconstruct necessary parts (such as rapidity renormalization constants) by demanding that the effective operator is well-defined. This logic is used in the collinear anomaly approach, see \cite{Becher:2010tm,Becher:2011dz}.
\end{itemize}
In the present NLP computation, we use the second way, because it leads to the correct result at NLP without additional computation. However, we expect that this approach does not hold at higher powers.

Let us stress, that all approaches should result into the same final expression, up to some finite terms. The fixation of finite terms is equivalent to the fixation of the scheme, and the definition of the parton distribution. The closest example is the difference between $\overline{\text{MS}}$ and DIS schemes, see e.g. \cite{CTEQ:1993hwr}. The only difference of the TMD case from the collinear is, that the coefficient of rapidity divergence is nonperturbative, and thus the finite terms added/subtracted from the physical distribution are also nonperturbative. Consequently, the scheme must be defined by a certain nonperturbative statement. At LP one fixes the scheme defining that cross-section for DY and SIDIS do not have extra nonperturbative functions except TMD distributions. The same definition can be applied to NLP, see sec.~\ref{sec:recombination}.

\subsection*{Determining the soft factors}

To determine soft factors for our operators, we use the following procedure. We split soft parts of collinear and anti-collinear fields
\begin{eqnarray}\label{split-soft}
q_{\bar n}(x)\to q_{\bar n}(x)+q_s(x),\qquad q_n(x)\to q_n(x)+q_s(x),\qquad A^\mu_{\bar n}(x)\to
A^\mu_{\bar n}(x)+A^\mu_{s}(x),
\end{eqnarray}
and similar for other components. Then we isolate the soft fields into a single factor dropping power suppressed contributions. For example, for the first term of the LP effective operator eq.~(\ref{tree:LP})
\begin{eqnarray}\label{soft:1}
&&\bar \xi^{(-)}_{\bar n}\gamma^\mu_T  \xi_n^{(-)}
\bar \xi^{(+)}_{n}\gamma^\nu_T \xi_{\bar n}^{(+)} \to 
\\\nn && 
(\bar \xi^{(-)}_{\bar n}+q^{(-)}_s)[y^-n+y_T,
L n +y_T]^{(-)}_{s}\gamma^\mu_T [\bar L \bar n+y_T,-y^+\bar n+y_T]^{(-)}_{s} (\xi_n^{(-)}+q^{(-)}_s)
\\\nn && \times 
(\bar \xi^{(+)}_{n}+q^{(+)}_s)[0,
\bar L \bar n]^{(+)}_{s}\gamma^\nu_T [L n,0]^{(+)}_{s} (\xi_{\bar n}^{(+)}+q^{(+)}_s)
\\\nn &&
=
\bar \xi^{(-)}_{\bar n}\gamma^\mu_T  \xi_n^{(-)}
\bar \xi^{(+)}_{n}\gamma^\nu_T \xi_{\bar n}^{(+)}\times 
\widetilde{\mathcal{S}}_{\text{LP}}(y_T)+\mathcal{O}(\lambda^4),
\end{eqnarray}
where $[a,b]_{s}$ is the Wilson line with the soft gluon field, and we omit transverse links for brevity. The operator for the LP soft factor is
\begin{eqnarray}\label{soft:2}
\widetilde{\mathcal{S}}_{\text{LP}}(y_T)&=&\frac{\Tr}{N_c}
[\bar L\bar n+y_T,y_T]^{(-)}[y_T,L n+y_T]^{(-)}[L n,0]^{(+)}[0,\bar L \bar n]^{(+)},
\end{eqnarray}
where the trace is taken with respect to color indices. The trace and the factor $1/N_c$ appears due to the fact that only gauge-invariant operators have non-zero matrix elements. To derive eq.~(\ref{soft:2}) we have used the counting rules for soft fields eq.~(\ref{s-counting-q}, \ref{s-counting-A}). 

The operator $\widetilde{\mathcal{S}}$ represents the soft part of LP effective operator. Therefore, the soft part to the functional integral eq.~(\ref{W-as-funInt1}) at LP is given by the vacuum (assuming that the hadrons do not carry soft partons) matrix element of eq.~(\ref{soft:2})
\begin{eqnarray}\label{soft:LP}
\mathcal{S}_{\text{LP}}(y_T)&=&\frac{\Tr}{N_c}\sum_X\langle 0|
[\bar L \bar n+y_T,y_T][y_T,L n+y_T] |X\rangle \langle X| 
[L n,0][0,\bar L \bar n]|0\rangle,
\end{eqnarray}
which is called the soft factor. To remove the double counting from the LP term, eq.~(\ref{tree:JLP}), we divide TMD operators by the soft factor
\begin{eqnarray}\label{SF:11}
\mathcal{O}^{li}_{11,\bar n}\overline{\mathcal{O}}^{jk}_{11,n}\to \frac{\mathcal{O}^{li}_{11,\bar n}\overline{\mathcal{O}}^{jk}_{11,n}}{\mathcal{S}_{\text{LP}}(y_T)}.
\end{eqnarray} 
The same structure follows from the region-separation method \cite{Ji:2004wu}. In SCET literature, this procedure is known as a zero-bin subtraction \cite{Manohar:2006nz}.
We remark that the problem of overlapping modes does not impact TMD factorization, and thus the replacement in eq.~(\ref{SF:11}) is valid to all orders in perturbation theory.

A similar computation can be done for operators contributing to $\mathcal{J}^{\mu\nu}_{\text{NLP}}$. We have two principal cases: the operators with derivatives (the first and the second line in eq.~(\ref{tree:NLP})), and operators with extra field $A_{\mu_T}$ (other lines in eq.~(\ref{tree:NLP})). In both cases, we obtain that the soft overlap contribution is equal to the LP soft factor eq.~(\ref{soft:LP}), since a derivative of a soft Wilson line, or an extra factor $A_s^\mu$ necessarily increase the power counting. Therefore, the subtraction of the soft region for NLP operators has the same form as for LP operator eq.~(\ref{SF:11}). Namely,
\begin{eqnarray}\label{SF:12}
\partial_\rho \mathcal{O}^{li}_{11,\bar n}\overline{\mathcal{O}}^{jk}_{11,n}\to \frac{\partial_\rho\mathcal{O}^{li}_{11,\bar n}\overline{\mathcal{O}}^{jk}_{11,n}}{\mathcal{S}_{\text{LP}}(y_T)},
\qquad
\mathcal{O}^{li}_{12,\bar n}\overline{\mathcal{O}}^{jk}_{11,n}\to \frac{\mathcal{O}^{li}_{11,\bar n}\overline{\mathcal{O}}^{jk}_{12,n}}{\mathcal{S}_{\text{LP}}(y_T)}
,
\end{eqnarray}
and similarly for other terms of effective operator eq.~(\ref{tree:NLP}). The LP soft factor is independent on $y^\pm$ and thus does modify convolutions in $\mathcal{J}_{eff}^{\mu\nu}$ at NLP.

Having the same soft factors for LP and NLP operators leads to the following consequences:
\begin{itemize}
\item LP and NLP operators must have the same rapidity divergence and, as the result, the same rapidity anomalous dimension.
\item LP and NLP operators must have the same collinear divergent part of the UV renormalization.
\end{itemize}
Indeed, these divergences arise in the interaction of soft modes, and (in the present approach) they are canceled by the soft factor. In sec.~\ref{sec:TMD}, we independently derive both statements, and we explicitly verify them at NLO.

A simple procedure described here allows to determine soft factors for LP and NLP terms. For higher power correction a more systematic procedure should be developed. 

\subsection*{$\delta$-regularization}

The soft factor has a complicated combination of divergences. Namely, it has UV divergence, rapidity divergences, and mass divergences (see ref.\cite{Echevarria:2013aca,Vladimirov:2017ksc} for detailed analysis). The mass  divergences cancel in the sum of all diagrams \cite{Vladimirov:2017ksc}, whereas rapidity and UV divergences remain. 

An essential feature of rapidity divergences is that they are not regularized by  dimensional regularization \cite{Collins:1992tv}. Therefore, an additional regularization must be implemented. There are many regularizations of rapidity divergences used in the literature, such as -- tilting of Wilson lines \cite{Collins:2011zzd}, analytic regularization \cite{Becher:2011dz,Chiu:2012ir}, exponential regulator \cite{Li:2016axz} and $\delta$-regularization \cite{Echevarria:2011epo,Echevarria:2016scs}. Each of these regularizations has been used in plenty of computations and has advantages and disadvantages. The final result after the recombination of divergences is independent on the rapidity regularizator. In this work we use the $\delta$-regularization, for the only simple reason that we are experienced in it.

The rapidity divergences arise due to the interaction with the far end of the half-infinite light-like Wilson line \cite{Vladimirov:2017ksc}. In the $\delta$-regularization these interactions are regularized by insertion of dumping factor into Wilson line,
\begin{eqnarray}\label{def:delta-regulator}
[zn,L n]=P\exp\(-ig\int_{L}^z d\sigma A_+(\sigma n)e^{-s\delta^+ \sigma}\),
\end{eqnarray}
where $\delta^+>0$ and $s=\sign(L)$. Similarly for the Wilson line in the direction $\bar n$
\begin{eqnarray}
[z\bar n,\bar L \bar n]=P\exp\(-ig\int_{\bar L}^z d\sigma A_-(\sigma \bar n)e^{-\bar s\delta^- \sigma}\),
\end{eqnarray}
with $\delta^->0$ and $\bar s=\sign(\bar L)$. Thus, there are two regulator parameters $\delta^+$ and $\delta^-$, which regularize divergences associated with different light-like directions.

\subsection*{Soft factor at NLO}

The calculation of the LP TMD soft factor has been performed in many papers, see e.g. NLO calculations \cite{Aybat:2011zv,Collins:2011zzd,Echevarria:2011epo,Chiu:2011qc}. The  expressions used here with $\delta$-regularization are taken from ref.~\cite{Echevarria:2015byo}. The bare soft factor at NLO reads
\begin{eqnarray}\label{SF:NLO}
\mathcal{S}_{\text{LP}}(b)=1-4a_sC_F\Gamma(-\epsilon)\(\frac{-b^2}{4}\)^\epsilon \(\ln\(\frac{-b^2 (2\delta^+\delta^-)}{4e^{-2\gamma_E}}\)-\psi(-\epsilon)-\gamma_E\)+O(a_s^2).
\end{eqnarray}
Here, $b$ is a transverse vector, so $-b^2>0$. This expression contains a product of rapidity divergences associated with different Wilson lines in the form of $\ln\delta^+$ and $\ln\delta^-$. Note, that some $1/\epsilon$ poles of eq.~(\ref{SF:NLO}) are not the UV divergence, but a part of rapidity divergence. The UV part of the soft factor should be computed separately. In $\overline{\text{MS}}$-scheme (see footnote \ref{def:MS}), it is \cite{Echevarria:2016scs}
\begin{eqnarray}
Z_{\mathcal{S}}(2\delta^+\delta^-)&=&1+4a_s C_F\(-\frac{1}{\epsilon^2}+\frac{1}{\epsilon}\ln\(\frac{2\delta^+\delta^-}{\mu^2}\)\)+O(a_s^2).
\end{eqnarray}
It contains $\ln(\delta^+\delta^-)$ which, in this case, is a remnant of the mass divergence.

In ref.~\cite{Vladimirov:2017ksc}, it is proven that the rapidity divergences of the LP soft factor can be written as a product of factors. The proof is made using the method of conformal transformation and it is valid to all orders in perturbation theory. Using it we present the LP soft factor in the form
\begin{eqnarray}\label{SF:S=RSR}
\mathcal{S}_{LP}(b)&=&Z_{\mathcal{S}}(2\delta^+\delta^-) R\(b^2,\frac{\delta^+}{\nu^+}\)S_{0}(b^2,\nu^2)R\(b^2,\frac{\delta^-}{\nu^-}\),
\end{eqnarray}
where $\nu^\pm$ are some scales, and $\nu^2=2\nu^+\nu^-$, and $S_0$ is free of UV and rapidity divergences. The NLO expression for $R$ can be deduced from eq.~(\ref{SF:NLO}). In $\overline{\text{MS}}$-scheme it reads
\begin{eqnarray}\label{def:R2}
R\(b^2,\frac{\delta^+}{\nu^+}\)=1-4a_s C_F\Big[
\Gamma(-\epsilon)\(-\frac{b^2 \mu^2}{4e^{-\gamma_E}}\)^\epsilon+\frac{1}{\epsilon}\Big] \ln\(\frac{\delta^+}{\nu^+}\)+O(a_s^2).
\end{eqnarray}

For future convenience, we introduce UV renormalization factor $Z_R$ for $\widetilde{R}$, and write
\begin{eqnarray}\label{def:R=ZR}
\widetilde{R}\(b^2,\frac{\delta^+}{\nu^+}\)=Z_{R}\(\frac{\delta^+}{\nu^+}\)R\(b^2,\frac{\delta^+}{\nu^+}\),
\end{eqnarray}
where
\begin{eqnarray}\label{def:Rtilde}
\widetilde{R}\(b^2,\frac{\delta^+}{\nu^+}\)&=&1-4a_s C_F\Gamma(-\epsilon)\(-\frac{b^2 \mu^2}{4e^{-\gamma_E}}\)^\epsilon \ln\(\frac{\delta^+}{\nu^+}\)+O(a_s^2),
\\
Z_{R}\(\frac{\delta^+}{\nu^+}\)&=&
1+\frac{4a_s C_F}{\epsilon}\ln\(\frac{\delta^+}{\nu^+}\)+O(a_s^2).
\end{eqnarray}
Note, that the identity
\begin{eqnarray}
Z_{\mathcal{S}}(2\delta^+\delta^-)=Z_{\mathcal{S}}(\nu^2)Z_R\(\frac{\delta^+}{\nu^+}\)Z_R\(\frac{\delta^-}{\nu^-}\)
\end{eqnarray}
is valid to all orders in perturbation theory.

\section{Divergences of TMD operators}
\label{sec:TMD}

The TMD operators are composed of two semi-compact light-cone operators, eq.~(\ref{UU->O}), separated by a transverse distance,
\begin{eqnarray}\label{TMD:op1}
\mathcal{O}_{NM,\bar n}(\{z^-\},b)=U_{N,\bar n}^{(-)}(\{z^-_1\},b)U_{M,\bar n}^{(+)}(\{z^-_2\},0).
\end{eqnarray}
The hadronic matrix element of the TMD operators in eq.~(\ref{TMD:op1}) defines the TMD distributions. It can be a TMD PDF(s)
\begin{eqnarray}\label{TMD:pdf1}
&&F_{NM}(\{z^-\},b)=
\sum_X \langle p|U_{N,\bar n}(\{z^-\},b)|X\rangle\langle X| U_{M,\bar n}(\{z^-\},0)|p\rangle,
\end{eqnarray}
or TMD FF(s)
\begin{eqnarray}\label{TMD:ff1}
&&D_{NM}(\{z^-\},b)=
\sum_X \langle 0|U_{N,\bar n}(\{z^-\},b)|p,X\rangle\langle p,X| U_{M,\bar n}(\{z^-\},0)|0\rangle.
\end{eqnarray}
The TMD distributions $F_{NM}$ and $D_{NM}$ can be decomposed over independent Lorentz components and reveal a plethora of TMD distributions. It is widely known that there are eight quark TMD PDFs defined by the LP operator $\mathcal{O}_{11}$ in eq.~(\ref{def:O11}), see e.g. \cite{Mulders:1995dh,Bacchetta:2006tn}. All these distributions (TMDPDF and TMDFF) obey the same evolution equations, because they are matrix elements of the same TMD operator. In this way, all these distributions are alike from the perspective of TMD operator expansion, despite the fact that they have very different partonic interpretation and are measured with different experimental set-ups.

In this work, we concentrate on the global properties of TMD factorization, and thus we do not systematize NLP TMD distribution (see \cite{Boer:2003cm,Bacchetta:2006tn}). This systematization as well as, the derivation of cross-section is the  object of a subsequent publication. Instead, we study the global properties of LP and NLP TMD operators, and write down the evolution equations for LP and NLP TMD distributions.

The operators $U_1$ and $U_2$ in eq.~(\ref{def:U1}, \ref{def:U2-i}) together with their C-conjugated versions $\overline{U}_1$ and $\overline{U}_2$, eq.~(\ref{def:U1-C}, \ref{def:U2bar-i}), set up all TMD operators at NLP eq.~(\ref{O=UU:11}-\ref{def:O12bar-position}). As we show in the following subsection the singularity and evolution properties of a TMD operator follow from the properties of each $U$ that composes it. Therefore, we concentrate on the studies of $U_1$ and $U_2$ rather then on studies of $\mathcal{O}$. In addition, we consider a more general operator $U_2^\mu$ eq.~(\ref{def:U2}), since it does not complicate the computation but allows us a simpler comparison with  known expressions.

\subsection{Rapidity divergences}
\label{sec:rapdiv}

\begin{figure}[t]
\begin{center}
\begin{tikzpicture}

\coordinate (A) at (1, 2);

\draw[very thick,blue,postaction={decorate, decoration={
    markings,
    mark=at position 0.6 with {\arrow{stealth}}}}] ($(A)+(0,-0.5)$)--($(A)$);
\draw[thick,postaction={decorate, decoration={
    markings,
    mark=at position 0.6 with {\arrow{stealth}}}}] ($(A)+(0,-2)$)--($(A)+(0,-0.5)$);
\draw[very thick, blue,style={decorate, decoration={snake,segment length=6.2}}] 
($(A)+(0,-0.5)$) to[out=0,in=140] ($(A)+(3.5,-1)$);

\draw[thick,dashed] ($(A)$) -- ($(A)+(4,0)$);
\draw[thick,dashed] ($(A)+(1.5,-1)$) -- ($(A)+(4,-1)$);

\draw[fill=white] ($(A)+(0,-2)$) circle (0.3cm);
\draw[thick,fill=white] ($(A)+(1.5,-1.2)$) circle (0.5cm);
\node[] at ($(A)+(0,-2)$) {$\Ds \xi_{\bar n}$};
\node[] at ($(A)+(1.5,-1.2)$) {$\Ds \hat U_N$};

\node[] at ($(A)+(2.5,-2)$) {r1};

\coordinate (A) at (6, 2);

\draw[very thick,blue,postaction={decorate, decoration={
    markings,
    mark=at position 0.6 with {\arrow{stealth}}}}] ($(A)+(0,-0.5)$)--($(A)$);
\draw[thick,postaction={decorate, decoration={
    markings,
    mark=at position 0.6 with {\arrow{stealth}}}}] ($(A)+(0,-2)$)--($(A)+(0,-0.5)$);
\draw[very thick, blue,style={decorate, decoration={snake,segment length=6.2}}] 
($(A)+(0,-0.5)$) to[out=0,in=140] ($(A)+(3.5,-1)$);
\draw[thick,style={decorate, decoration=snake}] ($(A)+(1,-2)$)--($(A)+(1,0)$);
\draw[fill=black]($(A)+(1,0)$) circle (0.1cm);

\draw[thick,dashed] ($(A)$) -- ($(A)+(4,0)$);
\draw[thick,dashed] ($(A)+(2,-1)$) -- ($(A)+(4,-1)$);

\draw[fill=white] ($(A)+(0,-2)$) circle (0.3cm);
\draw[fill=white] ($(A)+(1,-2)$) circle (0.3cm);
\draw[thick,fill=white] ($(A)+(2,-1.2)$) circle (0.5cm);
\node[] at ($(A)+(0,-2)$) {$\Ds \xi_{\bar n}$};
\node[] at ($(A)+(1,-2)$) {$\Ds A_{\bar n}$};
\node[] at ($(A)+(2,-1.2)$) {$\Ds \hat U_N$};

\node[] at ($(A)+(3,-2)$) {r2};

\coordinate (A) at (11, 2);

\draw[thick,postaction={decorate, decoration={
    markings,
    mark=at position 0.6 with {\arrow{stealth}}}}] ($(A)+(0,-2)$)--($(A)$);
\draw[very thick, blue,style={decorate, decoration={snake,segment length=6.2}}] 
($(A)+(1,-0.5)$) to[out=0,in=140] ($(A)+(3.5,-1)$);

\draw[thick,style={decorate, decoration=snake}] ($(A)+(1,-2)$)--($(A)+(1,-0.5)$);
\draw[very thick, blue,style={decorate, decoration={snake,segment length=6.2}}] 
($(A)+(1,-0.5)$)-- ($(A)+(1,0)$);
\draw[fill=black]($(A)+(1,0)$) circle (0.1cm);

\draw[thick,dashed] ($(A)$) -- ($(A)+(4,0)$);
\draw[thick,dashed] ($(A)+(2,-1)$) -- ($(A)+(4,-1)$);

\draw[fill=white] ($(A)+(0,-2)$) circle (0.3cm);
\draw[fill=white] ($(A)+(1,-2)$) circle (0.3cm);
\draw[thick,fill=white] ($(A)+(2,-1.2)$) circle (0.5cm);
\node[] at ($(A)+(0,-2)$) {$\Ds \xi_{\bar n}$};
\node[] at ($(A)+(1,-2)$) {$\Ds A_{\bar n}$};
\node[] at ($(A)+(2,-1.2)$) {$\Ds \hat U_N$};

\node[] at ($(A)+(3,-2)$) {r3};

\end{tikzpicture}
\caption{The rapidity divergent diagrams for operator $U_1$ (r1), and operator $U_{2}$ (r2,r3) that interacts with any other semi-compact operator $U_N$. The dynamical fields are shown in blue. Wilson lines are indicated as dashed.. The black dots are insertions of $F_{\mu+}$.}
\label{fig:loop-rapidity}    
\end{center}
\end{figure}
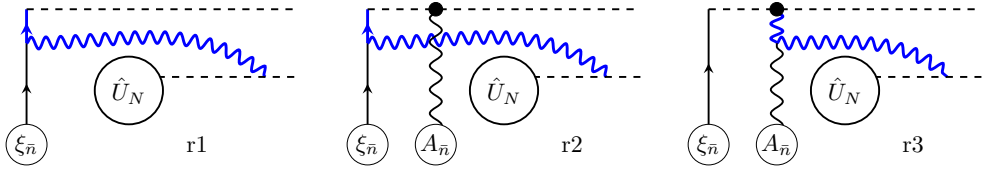

Rapidity divergences are specific of TMD operators, eq.~(\ref{TMD:op1}). They arise in the interaction of fields with the distant segments of light-like Wilson lines (accurate definition and properties rapidity divergences can be found in ref.~\cite{Vladimirov:2017ksc}). 

\subsection*{Renormalization of rapidity divergences}

Rapidity divergences are multiplicatively renormalizable. In ref.~\cite{Vladimirov:2017ksc} this statement is proven for the multi-parton scattering soft factors. The proof can be easily generalized to TMD operators $\mathcal{O}_{NM,\bar n}$ that appear at LP and NLP \footnote{
\label{footnote:conformal}
One needs to apply the conformal transformation $\mathcal{C}_{\bar n}$ (defined in eq.~(5.1) of \cite{Vladimirov:2017ksc}) to the TMD operator $\mathcal{O}_{NM,\bar n}$. The resulting operator $\mathcal{C}_{\bar n}\mathcal{O}_{NM,\bar n}$ is spatially-compact. It has the form of a Wilson line with two light-like segments that are joined at the origin of the light-cone.  The partonic fields are positioned along the Wilson line. The Wilson line has a light-like cusp, whose UV divergence corresponds to the rapidity divergence of the TMD operator before transformation. Using the fact that an UV divergence is multiplicatively renormalizable, and that the conformal invariance of QCD is restored in the Wilson-Fisher critical point \cite{Wilson:1971dc}, one derives that \textit{rapidity divergence is multiplicatively renormalizable}  as well. The derivation is made by iterations order-by-order in perturbative expansion, starting from the LO which respects conformal invariance. Practically, this derivation repeats the one given in sec.~5.2 of ref.~\cite{Vladimirov:2017ksc} for multi-parton soft factors.

The light-like cusp anomalous dimension associated with the cusp of $\mathcal{C}_{\bar n}\mathcal{O}_{NM,\bar n}$ (called the soft anomalous dimension, \cite{Dixon:2008gr,Gardi:2009qi}) corresponds to the rapidity anomalous dimension of the TMD operator. This correspondence is a simple equality at LO, but receives modifications beyond the LO due to the breaking of the conformal invariance in QCD. The modification terms can be derived using the same iterative method, which has been done in ref.~\cite{Vladimirov:2016dll} up to N$^2$LO. The result coincides with the three-loop brute force computation \cite{Li:2016ctv}, which non-trivially validate of the method of conformal transformation used to prove the renormalizability of rapidity divergences.

One of the important consequences of the derivation is that the rapidity divergence of LP operators and NLP operators are the same. It follows from the fact, that cusp divergences of $\mathcal{C}_{\bar n}O_{11,\bar n}$ and $\mathcal{C}_{\bar n}O_{21,\bar n}$ coincide. It independently confirms the same observation pointed out in sec.~\ref{sec:soft-overlap} after eq.~(\ref{SF:12}).
}.

Using the multiplicativity of the rapidity divergence we write the LP operators in eq~(\ref{def:O11-position}, \ref{def:O11bar-position}) as
\begin{eqnarray}\label{rap:O11*R}
\mathcal{O}^{ij}_{11,\bar n}(b)=R\(b^2,\frac{\delta^+}{\nu^+}\)\mathcal{O}^{ij}_{11,\bar n}(b;\nu^+),
\qquad
\overline{\mathcal{O}}^{ij}_{11,\bar n}(b)=R\(b^2,\frac{\delta^+}{\nu^+}\)\overline{\mathcal{O}}^{ij}_{11,\bar n}(b;\nu^+),
\end{eqnarray}
where we omit the argument $\{y^-,0\}$ of the TMD operators. The variable $\nu^+$ is a scale of rapidity divergences renormalization. The operators on the r.h.s. of eq.~(\ref{rap:O11*R}) are free from  rapidity divergences. We distinguish such operators by explicit indication of rapidity renormalization scale $\nu^+$. The rapidity-divergence renormalization factor $R$ is independent on $y^-$, and thus the Fourier transformed TMD operator eq.~(\ref{def:O11}, \ref{def:O11bar}) is renormalized in the same way. The \textit{same} renormalization factor absorbs rapidity divergences of NLP operators eq.~(\ref{def:O21-position}-\ref{def:O12bar-position}),
\begin{eqnarray}\label{rap:O12*R}
\mathcal{O}^{ij}_{21,\bar n}(b)=R\(b^2,\frac{\delta^+}{\nu^+}\)\mathcal{O}^{ij}_{21,\bar n}(b;\nu^+),
\qquad
\overline{\mathcal{O}}^{ij}_{21,\bar n}(b)=R\(b^2,\frac{\delta^+}{\nu^+}\)\overline{\mathcal{O}}^{ij}_{21,\bar n}(b;\nu^+),
\\\nn
\mathcal{O}^{ij}_{12,\bar n}(b)=R\(b^2,\frac{\delta^+}{\nu^+}\)\mathcal{O}^{ij}_{12,\bar n}(b;\nu^+),
\qquad
\overline{\mathcal{O}}^{ij}_{12,\bar n}(b)=R\(b^2,\frac{\delta^+}{\nu^+}\)\overline{\mathcal{O}}^{ij}_{12,\bar n}(b;\nu^+),
\end{eqnarray}
and similar for the Fourier-transformed operators eq.~(\ref{def:O11}-\ref{def:OMNbar}). 

For the  TMD-operators oriented along $\bar n$, rapidity divergences have the same structure but with $n\to\bar n$. Thus, one should replace $\delta^+\to \delta^-$ and $\nu^+\to \nu^-$ in formulas (\ref{rap:O11*R}-\ref{rap:O12*R}),
\begin{eqnarray}
\mathcal{O}^{ij}_{NM,n}(b)=R\(b^2,\frac{\delta^-}{\nu^-}\)\mathcal{O}^{ij}_{NM,n}(b;\nu^-),
\qquad
\overline{\mathcal{O}}^{ij}_{NM,n}(b)=R\(b^2,\frac{\delta^-}{\nu^-}\)\overline{\mathcal{O}}^{ij}_{NM,n}(b;\nu^-),
\end{eqnarray}
where $N+M$ is 2 or 3.

\subsection*{Rapidity divergence at NLO}

The factor $R$ that renormalizes the rapidity divergences of TMD operators is the same as for the soft factor eq.~(\ref{SF:S=RSR},~\ref{def:R2}) and here we check it at one-loop. For this purpose, we consider a TMD operator composed from $U_{1,\bar n}$ or $U^\mu_{2,\bar n}$ eq.~(\ref{def:U1}, \ref{def:U2}) and any other semi-compact operator $\overline{U}_{N,\bar n}$. We denote such operators as $\mathcal{O}_{N1,\bar n}=\overline{U}_{N,\bar n}U_{1,\bar n}$ and $\mathcal{O}_{N2,\bar n}=\overline{U}_{N,\bar n}U^\mu_{2,\bar n}$. These operators are more general than operators appearing at NLP, however, as we demonstrate below, their possibly complicated nature does not impact the rapidity divergences structure.

TMD operators $\mathcal{O}_{N1,\bar n}$ and $\mathcal{O}_{N2,\bar n}$ are color-neutral. Thus, the operator $U_{N,\bar n}$ has a Wilson line in the anti-fundamental representation pointing to $L n$. We split a far segment of this Wilson line from the rest of the operator,
\begin{eqnarray}
\overline{U}_{N,\bar n}(\{z^-\},b)=\overline{U}'_{N,\bar n}(\{z^-\},b;z_0)[z_0 n+b,L n+b],
\end{eqnarray}
where $z_0$ is such that $\max\{|z^-|\}<|z_0|$. The rapidity divergences arise only in the interaction of the Wilson line $[z_0 n,\pm \infty n]$ with $U_{1,2}$. One-loop diagrams that produce the rapidity divergence are shown in fig. \ref{fig:loop-rapidity}.

To extract the rapidity divergence, we follow the method developed in ref.~\cite{Scimemi:2019gge}. Let us describe it using the diagram $r1$, as an example. We write the diagram r1 using the background field method with a single background field in the $A_+=0$ gauge. It reads
\begin{eqnarray}\label{rap:loop1}
\text{diag}_{r1}&=&-ig^2C_F\frac{\Gamma(1-\epsilon)\Gamma(2-\epsilon)}{8\pi^d}
 \int_{L}^{z_0} d\sigma \int d^dy
\frac{\overline{U}_{N,\bar n}'(\{z^-\},b;z_0) P_+\fnot y \gamma^+ q(y)}{[-y^2+i0]^{2-\epsilon}[-(\sigma n+b-y)^2+i0]^{1-\epsilon}},
\end{eqnarray}
where $P_+=\gamma^-\gamma^+/2$ is the projector of the ``good'' component of the quark field, see eq.~(\ref{def:xi-eta+}). Here we have also used that 
the TMD operators in eq.~(\ref{TMD:op1}) can be written as a single T-ordered operator (or in terms of the functional integral, the superscripts $(\pm)$ can be omitted), since all fields in it are separated by light-like or space-like distances. Note, that the spinor indices of $U_N$ and the rest of the diagrams are not contracted. Joining the propagators by the Feynman parameter $\alpha$ and making a shift $y\to y+\alpha(n\sigma+b)$, we obtain
\begin{eqnarray}
\text{diag}_{r1}&=&-2ig^2C_F\frac{\Gamma(3-2\epsilon)}{8\pi^d}
 \overline{U}_{N,\bar n}'(\{z^-\},b;z_0)
 \\\nn && \times \int_0^1 d\alpha\, \alpha^{-\epsilon}\bar \alpha^{1-\epsilon} \int_{L }^{z_0} d\sigma \int d^dy
\frac{ y^+\xi(y+\alpha \sigma n+\alpha b)}{[-y^2-\alpha \bar \alpha b^2+i0]^{3-2\epsilon}},
\end{eqnarray}
where we used that $P_+\fnot y \gamma^+ q=2y^+\xi$ and $b^+=0$. The integral over $y$ can be computed in the sense of the generating functional. For that we expand $\xi$ in a Taylor series at $y=0$, and integrate this series term-by-term using eq.~(\ref{app:diag4:loopi}). The result reads
\begin{eqnarray}
\text{diag}_{r1}&=&2a_sC_F\(\frac{-b^2}{4}\)^\epsilon  \overline{U}_{N,\bar n}'(\{z^-\},b;z_0)
\\\nn && \times\sum_{n=0}^\infty \int_0^1 d\alpha \int_{L}^{z_0} d\sigma
\frac{(-1)^n\Gamma(-\epsilon-n)}{4^n\,n!}\bar \alpha (-\alpha \bar \alpha b^2)^{n} 
\partial^{2n}\partial_+\xi(\alpha \sigma n+\alpha b),
\end{eqnarray}
The $n=0$ term is rapidity divergent at $\alpha\to0$.  Indeed, in this limit the field $\xi$ is independent on $\sigma$ and the integral over $\sigma$ diverges at $\sigma\to L$. To reveal the divergence we make a change of variable $\tau=\alpha\sigma$, and obtain
\begin{eqnarray}\label{rap:loop3}
\text{diag}_{r1}&=&2a_sC_F \Gamma(-\epsilon)  \(\frac{-b^2}{4}\)^\epsilon \overline{U}_{N,\bar n}'(\{z^-\},b;z_0)
\int_0^1 \frac{d\alpha}{\alpha} \int_{L}^{0} 
d\tau \partial_+\xi(\tau n)+...\,,
\end{eqnarray}
where dots indicate the terms finite at $\alpha\to0$, and thus rapidity-divergence-free. In this expression the divergence is transparent. As eq.~(\ref{rap:loop3}) is independent of $z_0$,  the operator on $\overline{U}'_{N,\bar n}$ can be promoted to $\overline{U}_{N,\bar n}$ by limiting $z_0\to L$.

To regularize rapidity divergences we use the $\delta$-regularization eq.~(\ref{def:delta-regulator}). It gives the factor $e^{-L \delta \sigma}$ in the integral eq.~(\ref{rap:loop1}), and modifies eq.~(\ref{rap:loop3}) as
\begin{eqnarray}
\text{diag}_{r1}&=&2a_sC_F \Gamma(-\epsilon)\(\frac{-b^2}{4}\)^\epsilon \overline{U}_{N,\bar n}(\{z^-\},b)
\int_0^1 \frac{d\alpha}{\alpha} \int_{L}^{0} 
d\tau e^{- s \delta \frac{\tau}{\alpha}} \partial_+\xi(\tau n)+...\,,
\end{eqnarray}
where $s=\sign(L)$. Now, the integrals over $\alpha$ and $\tau$ can be computed. The result is
\begin{eqnarray}\label{rap:loop4}
\text{diag}_{r1}&=&-2a_sC_F\Gamma(-\epsilon) \(\frac{-b^2}{4}\)^\epsilon \ln\(\frac{\delta^+}{\hat p^+_\xi}\)\overline{U}_{N,\bar n}(\{z^-\},b)\xi(0)+...\,,
\end{eqnarray}
where $\hat p^+_\xi=-i\partial_+$ is the momentum of field $\xi$. The direction of the Wilson line does not impact the rapidity divergent part, but gives rise to a finite term $\sim is\pi$. The computation of similar diagrams for the C-conjugated operator, gives the same rapidity divergent part.

The most important observation is that the rapidity divergent term in diagram $r1$ is independent of the second part of the operator. In this sense, we can associate the rapidity divergence in eq.~(\ref{rap:loop4}) with the operator $U_{1,\bar n}$. In order to get the complete rapidity divergence of the TMD operator $\mathcal{O}_{N1,\bar n}=U_{N,\bar n}U_{1,\bar n}$, we should consider  also diagrams where the fields of $U_N$ interact with the Wilson line of $U_1$. 

Next, we study the rapidity divergence associated with $U^\mu_{2,\bar n}(\{z_1,0\},0_T)$, eq.~(\ref{def:U2}). For that we consider the TMD operator $\mathcal{O}_{N2,\bar n}$. There are two rapidity divergent diagrams shown in fig. \ref{fig:loop-rapidity}, $r2$ and $r3$. The computation is similar to eq.~(\ref{rap:loop1}-\ref{rap:loop4}). The rapidity divergent part of these diagrams is
\begin{eqnarray}\label{rap:r2}
\text{diag}_{r2}&=&-2a_s\(C_F-\frac{C_A}{2}\)\Gamma(-\epsilon)\(\frac{-b^2}{4}\)^\epsilon  \ln\(\frac{\delta^+}{\hat p^-_\xi}\)U_{N,\bar n}(\{z^-\},b) F_{\mu+}(z_1)\xi_{\bar n}(0)+...\,,
\\\label{rap:r3}
\text{diag}_{r3}&=&-2a_s\frac{C_A}{2}\Gamma(-\epsilon)\(\frac{-b^2}{4}\)^\epsilon  \ln\(\frac{\delta^+}{\hat p^+_A}\)U_{N,\bar n}(\{z^-\},b) F_{\mu+}(z_1)\xi_{\bar n}(0)+...\,,
\end{eqnarray}
where $\hat p_A$ is the momentum operator acting on the gluon field. Again the r.h.s. of eq.~(\ref{rap:r2}, \ref{rap:r3}) reproduces the original operator $\mathcal{O}_{N2,\bar n}$. Summing together diagrams r2 and r3 we obtain
\begin{eqnarray}\label{rap:r2+r3}
\text{diag}_{r2+r3}&=&-2a_s C_F\Gamma(-\epsilon)\(\frac{-b^2}{4}\)^\epsilon  \ln\(\frac{\delta^+}{\hat p^+}\)U_{N,\bar n}(\{z^-\},b) U^\mu_{2,\bar n}(\{z_1,0\},0_T)+...\,,
\end{eqnarray}
where $\hat p$ is some generic momentum.

We observe that the coefficient of the rapidity divergence for LP eq.~(\ref{rap:loop4}) and NLP eq.~(\ref{rap:r2+r3}) operators coincides, as it is predicted in previous sections. Using eq.~(\ref{rap:loop4}) or eq.~(\ref{rap:r2+r3}) (together with their corresponding charge-conjugated parts) we deduce the NLO expression for the (unrenormalized) factor $\widetilde{R}$ defined in eq.~(\ref{rap:O11*R}-\ref{rap:O12*R}). In $\overline{\text{MS}}$-scheme, we have
\begin{eqnarray}\label{def:R}
\widetilde{R}\(b^2,\frac{\delta^+}{\nu^+}\)=1-4a_s C_F\Gamma(-\epsilon)\(-\frac{b^2 \mu^2}{4 e^{-\gamma_E}}\)^\epsilon \ln\(\frac{\delta^+}{\nu^+}\)+O(a_s^2).
\end{eqnarray}
This expression coincides with eq.~(\ref{def:Rtilde}), which validates the factorization at one-loop. 

An important feature of the rapidity divergences is that they have a nonperturbative contribution. The computation presented in this section misses this part. It is clear that at large transverse distances $b$ the gluon propagator is modified by the confinement effects. Therefore, the perturbative result in eq.~(\ref{rap:r2+r3}) is valid only at small (but finite) values of $b^2$. At large values of $b^2$ the factor $R$ gets power corrections. In  dimensional regularization, the presence of these power corrections is indicated by  renormalon divergences \cite{Scimemi:2016ffw}. The renormalization theorem for rapidity divergences guarantees that the factors $R$ for TMD operators of twist-(1+1) and twist-(2+1) coincides to all powers of small-$b^2$ expansion. It allows us to assume that rapidity divergences for LP and NLP operators also coincides nonperturbatively.

\subsection*{Rapidity divergences of quasi-partonic operators}

The consideration on rapidity divergences and the soft factor presented above allows us a generalisation going far beyond NLP. That is, the rapidity divergences and hence the rapidity anomalous dimensions for quasi-partonic TMD operators coincides with the LP.

We define quasi-partonic TMD operators in analogy to collinear quasi-partonic operators. Namely, a quasi-partonic operator $U_N$ is the operator whose geometric twist equals the number of fields\footnote{We also assume that a quasi-partonic operator is continuously connected by Wilson line, and does not have disconnected parts.} (excluding Wilson lines), see e.g. \cite{Braun:2009vc,Braun:2011dg,Ji:2014eta}. Consequently, a quasi-partonic TMD operator is composed from two quasi-partonic operators $U$. It also implies that a quasi-partonic operator consists of ``good'' components of fields only. E.g. operators $U_{1,\bar n}$ and $U^\mu_{2,\bar n}$ are quasi-partonic. The first non-quasi-partonic semi-compact operator has twist-3.

The equality of rapidity divergent parts for all quasi-partonic TMD operators can be derived in several ways. First of all, one can observe that the soft part of such an operator coincides with a staple of Wilson lines. Indeed, since the operator already contains a maximum number of fields (and all of them are ``good'' fields), the replacement of any field by its soft part as in eq.~(\ref{split-soft}) increases the power counting. Therefore, the part of the soft factor, responsible for the cancellation of the rapidity divergence of a quasi-partonic operator, coincides with the LP soft factor. 

Another way is to use the method of conformal transformations (see footnote~\ref{footnote:conformal}). In this case, we utilize that ``good'' components preserve the projection properties after the conformal transformation $\mathcal{C}_{\bar n}$ \cite{Braun:2003rp}. Therefore, the soft anomalous dimension of the transformed operator coincides with the one of LP operator. It leads to the equality of factors $R$ for all quasi-partonic operators. In other words,
\begin{eqnarray}\label{rap:Oquasi*R}
\mathcal{O}^{\text{quasi-p.}}_{NM,\bar n}(\{z\},b)=\widetilde{R}\(b^2,\frac{\delta^+}{\nu^+}\)\mathcal{O}^{\text{quasi-p.}}_{NM,\bar n}(\{z\},b;\nu^+),
\end{eqnarray}
where $\widetilde{R}$ is the same as in eq.~(\ref{rap:O11*R}) and the operator on r.h.s. is rapidity-divergence-free.

Eq.~(\ref{rap:Oquasi*R}) is simple to confirm at one-loop. There are two types of diagrams contributing to the rapidity divergence of a quasi-partonic operator, the interaction with a gluon or a quark field. These diagrams are computed in eq.~(\ref{rap:r2}, \ref{rap:r3}), and they have the same expressions apart of color factors. In the quasi-partonic case, these color factors should be replaced by $[1]t^a[2]t^a$ (for interaction with quark) or $[1]if^{abc}t^b[2]t^c$ (for the interaction with gluon), where $[1]$ and $[2]$ are color matrices of the fields before and after the interacting field. Summing together all such structures, one obtains the original color structure multiplied by $t^ct^c$. I.e.
\begin{eqnarray}
\text{rap.div.}[U^{\text{quasi-p.}}_{N,\bar n}]=-2a_sC_K\Gamma(-\epsilon)\(\frac{-b^2}{4}\)^\epsilon \ln\(\frac{\delta^+}{\hat p^+}\),
\end{eqnarray}
where $C_K$ is the eigenvalue of the quadratic Casimir operator for color representation of $U^{\text{quasi-p.}}_N$.

\subsection{UV divergences}
\label{sec:UV}

The UV divergence of the TMD operator $\mathcal{O}_{NM,\bar n}=\overline{U}_{N,\bar n}U_{M,\bar n}$ consists of UV divergences of $U_{N,\bar n}$ and $U_{M,\bar n}$, which are independent of each other. It is obvious since the semi-compact operators are separated by a space-like distance $b^2$, and any interaction between them is UV finite. 

The UV divergence of a semi-compact operator contains also a collinear divergence from the interaction of  the distant segment of the light-like Wilson line. Despite some similarity this collinear divergence should not be confused with the rapidity divergences discussed in the previous section. They have different properties (e.g. collinear divergence does not receive power corrections), and they are treated differently within the factorization theorem. On another hand, they are regularized by a common regulator (the $\delta$-regulator in the present computation).

\subsection*{UV renormalization of $U_1$}

\begin{figure}[t]
\begin{center}
\begin{tikzpicture}

\coordinate (A) at (1, 2);

\draw[very thick,blue,postaction={decorate, decoration={
    markings,
    mark=at position 0.6 with {\arrow{stealth}}}}] ($(A)+(0,-1)$)--($(A)$);
\draw[thick,postaction={decorate, decoration={
    markings,
    mark=at position 0.6 with {\arrow{stealth}}}}] ($(A)+(0,-2)$)--($(A)+(0,-1)$);
\draw[very thick, blue,style={decorate, decoration={snake,segment length=6.2}}] 
($(A)+(0,-1)$) to[out=0,in=-90] ($(A)+(1.5,0)$);

\draw[thick,dashed] ($(A)$) -- ($(A)+(2.5,0)$);

\draw[fill=white] ($(A)+(0,-2)$) circle (0.3cm);
\node[] at ($(A)+(0,-2)$) {$\Ds \xi_{\bar n}$};

\node[] at ($(A)+(1,-2)$) {u1};

\coordinate (A) at (4, 2);

\draw[very thick,blue,postaction={decorate, decoration={
    markings,
    mark=at position 0.6 with {\arrow{stealth}}}}] ($(A)+(0,-1)$)--($(A)$);
\draw[thick,postaction={decorate, decoration={
    markings,
    mark=at position 0.6 with {\arrow{stealth}}}}] ($(A)+(0,-2)$)--($(A)+(0,-1)$);
\draw[very thick, blue,style={decorate, decoration={snake,segment length=6.2}}] 
($(A)+(0,-1)$) to[out=0,in=-90] ($(A)+(1.,0)$);

\draw[fill=black] ($(A)+(1,0)$) circle (0.1cm);

\draw[thick,dashed] ($(A)$) -- ($(A)+(2.5,0)$);

\draw[fill=white] ($(A)+(0,-2)$) circle (0.3cm);
\node[] at ($(A)+(0,-2)$) {$\Ds \xi_{\bar n}$};

\node[] at ($(A)+(1,-2)$) {u2};

\coordinate (A) at (7, 2);

\draw[very thick,blue,postaction={decorate, decoration={
    markings,
    mark=at position 0.6 with {\arrow{stealth}}}}] ($(A)+(0,-1)$)--($(A)$);
\draw[thick,postaction={decorate, decoration={
    markings,
    mark=at position 0.6 with {\arrow{stealth}}}}] ($(A)+(0,-2)$)--($(A)+(0,-1)$);
\draw[very thick, blue,style={decorate, decoration={snake,segment length=6.2}}] 
($(A)+(0,-1)$) to[out=0,in=-90] ($(A)+(2.,0)$);
\draw[thick,style={decorate, decoration=snake}] ($(A)+(1,0)$)--($(A)+(1,-2)$);

\draw[fill=black] ($(A)+(1,0)$) circle (0.1cm);

\draw[thick,dashed] ($(A)$) -- ($(A)+(2.5,0)$);

\draw[fill=white] ($(A)+(0,-2)$) circle (0.3cm);
\draw[fill=white] ($(A)+(1,-2)$) circle (0.3cm);
\node[] at ($(A)+(0,-2)$) {$\Ds \xi_{\bar n}$};
\node[] at ($(A)+(1,-2)$) {$\Ds A_{\bar n}$};

\node[] at ($(A)+(1.8,-2)$) {u3};

\coordinate (A) at (10, 2);

\draw[thick,postaction={decorate, decoration={
    markings,
    mark=at position 0.6 with {\arrow{stealth}}}}] ($(A)+(0,-2)$)--($(A)$);
\draw[very thick, blue,style={decorate, decoration={snake,segment length=6.2}}] 
($(A)+(1,-1)$) to[out=0,in=-90] ($(A)+(2.,0)$);
\draw[very thick, blue,style={decorate, decoration={snake,segment length=6.2}}] 
($(A)+(1,-1)$) -- ($(A)+(1.,0)$);
\draw[thick,style={decorate, decoration=snake}] ($(A)+(1,-1)$)--($(A)+(1,-2)$);

\draw[fill=black] ($(A)+(1,0)$) circle (0.1cm);

\draw[thick,dashed] ($(A)$) -- ($(A)+(2.5,0)$);

\draw[fill=white] ($(A)+(0,-2)$) circle (0.3cm);
\draw[fill=white] ($(A)+(1,-2)$) circle (0.3cm);
\node[] at ($(A)+(0,-2)$) {$\Ds \xi_{\bar n}$};
\node[] at ($(A)+(1,-2)$) {$\Ds A_{\bar n}$};

\node[] at ($(A)+(1.8,-2)$) {u4};

\coordinate (A) at (13, 2);

\draw[thick,postaction={decorate, decoration={
    markings,
    mark=at position 0.6 with {\arrow{stealth}}}}] ($(A)+(0,-2)$)--($(A)$);
\draw[very thick, blue,style={decorate, decoration={snake,segment length=6.2}}] 
($(A)+(1,0)$) to[out=45,in=-45] ($(A)+(1.,-1)$);
\draw[very thick, blue,style={decorate, decoration={snake,segment length=6.2}}] 
($(A)+(1,-1)$) to[out=135,in=-135] ($(A)+(1.,0)$);
\draw[thick,style={decorate, decoration=snake}] ($(A)+(1,-1)$)--($(A)+(1,-2)$);

\draw[fill=black] ($(A)+(1,0)$) circle (0.1cm);

\draw[thick,dashed] ($(A)$) -- ($(A)+(2.5,0)$);

\draw[fill=white] ($(A)+(0,-2)$) circle (0.3cm);
\draw[fill=white] ($(A)+(1,-2)$) circle (0.3cm);
\node[] at ($(A)+(0,-2)$) {$\Ds \xi_{\bar n}$};
\node[] at ($(A)+(1,-2)$) {$\Ds A_{\bar n}$};

\node[] at ($(A)+(1.8,-2)$) {u5};

\coordinate (A) at (1, -1);

\draw[very thick,blue,postaction={decorate, decoration={
    markings,
    mark=at position 0.6 with {\arrow{stealth}}}}] ($(A)+(0,-1)$)--($(A)$);
\draw[thick,postaction={decorate, decoration={
    markings,
    mark=at position 0.6 with {\arrow{stealth}}}}] ($(A)+(0,-2)$)--($(A)+(0,-1)$);
\draw[very thick, blue,style={decorate, decoration={snake,segment length=6.2}}] 
($(A)+(0,-1)$) to[out=0,in=-90] ($(A)+(0.6,0)$);
\draw[thick,style={decorate, decoration=snake}] ($(A)+(1,0)$)--($(A)+(1,-2)$);

\draw[fill=black] ($(A)+(1,0)$) circle (0.1cm);

\draw[thick,dashed] ($(A)$) -- ($(A)+(2.5,0)$);

\draw[fill=white] ($(A)+(0,-2)$) circle (0.3cm);
\draw[fill=white] ($(A)+(1,-2)$) circle (0.3cm);
\node[] at ($(A)+(0,-2)$) {$\Ds \xi_{\bar n}$};
\node[] at ($(A)+(1,-2)$) {$\Ds A_{\bar n}$};

\node[] at ($(A)+(1.8,-2)$) {u6};

\coordinate (A) at (4, -1);

\draw[very thick,blue,postaction={decorate, decoration={
    markings,
    mark=at position 0.6 with {\arrow{stealth}}}}] ($(A)+(0,-1)$)--($(A)$);
\draw[thick,postaction={decorate, decoration={
    markings,
    mark=at position 0.6 with {\arrow{stealth}}}}] ($(A)+(0,-2)$)--($(A)+(0,-1)$);
\draw[very thick, blue,style={decorate, decoration={snake,segment length=6.2}}] 
($(A)+(0,-1)$) -- ($(A)+(1,0)$);
\draw[thick,style={decorate, decoration=snake}] ($(A)+(1,0)$)--($(A)+(1,-2)$);

\draw[fill=black] ($(A)+(1,0)$) circle (0.1cm);

\draw[thick,dashed] ($(A)$) -- ($(A)+(2.5,0)$);

\draw[fill=white] ($(A)+(0,-2)$) circle (0.3cm);
\draw[fill=white] ($(A)+(1,-2)$) circle (0.3cm);
\node[] at ($(A)+(0,-2)$) {$\Ds \xi_{\bar n}$};
\node[] at ($(A)+(1,-2)$) {$\Ds A_{\bar n}$};

\node[] at ($(A)+(1.8,-2)$) {u7};

\coordinate (A) at (7, -1);

\draw[thick,postaction={decorate, decoration={
    markings,
    mark=at position 0.6 with {\arrow{stealth}}}}] ($(A)+(0,-2)$)--($(A)$);
\draw[very thick, blue,style={decorate, decoration={snake,segment length=6.2}}] 
($(A)+(1,-1)$) to[out=180,in=-90] ($(A)+(0.5,0)$);
\draw[very thick, blue,style={decorate, decoration={snake,segment length=6.2}}] 
($(A)+(1,-1)$) -- ($(A)+(1.,0)$);
\draw[thick,style={decorate, decoration=snake}] ($(A)+(1,-1)$)--($(A)+(1,-2)$);

\draw[fill=black] ($(A)+(1,0)$) circle (0.1cm);

\draw[thick,dashed] ($(A)$) -- ($(A)+(2.5,0)$);

\draw[fill=white] ($(A)+(0,-2)$) circle (0.3cm);
\draw[fill=white] ($(A)+(1,-2)$) circle (0.3cm);
\node[] at ($(A)+(0,-2)$) {$\Ds \xi_{\bar n}$};
\node[] at ($(A)+(1,-2)$) {$\Ds A_{\bar n}$};

\node[] at ($(A)+(1.8,-2)$) {u8};

\coordinate (A) at (10, -1);

\draw[thick,postaction={decorate, decoration={
    markings,
    mark=at position 0.6 with {\arrow{stealth}}}}] ($(A)+(0,-2)$)--($(A)+(0,-1)$);
\draw[thick,blue,postaction={decorate, decoration={
    markings,
    mark=at position 0.6 with {\arrow{stealth}}}}] ($(A)+(0,-1)$)--($(A)$);
\draw[very thick, blue,style={decorate, decoration={snake,segment length=6.2}}] 
($(A)+(1,-1)$) -- ($(A)+(0.,-1)$);
\draw[very thick, blue,style={decorate, decoration={snake,segment length=6.2}}] 
($(A)+(1,-1)$) -- ($(A)+(1.,0)$);
\draw[thick,style={decorate, decoration=snake}] ($(A)+(1,-1)$)--($(A)+(1,-2)$);

\draw[fill=black] ($(A)+(1,0)$) circle (0.1cm);

\draw[thick,dashed] ($(A)$) -- ($(A)+(2.5,0)$);

\draw[fill=white] ($(A)+(0,-2)$) circle (0.3cm);
\draw[fill=white] ($(A)+(1,-2)$) circle (0.3cm);
\node[] at ($(A)+(0,-2)$) {$\Ds \xi_{\bar n}$};
\node[] at ($(A)+(1,-2)$) {$\Ds A_{\bar n}$};

\node[] at ($(A)+(1.8,-2)$) {u9};

\coordinate (A) at (13, -1);

\draw[thick,blue,postaction={decorate, decoration={
    markings,
    mark=at position 0.6 with {\arrow{stealth}}}}] ($(A)+(0,-1)$)--($(A)$);
\draw[thick,blue,postaction={decorate, decoration={
    markings,
    mark=at position 0.6 with {\arrow{stealth}}}}] ($(A)+(1,-1)$)--($(A)+(0,-1)$);
\draw[thick,postaction={decorate, decoration={
    markings,
    mark=at position 0.6 with {\arrow{stealth}}}}] ($(A)+(1,-2)$)--($(A)+(1,-1)$);    

\draw[very thick, blue,style={decorate, decoration={snake,segment length=6.2}}] 
($(A)+(1,0)$) --($(A)+(1.,-1)$);

\draw[thick,style={decorate, decoration=snake}] ($(A)+(0,-1)$)--($(A)+(0,-2)$);

\draw[fill=black] ($(A)+(1,0)$) circle (0.1cm);

\draw[thick,dashed] ($(A)$) -- ($(A)+(2.5,0)$);

\draw[fill=white] ($(A)+(0,-2)$) circle (0.3cm);
\draw[fill=white] ($(A)+(1,-2)$) circle (0.3cm);
\node[] at ($(A)+(1,-2)$) {$\Ds \xi_{\bar n}$};
\node[] at ($(A)+(0,-2)$) {$\Ds A_{\bar n}$};

\node[] at ($(A)+(1.8,-2)$) {u10};

\end{tikzpicture}
\caption{The UV divergent diagrams for the operator $U_1$ (u1), and the operator $U_{2}$ (u2 -- u10). The dynamical fields are shown in blue. Wilson lines are indicated as dashed. The black dots are insertions of $F_{\mu+}$. Diagrams u1, u2, and u3 also contain collinear divergence.}
\label{fig:loop-UV}    
\end{center}
\end{figure}
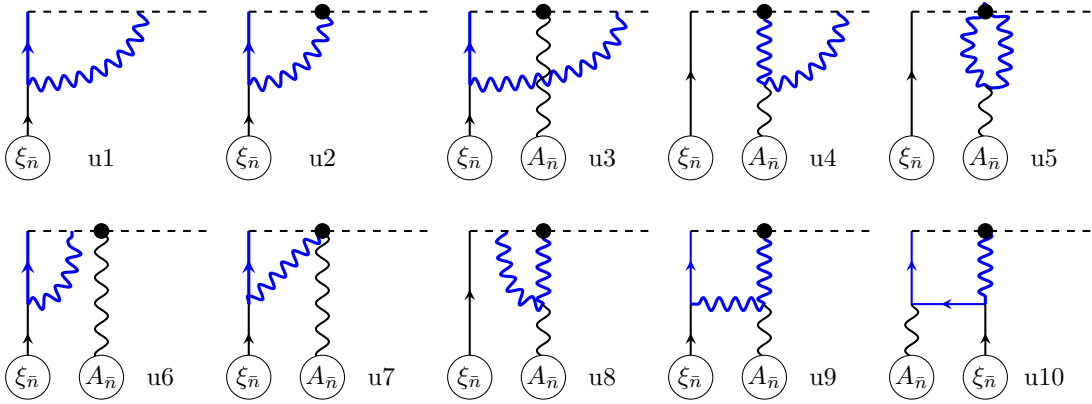

The renormalization factor for $U_{1,\bar n}$ is equal to the renormalization factor of the ``good'' component of the quark field in the light-cone gauge. It has an extra collinear divergence, which is proportional to the cusp anomalous dimension \cite{Ivanov:1985np,Korchemsky:1985xj}, and should be regularized by an additional regulator. In the present computation we use the $\delta$-regulator (\ref{def:delta-regulator}).

To get the renormalization factor in the $\delta$-regularization we compute the diagram $u1$ shown in fig. \ref{fig:loop-UV}. The expression for the diagram $u1$, is similar to eq.~(\ref{rap:loop1}) at $b=0$. Naively the loop integral is null due to the absence of the scale in the dimensional regularization, but it has an UV divergent part, which reads
\begin{eqnarray}
\text{diag}_{u1}&=&\frac{2a_sC_F}{\epsilon}\(1+\ln\(\frac{\delta^+}{is\hat p^+_\xi}\)\)U_{1,\bar n},
\end{eqnarray}
with $s=\sign(L)$. In addition to the diagram $u1$ one should take into account the renormalization of the quark wave function $Z_2^{1/2}=1-a_sC_F/(2\epsilon)$. 

Let us note, that the complex part of the diagram $\sim \ln(is)$ is canceled by the similar term in the charge-conjugated part of the TMD operator. Such cancellation takes place to all orders in perturbation theory. Therefore, we can safely eliminate these phase terms from the UV renormalization factors.
The renormalization of the operator is
\begin{eqnarray}\label{def:U1=Z1*U1}
U_{1,\bar n}(0)=Z_{U1}\(\frac{\delta^+}{\hat p_\xi^+}\)U_{1,\bar n}(0;\mu),
\end{eqnarray}
where
\begin{eqnarray}
Z_{U1}\(\frac{\delta^+}{\hat p_\xi^+}\)=1+\frac{a_sC_F}{\epsilon}\(\frac{3}{2}+2\ln\(\frac{\delta^+}{\hat p^+_\xi}\)\)+O(a_s^2).
\end{eqnarray}
We identify the UV renormalized operators by the presence of scaling argument $\mu$, similarly as we identify rapidity renormalized operators with scaling argument $\nu^+$. This is not a final expression for the renormalization of $U_1$, since  the variable $\delta^+$ is unspecified. This ambiguity disappears once we fix the value of $\delta^+$, which should be done coordinately with the rapidity divergent part, see sec.~\ref{sec:recombination}.

\subsection*{UV renormalization of $U_2$}

The UV divergent part of $U^\mu_{2,\bar n}$ is computed from the diagrams $u2$-$u10$ shown in fig.~\ref{fig:loop-UV}. The computation is straightforward, and in fact, it almost coincides with the computation of the $2\to2$ evolution kernels for collinear distributions made in ref.~\cite{Braun:2009vc,Ji:2014eta}. The only difference is the collinear divergence (and constant contributions) due to the half-infinite Wilson line.

The collinear divergences are present in the diagrams $u3$ and $u4$. Their expressions are
\begin{eqnarray}
\text{diag}_{u3}&=&\frac{2a_s}{\epsilon}\(C_F-\frac{C_A}{2}\)
\Big\{
\(1+\ln\(\frac{\delta^+}{is\hat p^+_\xi}\)\)U^\mu_{2,\bar n}(\{z,0\},0_T)
\\\nn && \qquad\qquad+\int_0^1 d\alpha \frac{\bar \alpha}{\alpha}
\(U^\mu_{2,\bar n}(\{z,0\},0_T)-U^\mu_{2,\bar n}(\{z,\alpha z\},0_T)\)\Big\}
,
\\
\text{diag}_{u4}&=&\frac{2a_s}{\epsilon}\frac{C_A}{2}
\(1+\ln\(\frac{\delta^+}{is\hat p^+_A}\)\)U^\mu_{2,\bar n}(\{z,0\},0_T),
\end{eqnarray}
where $\hat p_\xi$ and $\hat p_A$ are momenta of quark and gluon fields of the operator. Thus, the collinear singularity $\sim \ln\delta^+$ is proportional to the cusp anomalous dimension (just alike the $U_{1,\bar n}$ case), but the regulator $\delta^+$ is weighted by a different momentum in different diagrams. Different weighting produces a non-trivial contribution to the evolution kernel for TMD operators.

The computation of the remaining diagrams is straightforward. Let us only mention the diagram $u2$ that should be computed up to a derivative term, which after application of EOMs produces the operator $U_{2,\bar n}$. 
It is convenient to present
the sum of diagrams in the form
\begin{eqnarray}
&& \text{diag}_{u2}+...+\text{diag}_{u10}=
\frac{a_s}{\epsilon}\Bigg\{
\gamma_T^\mu\gamma_T^\nu
\mathbb{H}_1U^\nu_{2,\bar n}
+
\gamma_T^\nu\gamma^\mu_T \mathbb{H}_2U^\nu_{2,\bar n}
\\\nn && \qquad
+\Big[C_F\(2+2\ln\(\frac{\delta^+}{q^+}\)\)+2\(C_F-\frac{C_A}{2}\)\ln\(\frac{q^+}{is \hat p^+_\xi}\)
+C_A\ln\(\frac{q^+}{is \hat p^+_A}\)\Big]U^\mu_{2,\bar n}\Bigg\},
\end{eqnarray}
where $q^+=\hat p_A+\hat p_\xi$. The kernels $\mathbb{H}_{1,2}$ are quasi-partonic evolution kernels for quark-gluon pair \cite{Bukhvostov:1985rn}. In the notation of refs.~\cite{Braun:2009vc,Braun:2009mi}, they read
\begin{eqnarray}\label{def:H1}
\mathbb{H}_1&=&
\frac{C_A}{2}\widehat{\mathcal{H}}
-C_A\mathcal{H}^+
+2\(C_F-\frac{C_A}{2}\)\mathcal{H}^-,
\\\label{def:H2}
\mathbb{H}_2&=&
\frac{C_A}{2}\widehat{\mathcal{H}}
-\(C_F-\frac{C_A}{2}\)P_{12}\mathcal{H}^{e(1)}
.
\end{eqnarray}
Here, the $\mathcal{H}$ are elementary $2\to2$ kernels that are integral operators acting in  position space. In the present case, their explicit expressions are 
\begin{eqnarray}\label{def:Hhat}
\widehat{\mathcal{H}}U(z_1,z_2)&=&\int_0^1 \frac{d\alpha}{\alpha}
\(2U(z_1,z_2)-\bar \alpha^2 U(z_{12}^\alpha,z_2)-\bar \alpha U(z_1,z_{21}^\alpha)\),
\\
\mathcal{H}^+U(z_1,z_2)&=&\int_0^1 d\alpha \int^{\bar \alpha}_0 d\beta \,\bar \alpha\, U(z_{12}^\alpha ,z_{21}^\beta),
\\
\mathcal{H}^-U(z_1,z_2)&=&\int_0^1 d\alpha \int_{\bar \alpha}^1 d\beta \,\bar \alpha\, U(z_{12}^\alpha ,z_{21}^\beta),
\\\label{def:He}
P_{12}\mathcal{H}^{e(1)}U(z_1,z_2)&=&\int_0^1 d\alpha \,\bar \alpha \, U(z_{21}^\alpha,z_1),
\end{eqnarray}
where we use the shorthand notation
$$z_{ij}^\alpha=z_i (1-\alpha)+z_j \alpha.$$
These elementary kernels are invariant under the SL(2)-conformal transformation. For general expressions of $\mathcal{H}$, see appendix A in ref.~\cite{Braun:2009mi}. For the corresponding expressions in  momentum space, see \cite{Ji:2014eta}.

Adding the renormalization of the quark wave function, gluon wave function and the coupling constant (the latter two compensate each other) we get the renormalization factor
\begin{eqnarray}
U^\mu_{2,\bar n}(\{z,0\},0)=Z^{\mu\nu}_{U2}\(\frac{\delta^+}{q^+}\)\otimes U^\nu_{2,\bar n}(\{z,0\},0;\mu),
\end{eqnarray}
where
\begin{eqnarray}\label{U:Zmunu}
&&Z^{\mu\nu}_{U2}\(\frac{\delta^+}{q^+}\)=1+\frac{a_s}{\epsilon}\Bigg\{
\gamma_T^\mu\gamma_T^\nu \mathbb{H}_1
+
\gamma_T^\nu\gamma^\mu_T \mathbb{H}_2
\\\nn &&\qquad
+g^{\mu\nu}\Big[C_F\(\frac{3}{2}+2\ln\(\frac{\delta^+}{q^+}\)\)+2\(C_F-\frac{C_A}{2}\)\ln\(\frac{q^+}{\hat p^+_\xi}\)
+C_A\ln\(\frac{q^+}{\hat p^+_A}\)\Big]\Bigg\}+O(a_s^2).
\end{eqnarray}
The sign $\otimes$ indicates the integral convolution between kernels $\mathbb{H}_{1,2}$ and operator $U$. Here, we have eliminated $\sim \ln(is)$ terms, due to their cancellation with the charge-conjugated part of the TMD operator. The renormalization of the operator $\hat{U}_{2,\bar n}^i$ (defined in eq.~(\ref{def:U2-i})) is diagonal in the spinor indices
\begin{eqnarray}\label{def:U2=Z2*U2}
U^i_{2,\bar n}(\{z,0\},0)=Z_{U2}\(\frac{\delta^+}{q^+}\)\otimes U^i_{2,\bar n}(\{z,0\},0;\mu),
\end{eqnarray}
where
\begin{eqnarray}
Z_{U2}\(\frac{\delta^+}{q^+}\)&=&1+\frac{a_s}{\epsilon}\Bigg\{
2\mathbb{H}_1
+\Big[C_F\(\frac{3}{2}+2\ln\(\frac{\delta^+}{q^+}\)\)
\\\nn &&\qquad
+2\(C_F-\frac{C_A}{2}\)\ln\(\frac{q^+}{\hat p^+_\xi}\)
+C_A\ln\(\frac{q^+}{\hat p^+_A}\)\Big]\Bigg\}+O(a_s^2).
\end{eqnarray}
Note, that the kernel $\mathbb{H}_2$ vanishes, since $\gamma^\mu\gamma_T^\nu\gamma^\mu_T=0$. The momentum space representation for the kernel $\mathbb{H}_1$ is given in eq.~(\ref{app:H1-momentum}).

A simple structure of the renormalization factor (\ref{U:Zmunu}) allows us to guess the expressions for LO renormalization of many semi-compact operators, for example gluon or di-quark twist-(2+1) operators. In these cases, one should replace $\mathbb{H}$ by the corresponding parton evolution kernel, replace color coefficients in the last line, and take into account a different wave-function renormalization.

\subsection{Renormalization of unsubtracted TMD operators}
\label{sec:rap+UV}

Finally, we combine together the rapidity renormalization eq.~(\ref{rap:O11*R}, \ref{rap:O12*R}), and UV renormalization eq.~(\ref{def:U1=Z1*U1}, \ref{def:U2=Z2*U2}). For the TMD operators of twist-(1+1), eq.~(\ref{def:O11-position}, \ref{def:O11bar-position})  we obtain
\begin{eqnarray}\label{def:O11-renorm}
\mathcal{O}^{ij}_{11,\bar n}&=&R\(b^2,\frac{\delta^+}{\nu^+}\)
Z^*_{U1}\(\frac{\delta^+}{q^+}\)Z_{U1}\(\frac{\delta^+}{q^+}\)
\mathcal{O}^{ij}_{11,\bar n}(\nu^+,\mu),
\\
\overline{\mathcal{O}}^{ij}_{11,\bar n}&=&R\(b^2,\frac{\delta^+}{\nu^+}\)
Z_{U1}\(\frac{\delta^+}{q^+}\)Z^*_{U1}\(\frac{\delta^+}{q^+}\)\overline{\mathcal{O}}^{ij}_{11,\bar n}(\nu^+,\mu),
\end{eqnarray}
where we omit the common argument $(\{z_1,z_2\},b)$ of all TMD operators. For the TMD operators of twist-(1+2), eq.~(\ref{def:O21-position}-\ref{def:O12bar-position}) we have
\begin{eqnarray}\label{def:O12-renorm}
\mathcal{O}^{ij}_{21,\bar n}&=&
R\(b^2,\frac{\delta^+}{\nu^+}\)
Z^*_{U2}\(\frac{\delta^+}{q^+}\)Z_{U1}\(\frac{\delta^+}{q^+}\)\otimes\mathcal{O}^{ij}_{21,\bar n}(\nu^+,\mu),
\\\nn
\overline{\mathcal{O}}^{ij}_{21,\bar n}&=&
R\(b^2,\frac{\delta^+}{\nu^+}\)
Z_{U2}\(\frac{\delta^+}{q^+}\)Z^*_{U1}\(\frac{\delta^+}{q^+}\)\otimes\overline{\mathcal{O}}^{ij}_{21,\bar n}(\nu^+,\mu),
\\\nn
\mathcal{O}^{ij}_{12,\bar n}&=&
R\(b^2,\frac{\delta^+}{\nu^+}\)
Z^*_{U1}\(\frac{\delta^+}{q^+}\)Z_{U2}\(\frac{\delta^+}{q^+}\)\otimes\mathcal{O}^{ij}_{12,\bar n}(\nu^+,\mu),
\\\nn
\overline{\mathcal{O}}^{ij}_{12,\bar n}&=&
R\(b^2,\frac{\delta^+}{\nu^+}\)
Z_{U1}\(\frac{\delta^+}{q^+}\)Z^*_{U2}\(\frac{\delta^+}{q^+}\)\otimes\overline{\mathcal{O}}^{ij}_{12,\bar n}(\nu^+,\mu),
\end{eqnarray}
where we omit the common argument $(\{z_1,z_2,z_3\},b)$ of all TMD operators. The sign $\otimes$ indicates that the factor $Z_{U2}$ is an integral operator, which acts on the positions the of quark-gluon pair. These expressions also hold for Fourier transformed TMD operators, with the expression for the quasi-parton evolution kernels $\mathbb{H}$ given in app.~\ref{app:evol-momentum}. The expressions for the TMD operators composed from anti-collinear fields are analogous with $\{\delta^+,\nu^+,q^+\}\to\{\delta^-,\nu^-,q^-\}$.

In eq.~(\ref{def:O11-renorm}-\ref{def:O12-renorm}), we assume the forward kinematics, and $q^+$ being the momentum passing through semi-compact operators. Moreover, we set this momentum equal to the momentum passing though the EM current, since the corresponding combination ($\hat p_\xi^++\hat p_A^+$ for $Z_{U2}$ and $\hat p_\xi^+$ for $Z_{U1}$) equal to $q^+$ in the effective operator, due to the $\delta$-function in eq.~(\ref{NLO:J-momentum-deltas}).

Note, that in the eq.~(\ref{def:O11-renorm}-\ref{def:O12-renorm}) we are using the renormalized factor $R$, since its UV divergence is  part of the UV renormalization factor for TMD operator. Alternatively, one can use the unrenormalized factor $\widetilde{R}$ (\ref{def:Rtilde}). In this case,  eq.~(\ref{def:O11-renorm}) takes the form
\begin{eqnarray}
\mathcal{O}^{ij}_{11,\bar n}(y)&=&\widetilde{R}\(b^2,\frac{\delta^+}{\nu^+}\)
Z^*_{U1}\(\frac{\nu^+}{q^+}\)Z_{U1}\(\frac{\nu^+}{q^+}\)
\mathcal{O}^{ij}_{11,\bar n}(y;\nu^+,\mu),
\end{eqnarray}
and similar for other operators.

In eq.~(\ref{def:O11-renorm}-\ref{def:O12-renorm}), the asterisk denotes the complex conjugation. It affects the phase in the term $\sim \ln(iL)$ only. Due to this conjugation the complex parts of the factors $Z$ cancel. Since we already took this cancellation into account in the definitions eq.~(\ref{def:U1}, \ref{def:U2-i}) where the terms $\sim\ln(iL)$ are eliminated, the indication of conjugation is obsolete. However, we keep this indication in formulas eq.~(\ref{def:O11-renorm}, \ref{def:O12-renorm}) to keep track of terms' order.

\section{Recombination of divergences and scaling of TMD operators}
\label{sec:recombination}

Having computed all elements of  TMD factorization (UV and rapidity renormalization factors, hard coefficient functions and the soft factor), we can finally combine them into a divergence-free expression. This procedure defines the scheme of the rapidity and UV renormalization, and thus defines the physical TMD distributions.

\subsection*{Subtracted version of TMD operators}

The renormalized TMD operators eq.~(\ref{def:O11-renorm}-\ref{def:O12-renorm}) have the common form
\begin{eqnarray}\label{def:ONM-renorm}
\mathcal{O}_{NM,\bar n}(y)=R\(b^2, \frac{\delta^+}{\nu^+}\)Z^*_N\(\frac{\nu^+}{q^+}\)Z_M\(\frac{\nu^+}{q^+}\)\otimes \mathcal{O}_{NM,\bar n}(y;\nu^+,\mu).
\end{eqnarray}
Here, $N$ and $M$ can be 1 or 2 and the renormalization constants $Z_N$ and $Z_M$ are $Z_{U1}$ or $Z_{U2}$. A similar expression is valid for the TMD operator $\mathcal{O}_{KL,n}$ oriented along direction $\bar n$, 
\begin{eqnarray}
\mathcal{O}_{KL,n}(y)=R\(b^2, \frac{\delta^-}{\nu^-}\)Z_K\(\frac{\nu^-}{q^-}\)Z_L\(\frac{\nu^-}{q^-}\)\otimes \mathcal{O}_{KL,n}(y;\nu^-,\mu).
\end{eqnarray}
Combining the TMD operators with the soft factor, eq.~(\ref{SF:S=RSR}) we obtain
\begin{eqnarray}\label{def:OO/S=OO}
&&\frac{\mathcal{O}_{NM,\bar n}\mathcal{O}_{KL,n}}{\mathcal{S}_{\text{LP}}}
=
\[Z^{\text{sub}}_{N}(\zeta)Z^{\text{sub}}_{M}(\zeta)\otimes\mathcal{O}^{\text{sub}}_{NM,\bar n}(\zeta,\mu)\]
\[Z^{\text{sub}}_{K}(\bar \zeta)Z^{\text{sub}}_{L}(\bar \zeta)\otimes\mathcal{O}^{\text{sub}}_{KL,n}(\bar \zeta,\mu)\],
\end{eqnarray}
where the factors $R$ are canceled, and 
\begin{eqnarray}
Z^{\text{sub}}_{N}(\zeta)=\frac{Z_N\(\frac{\nu^+}{q^+}\)}{Z_{\mathcal{S}}^{\frac{1}{4}}(\nu^2)},
\qquad
Z^{\text{sub}}_{N}(\bar \zeta)=\frac{Z_N\(\frac{\nu^-}{q^-}\)}{Z_{\mathcal{S}}^{\frac{1}{4}}(\nu^2)},
\end{eqnarray}
and 
\begin{eqnarray}\label{def:TMDop-sub}
\mathcal{O}^{\text{sub}}_{NM,\bar n}(\zeta,\mu)=
\frac{\mathcal{O}^{}_{NM,\bar n}(\nu^+,\mu)}{\sqrt{S_0(b^2,\nu^2)}},
\qquad
\mathcal{O}^{\text{sub}}_{KL,n}(\bar \zeta,\mu)=
\frac{\mathcal{O}^{}_{KL,n}(\nu^-,\mu)}{\sqrt{S_0(b^2,\nu^2)}}.
\end{eqnarray}
Here, we have introduced the notation for Lorentz invariant combinations,
\begin{eqnarray}
\zeta=2 (q^+)^2\frac{\nu^-}{\nu^+},\qquad \bar \zeta=2 (q^-)^2\frac{\nu^+}{\nu^-}.
\end{eqnarray}
The variables $\zeta$ and $\bar \zeta$ naturally appear as the arguments of the logarithms in subtracted renormalization constants. Explicitly, the renomalization constants are
\begin{eqnarray}\label{def:ZU1-sub}
Z_{U1}^{\text{sub}}(\zeta)&=&1+\frac{a_sC_F}{\epsilon}\(\frac{1}{\epsilon}+\frac{3}{2}+
\ln\(\frac{\mu^2}{\zeta}\)\)+O(a_s^2),
\\\label{def:ZU2-sub}
Z^{\text{sub}}_{U2}(\zeta)&=&1+\frac{a_s}{\epsilon}\Bigg\{
2\mathbb{H}_1
+\Big[C_F\(\frac{1}{\epsilon}+\frac{3}{2}+\ln\(\frac{\mu^2}{\zeta}\)\)
\\\nn &&\qquad
+2\(C_F-\frac{C_A}{2}\)\ln\(\frac{q^+}{\hat p^+_\xi}\)
+C_A\ln\(\frac{q^+}{\hat p^+_A}\)\Big]\Bigg\}+O(a_s^2).
\end{eqnarray}
The factor $Z_{U1}^{\text{sub}}$ is a half of the TMD renormalization constant and coincides with it \cite{Echevarria:2016scs}. The factor $Z^{\text{sub}}_{U2}$ is a new one, to our best knowledge.

The rapidity divergent factor $R$ cancels between the soft factor and TMD operators\footnote{There is an important exception -- the transverse derivative acting to the twist-(1+1) operators also acts to the rapidity renormalization factor $R$. E.g. for the first term in the second line for $\mathcal{J}_{1111}^{\mu\nu}$ (\ref{def:J1111}) one has
\begin{eqnarray}
\frac{\partial_{\rho}\mathcal{O}_{11,\bar n}
\overline{\mathcal{O}}_{11,n}}{\mathcal{S}_{\text{LP}}}=
\partial_{\rho}\mathcal{O}_{11,\bar n}^{\text{sub}}(\zeta,\mu)\overline{\mathcal{O}}_{11,n}^{\text{sub}}(\zeta,\mu)
+
\mathcal{O}_{11,\bar n}^{\text{sub}}(\zeta,\mu)\overline{\mathcal{O}}_{11,n}^{\text{sub}}(\zeta,\mu)
\partial_{\rho}\ln\(R\(b^2,\frac{\delta^+}{\nu^+}\)\sqrt{S_0(b^2,\nu^2)}\),
\end{eqnarray}
where we omit the factors $Z$ (which commute with the derivative) for brevity. The second term of this expression is divergent. This divergence is compensated by the end-point divergence of $\sim n^\mu$ part of the operator $\mathcal{J}_{2111}^{\mu\nu}$ (\ref{def:2111}) at $x_2\to0$, which can be easily confirmed at LO. In this way, derivatives of rapidity rapidity renormalization factors cancels in-between terms related by EOMs.

The end-point divergence of $\sim \bar n^\mu$ part of the operator $\mathcal{J}_{2111}^{\mu\nu}$ remains uncompensated. However, this part of the effective operator is a result of the direct interaction with background field. The limit $x_2\to0$ pushes the gluon field to the overlap region and
thus should be subtracted.  

The same applies for other terms of the effective operator. The explicit realisation of this procedure will be presented in a different publication.
}. The leftover of the rapidity divergences are the scaling parameters $\nu^\pm$. In the final definition of the TMD operator in eq.~(\ref{def:TMDop-sub}), we replace $\nu^\pm$ by $\zeta$ and $\bar \zeta$ using that
\begin{eqnarray}
\nu^+=q^+\sqrt{\frac{\nu^2}{\zeta}},\qquad \nu^-=q^-\sqrt{\frac{\nu^2}{\bar \zeta}}.
\end{eqnarray}
The $\nu^2=2\nu^+\nu^-$ is a low-energy parameter related to the definition of soft modes, and it can be hidden in the definition of TMD distribution. The scaling variables $\zeta$ and $\bar \zeta$ satisfy
\begin{eqnarray}\label{def:zeta*zeta=q+q-}
\zeta \bar \zeta=(2q^+q^-)^2.
\end{eqnarray}

The definition of the subtracted TMD operator incorporates the remnant of the soft factor $S_{0}$, which is a finite number that depends on $b^2$. We recall that the function $S_0$ is nonperturbative. The absorption of $S_0$ into TMD distributions is a part of a scheme definition for the rapidity renormalization. Effectively it adds finite terms to the ``minimal subtraction scheme'' factor $R$. The equation (\ref{def:OO/S=OO}) serves as the definition of the scheme, i.e. the final expression for the effective operator in DY, SIDIS, and SIA does not contain any extra factors. This statement defines the commonly used physical TMD distributions. It also implies that the TMD factorization formula for processes, which contain a soft factor different from $\mathcal{S}_{\text{LP}}$, would have an extra nonperturbative function composed from the remnants of the soft factors. For example, such function occurs in the factorization theorem for quasi-TMD distribution \cite{Vladimirov:2020ofp}.

\subsection*{Evolution equations for TMD operators}

The TMD operator $\mathcal{O}_{NM,\bar n}$ has two scaling parameters $\mu$ and $\zeta$. The evolution equations with respect to these parameters follow from the scaling invariance of bare TMD operator eq.~(\ref{def:ONM-renorm}).

The evolution equation with respect to $\mu$ is
\begin{eqnarray}\label{evol:mu}
\mu^2 \frac{d}{d\mu^2}\mathcal{O}^{\text{sub}}_{NM,\bar n}(\zeta,\mu)
=\(\gamma_N(\mu,\zeta)+\gamma_M(\mu,\zeta)\)\otimes \mathcal{O}^{\text{sub}}_{NM,\bar n}(\zeta,\mu),
\end{eqnarray}
where
\begin{eqnarray}
\gamma_N(\mu,\zeta)=-\mu^2\frac{dZ^{\text{sub}}_{N}(\zeta)}{d\mu^2},
\end{eqnarray}
The symbol $\otimes$ indicate a possible integral convolution. Each anomalous dimension acts on corresponding arguments. 

The evolution equation (\ref{evol:mu}) is general in the sense that a TMD operator of any twist-(N+M) satisfies it. The value of twist is preserved by the evolution, which is a part of geometrical twist definition. However, if there are several operators $U_N$ of the same twist and other quantum numbers they can mix with each other.

In the present case, we have deal on with twist-1 and twist-2 operators. The corresponding anomalous dimensions are
\begin{eqnarray}
\gamma_1(\mu,\zeta)&=&a_sC_F\(\frac{3}{2}+\ln\(\frac{\mu^2}{\zeta}\)\)+O(a_s^2),
\\
\gamma_2(\mu,\zeta)&=&a_s\Bigg\{
2\mathbb{H}_1
+\Big[C_F\(\frac{3}{2}+\ln\(\frac{\mu^2}{\zeta}\)\)
\\\nn &&\qquad
+2\(C_F-\frac{C_A}{2}\)\ln\(\frac{q^+}{\hat p^+_\xi}\)
+C_A\ln\(\frac{q^+}{\hat p^+_A}\)\Big]\Bigg\}+O(a_s^2).
\end{eqnarray}
In momentum space, the expression for $\mathbb{H}$ is given in app.~\ref{app:evol-momentum} and $\hat p^+/q^+$ should be replaced by corresponding momentum fractions. E.g. for $\gamma_2\otimes \mathcal{O}_{21,\bar n}(x_{1,2,3},b)$ one replaces
\begin{eqnarray}
\ln\(\frac{q^+}{\hat p^+_\xi}\)\to -\ln x_1,\qquad \ln\(\frac{q^+}{\hat p^+_A}\)\to -\ln x_2.
\end{eqnarray}

These anomalous dimensions can be written in a more general form
\begin{eqnarray}\label{def:gamma1}
\gamma_1(\mu,\zeta)&=&\frac{\Gamma_{\text{cusp}}}{4}\ln\(\frac{\mu^2}{\zeta}\)-\frac{\gamma_V}{4},
\\\label{def:gamma2}
\gamma_2(\mu,\zeta)&=&\mathbb{H}_{A\xi}+
\frac{\Gamma_{\text{cusp}}}{4}\ln\(\frac{\mu^2}{\zeta}\)
+\frac{\Gamma_{\text{cusp}}-\Gamma_g}{2} \ln\(\frac{q^+}{\hat p^+_\xi}\)
+\frac{\Gamma_g}{2}\ln\(\frac{q^+}{\hat p^+_A}\),
\end{eqnarray}
where $\Gamma_{\text{cusp}}$ is the (quark) light-like cusp anomalous dimension (known up to N$^3$LO \cite{Henn:2019rmi,vonManteuffel:2020vjv}), $\gamma_V$ is the anomalous dimension of the quark vector form factor (known up to NNLO, see e.g. \cite{Gehrmann:2010ue}). The integral kernel $\mathbb{H}_{A\xi}$ is the evolution kernel for the quark-gluon pair. And $\Gamma_g$ is some constant $\Gamma_g=2a_sC_A+O(a_s^2)$. The sum of the last two terms in eq.~(\ref{def:gamma2}) must be $\Gamma_{\text{cusp}}$, since the sum of $\ln(q^+)$ generates $\ln(\zeta)$ whose coefficient is fixed by the integrability condition eq.~(\ref{def:integrability}).

The evolution with respect to the rapidity parameter follows from the independence of bare TMD operators eq.~(\ref{def:O11-renorm}, \ref{def:O12-renorm}) on the parameter $\nu^+$. It gives
\begin{eqnarray}\label{evol:zeta}
\zeta \frac{d\mathcal{O}^{\text{sub}}_{NM,\bar n}(\zeta,\mu)}{d\zeta}=-2\nu^+\frac{d\mathcal{O}^{\text{sub}}_{NM,\bar n}(\zeta,\mu)}{d\nu^+}=-\mathcal{D}(b,\mu)\mathcal{O}^{\text{sub}}_{NM,\bar n}(\zeta,\mu),
\end{eqnarray}
with
\begin{eqnarray}
\mathcal{D}(b,\mu)=-\frac{\nu^+}{2}\frac{d}{d\nu^+}R\(b^2,\frac{\delta^+}{\nu^+}\).
\end{eqnarray}
The function $\mathcal{D}(b,\mu)$ is the rapidity anomalous dimension. It is also known as the Collins-Soper kernel ($\tilde K=-2\mathcal{D}$), originally introduced in ref.~\cite{Collins:1981uk}. For a relation among different definitions of the CS-kernel, see ref.~\cite{Scimemi:2018xaf}. The LO expression at small $b$ is
\begin{eqnarray}\label{RAD-LO}
\mathcal{D}(b,\mu)&=&-2a_s C_F\Bigg[
\Gamma(-\epsilon)\(-\frac{b^2 \mu^2}{4e^{-\gamma_E}}\)^\epsilon+\frac{1}{\epsilon}\Bigg]+O(a_s^2)
\\\nn &=&2a_SC_F \ln\(\frac{-b^2 \mu^2}{4 e^{-2\gamma_E}}\)+O(a_s^2),
\end{eqnarray}
where in the second line took the $\epsilon\to0$ limit. The rapidity anomalous dimension is known up to NNLO \cite{Vladimirov:2016dll,Li:2016ctv}. The important feature of the rapidity anomalous dimension is its nonperturbative nature. In that respect, eq.~(\ref{RAD-LO}) is valid only at small values of $b$ and gets  power corrections once $b$ increases. The $\sim b^2$ power correction has been computed in ref.~\cite{Vladimirov:2020umg} from the analysis of the soft factor. Phenomenologically, the Collins-Soper kernel can be extracted from the analysis of TMD factorization at different scales. The most recent extraction of it can be found in refs.~\cite{Scimemi:2019cmh,Bacchetta:2019sam} (see also \cite{Vladimirov:2020umg} for comparison of different extractions).

The TMD evolution is given by a pair of equations, eq.~(\ref{evol:mu}, \ref{evol:zeta}). The existence of a common solution is guaranteed by the integrability condition (known also as the Collins-Soper equation \cite{Collins:1981va}). It relates UV and rapidity anomalous dimensions
\begin{eqnarray}\label{def:integrability}
\zeta\frac{d}{d\zeta}\(\gamma_{N}+\gamma_M\)=\mu^2 \frac{d}{d\mu^2}\mathcal{D}(b,\mu)=\frac{\Gamma_{\text{cusp}}}{2},
\end{eqnarray}
where the last equality follows from eq.~(\ref{def:gamma1}, \ref{def:gamma2}). The integrability condition guarantees the path-independence of the evolution, see ref.~\cite{Scimemi:2018xaf} for an extended discussion.

\subsection*{Cancellation of divergences in the factorized expression}

Finally, we should check that the UV poles of the TMD operators cancel the IR poles of the hard coefficient function. For the LP term (the first line of eq.~(\ref{NLO:J-momentum-deltas})), it implies that
\begin{eqnarray}
C_1&=&\tilde C_1Z^{\text{sub}}_{U1}(\zeta)Z^{\text{sub}}_{U1}(\bar \zeta),
\end{eqnarray}
is finite. Indeed, the pole part of $\tilde C_1$, eq.~(\ref{NLO:C1}) is
\begin{eqnarray}\label{pole-cancel:C1}
\text{pole}[\tilde C^{\text{NLO}}_1]&=&-\frac{a_sC_F}{\epsilon}\(\frac{2}{\epsilon}+3+2\ln\(\frac{\mu^2}{|2q^+q^-|}\)\)\!
=\!-\text{pole}\[Z^{\text{sub};{\text{NLO}}}_{U1}(\zeta)\!+\!Z^{\text{sub};{\text{NLO}}}_{U1}(\bar \zeta)\]\!,
\end{eqnarray}
where we use the relation (\ref{def:zeta*zeta=q+q-}) between $\zeta$ and $\bar \zeta$.

The relation for NLP part (the second and the third lines of eq.~(\ref{NLO:J-momentum-deltas})) is a bit more complicated, because it involves a convolution. Using the definition of the operators eq.~(\ref{def:J21-momentum}, \ref{def:J12-momentum}) and the form of the convolution in eq.~(\ref{NLO:J-momentum-deltas}), one shows that the cancellation of poles requires that
\begin{eqnarray}
&&\int \frac{d x_1 dx_2}{x_2}\delta(x_1+x_2+x_3) C_2(x_{1,2})U(x_2,x_1)
\\\nn && \qquad\qquad=
\int \frac{d x_1 dx_2}{x_2}\delta(x_1+x_2+x_3)
\tilde C_2(x_{1,2})\otimes Z^{\text{sub}}_{U2}(\zeta)Z^{\text{sub}}_{U1}(\bar \zeta)\otimes U(x_2,x_1),
\end{eqnarray}
where $U$ is a test function, and $x_1$ is the momentum fraction associated with the gluon. Inserting the momentum space representation for the $Z_{U2}$ renormalization constant eq.~(\ref{app:H1-momentum}) and performing changes of variables, we confirm that
\begin{eqnarray}\label{pole-cancel:C2}
&& \text{pole}\Big[\int \frac{d x_1 dx_2}{x_2}\delta(x_1+x_2+x_3)
\tilde C^{\text{NLO}}_2(x_{1,2})U(x_2,x_1)\Bigg] 
\\\nn &&\qquad= -\frac{a_s}{\epsilon}
\int \frac{d x_1 dx_2}{x_2}\delta(x_1+x_2+x_3)
\Big[C_F\(\frac{2}{\epsilon}+1+2\ln\(\frac{\mu^2}{|2q^+q^-|}\)\)
\\\nn &&\qquad\qquad\qquad
+2\(C_F-\frac{C_A}{2}\)\frac{x_1+x_2}{x_1}\ln\(\frac{x_1+x_2}{x_2}\)\Big]U(x_2,x_1)
\\\nn &&\qquad
=-\text{pole}\[
\int \frac{d x_1 dx_2}{x_2}\delta(x_1+x_2+x_3)
\(Z^{\text{sub};{\text{NLO}}}_{U2}(\zeta)+Z^{\text{sub};{\text{NLO}}}_{U1}(\bar \zeta)\)\otimes U(x_2,x_1)\].
\end{eqnarray}
This is an independent check of the computations made in sec.~\ref{sec:NLO}. We also stress that most part of the renormalization constants $Z_{U2}$ is equal to the quasi-parton-pair evolution kernel, and it can be cross-checked with the literature e.g. \cite{Bukhvostov:1985rn,Braun:2009vc,Ji:2014eta}. Altogether, these checks strongly support the results presented in this work.

\section{Conclusion}

In this work, we have developed a method to derive the TMD factorization theorem. It is based on the background field method and has similarities with the high energy expansions \cite{Balitsky:1995ub}, SCET, and operator product expansion. The method, to which we refer to as {\it  TMD operator expansion}, allows a systematic derivation of the TMD factorization formulas at operator level and at any order of power series.  As a demonstration, we compute the TMD factorization to NLP at NLO and confirm our computation by comparison with the well-known LP and partially-known NLP expressions. The expression for effective operator is given in eq.~(\ref{NLO:J-momentum-deltas}), and evolution equations for NLP TMD operators are given in eqs.~(\ref{evol:mu}, \ref{evol:zeta}). We recover many results found in the literature in one or another form even without reference to TMD factorization. For example, the evolution kernel for TMD operators at NLP incorporates the standard quasi-parton evolution kernels \cite{Bukhvostov:1985rn}, which we successfully reproduce. To our best knowledge, the NLO perturbative correction to the NLP TMD factorization theorem is a new result. The main goal of this work is the formulation of a general approach to power corrections in TMD factorization. Let us remark some of the most important lessons.

The TMD operator expansion is derived starting from the definition of QCD in the functional integral form. As a part the of derivation, we suppose that hadron states are built from collinear and anti-collinear fields and perform a functional integration of the remaining components. Due to it, the derivation of a factorized expression is automatic, as all operators and their coefficient functions arise from a single initial definition, avoiding any matching procedure, typical for many other derivations. The computations are done in position space, which is the natural language for power corrections since the momentum space expressions are complicated due to multiple momentum ranges. To our best knowledge, it is the first computation of TMD factorization solely in position space.

The famous process dependence of TMD factorization \cite{Collins:2002kn,Boer:2003cm} (which consists in the orientation of gauge links) appears due to the boundary conditions that are imposed on the background fields. These boundary conditions follow from the request that (matrix elements of) operators are analytic in a proper part of momentum space in accordance with the process. The purely operator-level derivation of TMD factorization is an important step forward for the future development of the TMD approach. With minimal modifications our expressions can be used for a description of processes with jets, e.g. \cite{Neill:2016vbi,Buffing:2018ggv,Liu:2020dct}, or involving generalized TMD distributions (GTMDs).

The TMD operators are built from two semi-compact light-ray operators $\mathcal{O}_{MN}=U_M U_N$, separated by a transverse distance $b$. The operators $U_N$ can be sorted with respect to geometrical twists, alike ordinary compact operators. They are subject to independent UV renormalization. Therefore, the TMD operators can be classified with respect to TMD-twist, which is given by a pair of numbers (M+N), where M, N  are geometrical twists of semi-compact operators $U$. The TMD distributions derived from operators with different TMD-twist have independent UV scaling, and thus they are independent observables. Let us mention that, in principle, operators of higher TMD-twists can mix with operators of lower TMD-twist via the rapidity scale evolution. This question is to be studied in the future.

In the small-b limit, a TMD operator of twist-(M+N) matches onto collinear operators of twist-$(M+N)$ or higher. This is an essential property because it ensures that the terms of TMD factorization with TMD twist-(1+1) (that is, the LP term and its kinematic power corrections) turn into the fixed order expression with twist-2 parton distributions in the high-energy limit. This observation gives hope to describe the whole $q_T$ spectrum of DY, SIDIS, and SIA processes within the TMD factorization approach.

We demonstrate in three independent ways (from the side of the soft factor, from the side of the rapidity divergences renormalization theorem, and by the direct NLO computation) that the rapidity-scale evolution is the same for LP and NLP operators. We also show that it is the same for a part of N$^2$LP TMD operators (namely, for all operators of TMD-twist-(2+2) and TMD-twist-(3+1) quasi-partonic operators). The equality of rapidity evolution for different power TMD operators opens new possibilities to measure the Collins-Soper kernel, which is one of the fundamental functions in QCD \cite{Vladimirov:2020umg}. For example, studying the ratios of sub-leading quasi-TMD distributions on the lattice, similar to the existing cases \cite{Shanahan:2020zxr,LatticeParton:2020uhz,Schlemmer:2021aij}. The important discussion on the derivatives of rapidity renormalization factors, the end-point divergences and their mutual cancellation is left for the future publication.

The derivation allows an accurate separation of sources of power corrections, which are listed in the introduction. In particular, we demonstrate that the twist-(1+1)$\times$(1+1) part of the NLP term extends the LP term and is the kinematic power correction. It restores the EM gauge invariance up the N$^2$LP order and has the same hard coefficient function. Another important observation, which was overlooked or ignored in previous studies (except \cite{Inglis-Whalen:2021bea}), is that the actual expansion parameter is $\tau_T^2=q_T^2/|2q^+q^-|$ ($=q_T^2/(Q^2+q_T^2)$ for DY ans SIA, and $(=q_T^2/(Q^2-q_T^2)$ for SIDIS) rather than $q_T^2/Q^2$. In fact, the value $Q^2$ does not appear in any formula, but all kinematic variables are expressed via $\tau_T^2$. Accounting for this simple fact can be important for phenomenology (see some studies in ref.~\cite{Scimemi:2019cmh}).

In the present work, we consider only the theoretical aspects of the TMD factorization at NLP. The relevant cross-section, systematization and relations to earlier defined NLP TMD distributions \cite{Boer:2003cm,Bacchetta:2006tn} will be performed in future publications. It is known that many observables are not sensitive to NLP $\sim q_T/Q$ corrections but have the first non-vanishing correction at N$^2$LP $\sim q_T^2/Q^2$. Even so, our work will improve the present LP picture due to kinematic corrections and helps to clear up the definitions of some observables linear in $q_T$, such as the Cahn effect \cite{Cahn:1978se}.

\acknowledgments Authors are very thankful to V.Braun and A.Manashov for numerous discussions and helpful comments, and to M.Beneke and collaborators for the sharing of preliminary results of their computation. This study was supported by Deutsche Forschungsgemeinschaft (DFG) through the research Unit FOR 2926, “Next Generation pQCD for Hadron Structure: Preparing for the EIC”, project number 30824754. I.S. is supported by the Spanish Ministry grant PID2019-106080GB-C21. 
This project has received funding from the European Union Horizon 2020 research and innovation program under grant agreement Num. 824093 (STRONG-2020).

\appendix

\section{QCD Lagrangian with composite background field}
\label{app:Sint}

The rules for the interactions of dynamical QCD fields with a background QCD field have been derived a long time ago \cite{DeWitt:1980jv,Boulware:1980av,Abbott:1980hw}.  They should be updated, since we operate with two independent copies of background fields, which modifies the structure of the interaction Lagrangian. Although we only need a single extended vertex in the present work, we revisit the background field method for the case of composite background fields and derive the full structure for completeness.

The starting point is the QCD Lagrangian with the ordinary background field. It is derived with the replacement of ordinary gluon and quark fields in the QCD Lagrangian $\mathcal{L}[q,A]$ by
\begin{eqnarray}
A_\mu \to A_\mu+B_\mu,\qquad q\to q+\psi,
\end{eqnarray}
where $A_\mu$ ($B_\mu$) and $q$ ($\psi$) are background (dynamical) gluon and quark fields. The Lagrangian $\mathcal{L}[q+\psi,A+B]$ is invariant under gauge transformations of the dynamical fields 
\begin{eqnarray}
B_\mu\to B_\mu+D_\mu[A+B] \alpha,
\end{eqnarray}
where $D_\mu[A]$ is the covariant derivative with the gauge field $A$. The background-gauge-fixing condition for dynamical fields is given by 
\begin{eqnarray}
G[B]=D^\mu[A]B_\mu.
\end{eqnarray}
After application of the Faddeev-Popov trick one arrives at the action \cite{Abbott:1980hw,Abbott:1981ke}
\begin{eqnarray}
S=\int d^4x \(\mathcal{L}_{QCD}[q+\psi,A+B]+\mathcal{L}_{gh}[A,B]+\mathcal{L}_{fix}[A,B]\),
\end{eqnarray}
where
\begin{eqnarray}
\mathcal{L}_{QCD}[q,A]=i \bar q\fnot D q-\frac{1}{4}\(F_{\mu\nu}^a[A]\)^2,
\end{eqnarray}
\begin{eqnarray}
\mathcal{L}_{gh}[A,B]=-\eta^\dagger D_\mu[A+B]D^\mu[A]\eta,\qquad
\mathcal{L}_{fix}[A,B]=\frac{1}{2\alpha}B^\mu D_\mu[A]D_\nu[A]B^\nu,
\end{eqnarray}
and where $\eta$ is the ghost field, and $\alpha$ is residual gauge parameter. The gluon-field strength tensor is $F_{\mu\nu}^a[A]=\partial_\mu A_\nu^a-\partial_\nu A_\mu^a+gf^{abc}A^b_\mu A^c_\nu$. The effective action revealed after the integration over the dynamical fields is invariant under the gauge transformation of the background fields.

We split the background field into two non-overlapping components
\begin{eqnarray}
A_\mu\to A_{1\mu}+A_{2\mu},\qquad q\to q_1+q_2.
\end{eqnarray}
Then the action can be presented as
\begin{eqnarray}
S=S_{QCD\text{(cov.g.)}}[B,\psi]+S_{QCD}[A_1,q_1]+S_{QCD}[A_2,q_2]+S_{int},
\end{eqnarray}
where $S_{int}$ involves all fields, and $S_{QCD\text{(cov.g.)}}$ is the QCD action in the covariant gauge. In turn, the interaction term can be conveniently split into
\begin{eqnarray}\label{app:Sint:main}
S_{int}&=&S_{1h}+S_{2h}+S_{12}+S_{12h},
\end{eqnarray}
where $S_{1(2)h}$ contains only fields $\{A_{1(2)},q_{1(2)},B,\psi\}$, $S_{12}$ contains only background fields, and $S_{12h}$ contains combinations of both components of background field and dynamical fields. Since the background fields are external, one can eliminate all terms in the interaction Lagrangians that satisfy EOMs. These are all terms $\sim B^1$ in $S_{1h}$ and $S_{2h}$, and some terms in $S_{12}$. The resulting actions are
\begin{eqnarray}\nn
S_{1h(2h)}&=&g\int d^4x\Big\{
\bar q \fnot B \psi+\bar \psi \fnot A\psi+\bar \psi \fnot B q
\\\label{app:Sint:S1q} &&
+f^{abc}A_\mu^a\eta^{b\dagger}(\overrightarrow{\partial}_\mu-\overleftarrow{\partial}_\mu)\eta^c
-f^{abr}f^{rcd}\eta^{a\dagger}(A^b_\mu+B^b_\mu)A_\mu^c \eta^d
\\ && \nn
-f^{abc}A_\mu^a\Big[2 (\partial_\nu B_\mu^b)B_\nu^c-(\partial_\mu B_\nu^b)B_\nu^c-\frac{1+\alpha}{\alpha}(\partial_\nu B_\nu^b)B_\mu^c\Big]
\\\nn && -\frac{g}{2}f^{abr}f^{rcd}\Big[A^a_\mu A^b_\nu B^c_\mu B^d_\nu+A^a_\mu B^b_\nu A^c_\mu B^d_\nu+A^a_\mu B^b_\nu B^c_\mu A^d_\nu+\frac{1}{\alpha}A^a_\mu B^b_\mu A^c_\nu B^d_\nu\Big]\Big\},
\end{eqnarray}
here $q=q_{1(2)}$ and $A=A_{1(2)}$,
\begin{eqnarray}\nn
S_{12h}&=&g\int d^4x\Big\{
\bar q_1 \fnot A_2 \psi
+\bar \psi \fnot A_1 q_2
+\bar q_1 \fnot B q_2
+\{1\leftrightarrow 2\}
\\\nn &&
-f^{abr}f^{rcd}\eta^{a\dagger}A^b_{1\mu} A_{2\mu}^c \eta^d
-
f^{abc} [
(\partial_\mu A_{1\nu}^a-\partial_\nu A_{1\mu}^a)A_{2\mu}^b B_{\nu}^c+A_{1\mu}^aA_{2\nu}^b\partial_\mu B^c_\nu]+\{1\leftrightarrow 2\}
\\\label{app:Sint:S12q} && 
- gf^{abr}f^{rcd}(
A^a_{1\mu}A^b_{1\nu}A^c_{2\mu}B^d_{\nu}
+A^a_{1\mu}B^b_{\nu}A^c_{1\mu}A^d_{2\nu}
+B^a_{\mu}A^b_{1\nu}A^c_{1\mu}A^d_{2\nu})+\{1\leftrightarrow 2\}
\\\nn && - gf^{abr}f^{rcd}(
B^a_{\mu}B^b_{\nu}A^c_{1\mu}A^d_{2\nu}
+A^a_{1\mu}B^b_{\nu}A^c_{2\mu}B^d_{\nu}
+B^a_{\mu}A^b_{1\nu}A^c_{2\mu}B^d_{\nu}
+\frac{1}{\alpha}A^a_{1\mu}B^b_{\nu}A^c_{2\mu}B^d_{\nu})\Big\},
\\\nn
S_{12}&=&\frac{g}{2}\int d^4x\Big\{\bar q_1(\fnot A_1+\fnot A_2)q_2+\bar q_2(\fnot A_1+\fnot A_2)q_1
+\bar q_1\fnot A_2q_1+\bar q_2\fnot A_1q_2
\\\label{app:Sint:S12} && -f^{abc}\Big[
(\partial_\mu A_{1\nu}^a)A_{2\mu}^b A_{2\nu}^c
+(\partial_\mu A_{2\nu}^a)A_{1\mu}^b A_{2\nu}^c
+(\partial_\mu A_{2\nu}^a)A_{2\mu}^b A_{1\nu}^c\Big]+\{1\leftrightarrow 2\}
\\\nn&&-gf^{abr}f^{rcd}\Big[
A_{1\mu}^aA_{1\nu}^bA_{2\mu}^cA_{2\nu}^d
+A_{1\mu}^aA_{1\nu}^bA_{1\mu}^cA_{2\nu}^d
\\\nn && \qquad\qquad
+A_{2\mu}^aA_{2\nu}^bA_{2\mu}^cA_{1\nu}^d
+A_{1\mu}^aA_{2\nu}^bA_{1\mu}^cA_{2\nu}^d
+A_{1\mu}^aA_{2\nu}^bA_{2\mu}^cA_{1\nu}^d
\Big]\Big\}.
\end{eqnarray}
The terms $\{1\leftrightarrow 2\}$ represent the previous terms in the line with the replacement $q_1\leftrightarrow q_2$ and $A_1\leftrightarrow A_2$. The factor $1/2$ in $S_{12}$  results from the symmetric  elimination of EOMs (with respect to $A_1\leftrightarrow A_2$). The Feynman rules for $S_{1q}$ and $S_{2q}$ interaction can be found in ref.~\cite{Abbott:1981ke}.

\section{Computation of the hard coefficient function}
\label{app:diag4}

In some cases, the computation in coordinate space is simpler than in momentum space. In particular, it is known that computations for higher-twists operators are always simpler in  position space (cf. e.g. \cite{Ji:2014eta} and \cite{Braun:2009vc}, sec. 4 and 5 of ref.~\cite{Braun:2021aon}). This is due to the fact that  momentum fractions are not sign-definite for higher twist observables, which leads to the necessity of a separate evaluation of loop integrals for each domain. The coordinate space does not have this problem, which is reflected in the different relative positioning of fields. 

There are many examples of computations with the background field for collinear factorization. The pedagogical introduction into such computation is given in ref.~\cite{Scimemi:2019gge}, and also in refs.~\cite{Balitsky:1987bk,Braun:2021aon}. However, the case of composite background fields seems to be completely new. The main problem here is that the presence of two (or more) directions makes loop-integrals very asymmetric and thus cumbersome to evaluate. In this appendix, we present a simple technique that bypasses these problems. The main structure of the computation follows the one described in app.~B of ref.~\cite{Scimemi:2019gge}. The technique is universal and can be applied to NLO (and with minimal modification to higher orders) computations of diagrams with an arbitrary number of external fields. We exemplarily show the computation of diagram 4. All other diagrams, including two-point diagrams and the computation of UV divergence in sec.~\ref{sec:UV}, are computed in the same manner.

Diagram 4 in the dimensional regularization (with $d=4-2\epsilon$) reads
\begin{eqnarray}\label{app:diag4:1}
\text{diag}_4&=&
-\(C_F-\frac{C_A}{2}\)\frac{ga_s}{2\cdot 4^\epsilon \pi^{3d/2}}\Gamma(1-\epsilon)\Gamma^3(2-\epsilon)
\\\nn &&
\int d^dx d^dy d^dz \frac{
\bar \xi_{\bar n}(x^- n)\gamma^\alpha (\fnot x-\fnot y)\fnot A_{\bar n}(y^-n) \fnot y \gamma^\mu \fnot z \gamma_{\alpha}\xi_{n}(z^+ \bar n)}{
[-(x-y)^2]^{2-\epsilon}
[-y^2]^{2-\epsilon}
[-z^2]^{2-\epsilon}
[-(x-z)^2]^{1-\epsilon}},
\end{eqnarray}
where only algebraic simplifications are made, and $+i0$ terms in propagators are omitted for brevity. To write this expression we used that the Feynman propagators in coordinate space are
\begin{eqnarray}\label{app:prop-q}
\Delta_q(x-y)&=&
\contraction{}{\bar \psi}{(x)}{\psi}
\psi(x)\bar \psi(y)=\frac{i\Gamma(2-\epsilon)}{2\pi^{d/2}}
\frac{\fnot x-\fnot y}{[-(x-y)^2+i0]^{2-\epsilon}},
\\
\Delta^{\mu\nu}_g(x-y)&=&
\contraction{}{A_\mu}{(x)}{A}
A_\mu(x)A_\nu(y)
=\frac{-\Gamma(1-\epsilon)}{4\pi^{d/2}}
\frac{g^{\mu\nu}}{[-(x-y)^2+i0]^{1-\epsilon}}.
\end{eqnarray}
The integration over each position in eq.~(\ref{app:diag4:1}) is asymmetric, in the sense that one of the components enters a background field and can not be integrated. To resolve this issue, we use the observation that the loop-integral with a spherically symmetric (over the integration variable) denominator automatically symmetrizes the numerator. 

Let us consider the part of the diagram which has the integral over $y$. It reads
\begin{eqnarray}
I_x&=& \Gamma^2(2-\epsilon) \int d^dy \frac{A_{\bar n}(y^- n)y^\alpha(x-y)^\beta}{[-(x-y)^2+i0]^{2-\epsilon}[-y^2+i0]^{2-\epsilon}}.
\end{eqnarray}
Joining the propagators by the Feynman trick, and shifting $y\to y+\alpha x$ we obtain
\begin{eqnarray}
I_x&=& \Gamma(4-2\epsilon)\int_0^1 d\alpha (\alpha \bar \alpha)^{1-\epsilon} \int d^dy \frac{A_{\bar n}(y^- n+\alpha x^- n)(y+\alpha x)^\alpha(\bar \alpha x-y)^\beta}{[-y^2-\alpha\bar \alpha x^2+i0]^{4-2\epsilon}},
\end{eqnarray}
where $\alpha$ is the Feynman variable. The denominator is spherically symmetric over $y$. The integral with such a denominator is
\begin{eqnarray}\label{app:diag4:loopi}
\int d^dy \frac{y^{\mu_1}...y^{\mu_{2n}}}{[-y^2-M^2+i0]^\lambda}=-i\pi^{d/2}\frac{\Gamma(\lambda-n+\epsilon-2)}{\Gamma(\lambda)}\frac{(-1)^n g_s^{\mu_1...\mu_{2n}}}{2^n[-M^2+i0]^{\lambda-n+\epsilon-2}},
\end{eqnarray}
where $g_s^{\mu_1...\mu_{2n}}$ is a totally symmetric combination of metric tensors. Next, we expand the background field in a series at $y=0$, and integrate this series term-by-term using eq.~(\ref{app:diag4:loopi}). The point is that the higher order terms integrate to null, because they are contracted with too many light-cone vectors, and only a few terms remains. We obtain 
\begin{eqnarray}\label{app:diag4:2}
I_x&=&-i\pi\int_0^1 d\alpha \Big(
\Gamma(2-\epsilon)\frac{x^\alpha x^\beta}{[-x^2+i0]^{2-\epsilon}}
\\\nn &&
+\frac{\Gamma(1-\epsilon)}{2}\frac{g^{\alpha\beta}+\alpha x^\alpha \bar n^\beta \partial_+-\bar \alpha \bar n^\alpha x^\beta \partial_+}{[-x^2+i0]^{1-\epsilon}}
-\frac{\Gamma(-\epsilon)}{4}\frac{\alpha \bar \alpha \bar n^\alpha \bar n^\beta \partial_+^2}{[-x^2+i0]^{-\epsilon}}\Big)A_{\bar n}(\alpha x^- n).
\end{eqnarray}
Note, that the fractional powers of $\alpha$ cancel. This property holds for all integrals of one-loop topology. The integration over $\alpha$ cannot be computed, and it is a part of integral convolution between the operator and the coefficient function.

Next, using the same trick, we integrate over the position $x$. The only difference is that we should expand the fields $\xi_{\bar n}$ and $A_{\bar n}$ in series. The resulting expression has the form
\begin{eqnarray}
\text{diag}_4&=& 
\(C_F-\frac{C_A}{2}\)\frac{ga_s}{4^\epsilon \pi^{d/2}}\Gamma(-\epsilon)\Gamma^2(2-\epsilon)\int d^dz \int_0^1 d\alpha \,d\beta \,\beta \, 
\bar \xi_{\bar n}(\beta z^- n)\Bigg\{
\\\nn &&
\frac{\bar n^\mu \gamma^\nu \overrightarrow{\partial_+}}{[-z^2+i0]^{1-2\epsilon}}
+4\epsilon(1-\epsilon)\frac{ z^\mu z^\nu \fnot z}{[-z^2+i0]^{3-2\epsilon}}+...
\Bigg\}
A_{\bar n}^\nu(\alpha \beta z^- n)\xi_n(z^+ \bar n),
\end{eqnarray}
where dots indicate a long term with different combinations of derivatives and vectors. In this expression we can integrate over the transverse components of $z$, using the integral
\begin{eqnarray}\label{app:diag4:loopit}
\int d^{d-2}z_T \frac{z_T^{\mu_1}...z_T^{\mu_{2n}}}{[-z^2+i0]^\lambda}=\pi^{\frac{d-2}{2}}\frac{\Gamma(\lambda-n+\epsilon-1)}{\Gamma(\lambda)}\frac{(-1)^n g_{T,s}^{\mu_1...\mu_{2n}}}{2^n[-2z^+z^-+i0]^{\lambda-n+\epsilon-1}},
\end{eqnarray}
where $g_{T,s}$ is the fully symmetric composition of $g_T$-tensors, defined as
\begin{eqnarray}\label{def:gT}
g_T^{\mu\nu}=g^{\mu\nu}-n^\mu \bar n^\mu-\bar n^\mu n^\nu.
\end{eqnarray}
After algebraic simplifications we obtain
\begin{eqnarray}\label{app:transverse-integral}
\text{diag}_4&=& g a_s \(C_F-\frac{C_A}{2}\)
\frac{\Gamma(-\epsilon)\Gamma(1-\epsilon)\Gamma(2-\epsilon)}{\Gamma(2-2\epsilon)}\int \frac{dz^+dz^-}{4^\epsilon\pi}\int_0^1 d\alpha \,d\beta 
\\\nn &&
\frac{\bar \xi_{\bar n}(\beta z^- n)\fnot A_{\bar n,T}(\alpha \beta z^- n)\xi_n(z^+ \bar n)}{[-2z^+z^-+i0]^{1-\epsilon}}\Bigg\{
z^- n^\mu \beta [2-\epsilon+2 z^-(\alpha \beta \partial_+^A-\bar \beta \partial_+^\xi)]
\\\nn &&
-z^+n^\mu \frac{\beta}{\epsilon}[3\epsilon^2+2 \bar \beta (1-\epsilon^2)z^-\partial_+^\xi
+2 (1-\epsilon)(\epsilon -\alpha \beta (1+\epsilon))z^-\partial_+^A]\Bigg\},
\end{eqnarray}
where $\partial_+^A$ ($\partial_+^\xi$) is the derivative acting on $A_{\bar n}$ $(\bar \xi_{\bar n})$. These derivatives can be rewritten as derivatives with respect to $\alpha$ and $\beta$, and then eliminated, integrating by parts. It leads to
\begin{eqnarray}
\text{diag}_4&=& g a_s \(C_F-\frac{C_A}{2}\)
\frac{\Gamma(-\epsilon)\Gamma(1-\epsilon)\Gamma(2-\epsilon)}{\Gamma(2-2\epsilon)}\int \frac{dz^+dz^-}{4^\epsilon\pi}\int_0^1 d\alpha \,d\beta 
\\\nn &&
\frac{1}{[-2z^+z^-+i0]^{1-\epsilon}}\Bigg\{
z^- n^\mu [2 \mathcal{K}(\beta,\beta)-\beta(2+\epsilon)\mathcal{K}(\beta,\alpha\beta)]
\\\nn &&
+z^+n^\mu\Big[
2(1-\epsilon) \mathcal{K}(\beta,0)+2 \frac{1-\epsilon}{\epsilon}\mathcal{K}(\beta,\beta)-\beta\frac{4-\epsilon^2}{\epsilon}\mathcal{K}(\beta,\alpha\beta)
\Big]\Bigg\},
\end{eqnarray}
where
\begin{eqnarray}
\mathcal{K}(s,t)=\bar \xi_{\bar n}(s z^- n)\fnot A_{\bar n,T}(tz^- n )\xi_{n}(z^+\bar n).
\end{eqnarray}
Finally, this expression can be turned into the universal form by rescaling $z^-$ such that $\bar \xi$ of $A_{\bar n}$ is located at $z^-n$. It allows to integrate over one of the Feynman variables and the result is
\begin{eqnarray}\label{app:diag4:3}
\text{diag}_4&=& g a_s \(C_F-\frac{C_A}{2}\)
\frac{\Gamma(-\epsilon)\Gamma(1-\epsilon)\Gamma(2-\epsilon)}{\Gamma(2-2\epsilon)}\int \frac{dz^+dz^-}{4^\epsilon\pi}\frac{1}{[-2z^+z^-+i0]^{1-\epsilon}}\int_0^1 ds
\\\nn &&
\Bigg\{
z^- n^\mu [-\frac{2}{\epsilon} \mathcal{K}(1,1)-\frac{2+\epsilon}{1-\epsilon}\mathcal{K}(1,s)]
+z^+n^\mu\Big[
\frac{2}{\epsilon}\mathcal{K}(1,1)+2\mathcal{K}(1,0)-\frac{2+\epsilon}{\epsilon}\mathcal{K}(1,s)
\Big]\Bigg\}.
\end{eqnarray}
Note, that the order of integration can be different. It leads to different intermediate expressions, but the final representation eq.~(\ref{app:diag4:3}) coincides.

Computing all other diagrams for the coefficient function of the effective operator and summing them together, we obtain eq.~(\ref{NLO:3point-diagrams}). The same technique, but with one light-cone direction, has been used to compute UV part of the TMD operators in sec.~\ref{sec:UV}.

\section{Evolution kernels in  momentum space}
\label{app:evol-momentum}

In this appendix, we present the evolution kernels eq.~(\ref{def:Hhat}-\ref{def:He}) in  momentum space. In the formulas below the variables $x_1$ and $x_2$ are Fourier conjugated to $z_1$ and $z_2$, as it is defined in eq.~(\ref{def:U2-momentum}). We obtain the following expressions
\begin{eqnarray}\nn
&&\widehat{\mathcal{H}}U(x_1,x_2)=
\int dv \Bigg\{\frac{x_1}{x_1+v}\frac{1}{v}\(U(x_1,x_2)-\frac{x_1}{x_1+v}U(x_1+v,x_2-v)\)
\(\theta(v,x_1)-\theta(-v,-x_1)\)
\\ &&
\qquad
+ \frac{x_2}{x_2+v}\frac{U(x_1,x_2)-U(x_1-v,x_2+v)}{v}
\(\theta(v,x_2)-\theta(-v,-x_2)\)\Bigg\}+\delta_{x_1}U(0,x_2),
\\
&&\mathcal{H}^+U(x_1,x_2)=
\int \frac{dv}{2(x_2+v)} \Bigg\{
\(1-\(\frac{x_1}{x_1+x_2}\)^2\)
\(\theta(v,x_2)-\theta(-v,-x_2)\)
\\\nn &&\qquad
+
\(\(\frac{x_1}{x_1+x_2}\)^2-\(\frac{x_1}{v-x_1}\)^2\)
\(\theta(v,-x_1)-\theta(-v,x_1)\)\Bigg\}U(x_1-v,x_2+v),
\\
&&\mathcal{H}^-U(x_1,x_2)=
\int \frac{dv}{2(x_1+v)} \Bigg\{
\frac{-x_1^2}{(x_1+x_2)^2}\(\theta(v,x_1)-\theta(-v,-x_1)\)
\\\nn &&\qquad
+
\(\(\frac{x_1}{x_1+x_2}\)^2-\(\frac{v}{v-x_2}\)^2\)
\(\theta(v,-x_2)-\theta(-v,x_2)\)\Bigg\}U(x_2-v,x_1+v),
\\
&&P_{12}\mathcal{H}^{e(1)}U(x_1,x_2)=
\int dv \
\frac{x_2}{(x_2+v)^2}
\(\theta(v,x_2)-\theta(-v,-x_2)\)U(x_2+v,x_1-v),
\end{eqnarray}
where 
\begin{eqnarray}
\theta(a,b)=\left\{
\begin{array}{ll}
    1, & a>0 \text{~and~} b>0, \\
    0, & a\leqslant 0 \text{~or~} b\leqslant 0,
\end{array}\right.
\qquad
\delta_a=
\left\{
\begin{array}{ll}
    1, & a=0,\\
    0, & a\neq 0.
\end{array}\right.
\end{eqnarray}
The evolution kernels preserve the total momentum passing though the diagrams (i.e. $x_1+x_2$). However, they do not preserve the sign of individual components. Summing together these kernels we obtain the expression for the kernels $\mathbb{H}_1$ and $\mathbb{H}_2$ (\ref{def:H1}, \ref{def:H2}),
\begin{eqnarray}\label{app:H1-momentum}
2\mathbb{H}_1U(x_1,x_2)&=&C_A \delta_{x_1}U(0,x_2)+\int dv 
\Bigg\{
C_A(\theta(v,x_1)-\theta(-v,-x_1))\frac{x_1}{x_1+v}
\Big[
\\\nn &&\frac{U(x_1,x_2)-U(x_1+v,x_2-v)}{v} +\(\frac{x_2}{x_1+v}-\frac{x_1}{x_1+x_2}\)\frac{U(x_1+v,x_2-v)}{x_1+x_2}\Big]
\\\nn &&
+C_A(\theta(v,x_2)-\theta(-v,-x_2))\frac{x_2}{x_2+v}
\Big[
\\\nn && \frac{U(x_1,x_2)-U(x_1-v,x_2+v)}{v}-\frac{2x_1+x_2}{(x_1+x_2)^2}U(x_1-v,x_2+v)\Big]
\\\nn &&
+2\(C_F-\frac{C_A}{2}\)\frac{1}{x_1+v}\Big[\frac{-x_1^2}{(x_1+x_2)^2}
\(\theta(v,x_1)-\theta(-v,-x_1)\)
\\\nn &&
+
\(\(\frac{x_1}{x_1+x_2}\)^2-\(\frac{v}{x_2-v}\)^2\)
\(\theta(v,-x_2)-\theta(-v,x_2)\)\Big]U(x_2-v,x_1+v)\Bigg\},
\\
\label{app:H2-momentum}
2\mathbb{H}_2U(x_1,x_2)&=&C_A \delta_{x_1}U(0,x_2)+\int dv 
\Bigg\{
C_A(\theta(v,x_1)-\theta(-v,-x_1))\frac{x_1}{x_1+v}
\Big[
\\\nn &&\frac{U(x_1,x_2)-U(x_1+v,x_2-v)}{v} +\frac{U(x_1+v,x_2-v)}{x_1+v}\Big]
\\\nn &&
+C_A(\theta(v,x_2)-\theta(-v,-x_2))\frac{x_2}{x_2+v}
\frac{U(x_1,x_2)-U(x_1-v,x_2+v)}{v}
\\\nn &&
-\(C_F-\frac{C_A}{2}\)\frac{x_2}{(x_2+v)^2}\(\theta(v,x_2)-\theta(-v,-x_2)\)U(x_2+v,x_1-v)
\Bigg\}.
\end{eqnarray}

In the literature, one can find a different set of variables used for the momentum fractions. Namely, $x_1=r x$ and $x_2=r \bar x$. The value of $r=x_1+x_2$ is preserved by the evolution kernels, and drops out of the expressions. In these variables, the evolution kernels have the form (for $r>0$)
\begin{eqnarray}\nn
2\mathbb{H}_1\widehat{U}(x)&=&\int dy 
\Bigg\{
C_A(\theta(y-x,x)-\theta(x-y,-x))
\frac{x}{y}\Big[
\frac{\widehat{U}(x)-\widehat{U}(y)}{y-x}
+\frac{1-x-xy}{y}\widehat{U}(y)\Big]
\\ &&\label{app:H1-beneke}
+C_A(\theta(x-y,\bar x)-\theta(y-x,-\bar x))\frac{\bar x}{\bar y}\Big[
\frac{\widehat U(x)-\widehat U(y)}{x-y}-(1+x)\widehat{U}(y)\Big]
\\\nn &&
+2\(C_F-\frac{C_A}{2}\)\Big[-\frac{x^2}{\bar y}
\(\theta(\bar y-x,x)-\theta(x-\bar y,-x)\)
\\\nn &&
+\frac{\bar x(1-x-y-xy)}{y^2}
\(\theta(x-\bar y,\bar x)-\theta(\bar y-x,-\bar x)\)\Big]\widehat{U}(y)\Bigg\}+C_A \delta_{x}\widehat{U}(0),
\\\label{app:H2-beneke}
2\mathbb{H}_2\widehat{U}(x)&=&\int dy 
\Bigg\{
C_A(\theta(y-x,x)-\theta(x-y,-x))\frac{x}{y^2}
\frac{y\widehat{U}(x)-x\widehat{U}(y)}{y-x}
\\\nn &&
+C_A(\theta(x-y,\bar x)-\theta(y-x,-\bar x))\frac{\bar x}{\bar y}
\frac{\widehat{U}(x)-\widehat{U}(y)}{x-y}
\\\nn &&
-\(C_F-\frac{C_A}{2}\)\frac{\bar x}{y^2}\(\theta(x-\bar y,\bar x)-\theta(\bar y-x,-\bar x)\)\widehat{U}(y)
\Bigg\}+
C_A \delta_{x}\widehat{U}(0).
\end{eqnarray}
Here we denote
\begin{eqnarray}
\widehat U(x)=U(r x,r\bar x).
\end{eqnarray}
The expressions (\ref{app:H1-beneke}, \ref{app:H2-beneke}) coincide with the one computed in ref.~\cite{Beneke:2017ztn}, cf. eq.~(83). We stress that the value of $y$ is not restricted to $[0,1]$ as it is supposed in ref.~\cite{Beneke:2017ztn}.

\begin{eqnarray}
\end{eqnarray}

\normalbaselines 
\bibliography{bibFILE}
\end{document}